\documentclass[12pt]{spieman}  % 12pt font required by SPIE;
\usepackage{amsmath,amsfonts,amssymb}
\usepackage{graphicx}
\usepackage{setspace}
\usepackage{tocloft}
\usepackage{caption}
\usepackage{ifsym}
\usepackage{wasysym}
\usepackage{fontawesome}
\usepackage{tcolorbox}
\usepackage{enumitem}

\usepackage{natbib}
\usepackage{hyperref}
\usepackage{color}
\usepackage{pdfcomment}
\usepackage{ulem}
\usepackage{contour}
\usepackage{soul}
\usepackage{longtable}

\usepackage{array,etoolbox}
\preto\tabular{\setcounter{magicrownumbers}{0}}
\newcounter{magicrownumbers}

\contourlength{0.8pt}

\title{Maximizing LSST Solar System Science: Approaches, Software Tools, and Infrastructure Needs}

\author[a,b]{Henry H.\ Hsieh}     % ORCID: 0000-0001-7225-9271
\author[c]{Michele T.\ Bannister} % ORCID: 0000-0003-3257-4490
\author[d,e]{Bryce T.\ Bolin} %ORCID: 0000-0002-4950-6323
\author[f]{Josef \v{D}urech}
\author[d]{Siegfried Eggl} %ORCID: 0000-0002-1398-6302
\author[g]{Wesley C.\ Fraser}     % ORCID: 0000-0001-6680-6558
\author[h,i]{Mikael Granvik} % ORCID: 0000-0002-5624-1888
\author[j]{Michael S.\ P.\ Kelley}    % ORCID: 0000-0002-6702-7676
\author[j]{Matthew M.\ Knight}    % ORCID: 0000-0003-2781-6897
\author[k]{Rodrigo Leiva}  % ORCID: 0000-0002-6477-1360
\author[l,m]{Marco Micheli}   % ORCID: 0000-0001-7895-8209
\author[d]{Joachim Moeyens} % ORCID: 0000-0001-5820-3925
\author[n]{Michael Mommert}  % ORCID: 0000-0002-8132-778X
\author[o]{Darin Ragozzine}   % ORCID: 0000-0003-1080-9770
\author[p]{Cristina A.\ Thomas}   % ORCID: 0000-0003-3091-5757
\affil[a]{Planetary Science Institute, 1700 East Fort Lowell Rd., Suite 106, Tucson, AZ 85719, USA}
\affil[b]{Institute of Astronomy and Astrophysics, Academia Sinica, P.O.\ Box 23-141, Taipei 10617, Taiwan}
\affil[c]{Astrophysics Research Centre, Queen's University Belfast, Belfast BT7 1NN, United Kingdom}
\affil[d]{Department of Astronomy and the DIRAC Institute, University of Washington, 3910 15th Avenue NE, Seattle, WA 98195, USA}
\affil[e]{B612 Asteroid Institute, 20 Sunnyside Ave, Suite 427, Mill Valley, CA 94941, USA}
\affil[f]{Astronomical Institute, Faculty of Mathematics and 
Physics, Charles University, V Hole\v{s}ovi\v{c}k\'ach 2, 180\,00 Prague 8, Czech Republic}
\affil[g]{Herzberg Institute of Astrophysics, National Research Council of Canada, 5071 West Saanich Road, Victoria, BC V9E 2E7, Canada}
\affil[h]{Department of Physics, P.O. Box 64, FI-00014 University of Helsinki, Finland}
\affil[i]{Division of Space Technology, Lule\aa{} University of Technology, Box 848, S-981 28 Kiruna, Sweden}
\affil[j]{Department of Astronomy, University of Maryland, 1113 Physical Sciences Complex, Building 415, College Park, MD 20742, USA}
\affil[k]{Southwest Research Institute, 1050 Walnut Street, Suite 300, Boulder, CO 80302, USA}
\affil[l]{ESA NEO Coordination Centre, Largo Galileo Galilei, 1, 00044 Frascati (RM), Italy}
\affil[m]{INAF - Osservatorio Astronomico di Roma, Via Frascati, 33, 00040 Monte Porzio Catone (RM), Italy}
\affil[n]{Lowell Observatory, 1400 W.\ Mars Hill Rd, Flagstaff, AZ 86001, USA}
\affil[o]{Brigham Young University, Department of Physics and Astronomy, N283 ESC, Provo, UT 84602, USA}
\affil[p]{Northern Arizona University, Department of Physics and Astronomy, PO Box 6010, Flagstaff, AZ 86011, USA}

\cftpagenumbersoff{figure}
\cftpagenumbersoff{table} 
\begin{document}
\maketitle

%\correspondingauthor{Henry Hsieh}
%\email{hhsieh@psi.edu}

%\author[0000-0001-7225-9271]{Henry H.\ Hsieh}
%\affil{Planetary Science Institute, 1700 East Fort Lowell Rd., Suite 106, Tucson, AZ 85719, USA}
%\affil{Institute of Astronomy and Astrophysics, Academia Sinica, P.O.\ Box 23-141, Taipei 10617, Taiwan}

%\author[0000-0003-3257-4490]{Michele T. Bannister}
%\affiliation{Astrophysics Research Centre, Queen's University Belfast, Belfast BT7 1NN, United Kingdom}

%\author[0000-0001-6680-6558]{Wesley C. Fraser} 
%\affiliation{Astrophysics Research Centre, Queen's University Belfast, Belfast BT7 1NN, United Kingdom}

%\author{Michael S.\ Kelley}
%\affil{Department of Astronomy, University of Maryland, 1113 Physical Sciences Complex, Building 415, College Park, MD 20742, USA}

%\author{Matthew M.\ Knight}
%\affil{Department of Astronomy, University of Maryland, 1113 Physical Sciences Complex, Building 415, College Park, MD 20742, USA}

\begin{abstract}
The Large Synoptic Survey Telescope (LSST) is expected to increase known small solar system object populations by an order of magnitude or more over the next decade, enabling a broad array of transformative solar system science investigations to be performed.  In this white paper, we discuss software tools and infrastructure that we anticipate will be needed to conduct these investigations and outline possible approaches for implementing them.
Feedback from the community or contributions to future updates of this work are welcome. Our aim is for this white paper to encourage further consideration of the software development needs of the LSST solar system science community, and also to be a call to action for working to meet those needs in advance of the expected start of the survey in late 2022.
\looseness=-1
\end{abstract}

%\keywords{comets: general; Kuiper belt: general; minor planets, asteroids: general; methods: data analysis; methods: observational; techniques: image processing; techniques: photometric; telescopes; catalogs; surveys}

\section{Background}\label{section:intro}

\subsection{The LSST Survey and the LSST Solar System Science Collaboration}\label{subsection:intro_lsst}

The Large Synoptic Survey Telescope (LSST) is a 8.4~m optical survey telescope currently being constructed at Cerro Pach\'on in Chile \citep[][]{ivezic2008_lsst}.  
Science verification is expected to begin in mid-2021, with the main survey starting in late 2022. % MB: first minor planets arrive earlier thanks to the commissioning data + auxiliary telescope 
%\hhh{define survey components, including wide-fast-deep, etc.} MB: there's thinking it may not be called the wide-fast-deep, just Main Survey
The LSST's main survey will repeatedly cover 18,000 deg$^{2}$ of the southern sky (declinations of $\delta<0^{\circ}$) in six filters ($ugrizy$) over the course of a decade. About 10\% of the telescope's time will be used for other programs, such as a potential survey of the portion of the ecliptic that lies north of $\delta$$\,=\,$0$^{\circ}$ (``the North Ecliptic Spur'').
\looseness=-1
%, and various individual 9.6 deg$^{2}$ Deep Drilling Fields.
%LSST survey data will be immediately available with no proprietary period to the United States and Chilean astronomy communities as well as various other international and institutional partners.  In addition, certain data such as transient detections released as alerts by the LSST pipeline will be immediately available for use by the worldwide scientific community with no proprietary period.
%\hhh{modify/supplement last two sentences to summarize latest data rights policy document when available}

The LSST survey is expected to increase known populations of small solar system objects by an order of magnitude or more, and to enable a broad array of transformative solar system science investigations\footnote{LSST Science Book, v2.0; {\tt http://www.lsst.org/scientists/scibook}}. 
The discovery of new objects, computation of orbits, and basic data analysis should be largely handled by the LSST's Moving Object Processing System (MOPS) and the Minor Planet Center (MPC).  
To realize the LSST's full potential for solar system science, however, substantial development of additional solar-system-specific software is also needed.
\looseness=-1

%\subsection{The }\label{subsection:intro_sssc}

The LSST Solar System Science Collaboration (SSSC) has been organized to coordinate activities among solar system scientists in the community of LSST data rights holders.  Within the SSSC, a Community Software and Infrastructure Development working group has been created to coordinate community efforts specifically related to the development of computational tools necessary to do solar system science with LSST survey data. New members to the SSSC and the software working group are welcome at any time, where applications to join can be made via the SSSC website\footnote{\tt http://lsst-sssc.github.io/}.
\looseness=-1
%and indicating interest in the software working group on the membership application.

\subsection{Planned Project-Provided Data Products}\label{subsection:intro_promptproducts_drproducts}

The LSST will deliver three levels of data products and services, as described in the project's Data Products Definition Document\footnote{{\tt http://www.lsst.org/content/data-products-definition-document}}.  Prompt Products will be produced nightly and will include images, difference images, catalogs of sources and objects detected in difference images, and catalogs of solar system objects, and are aimed at enabling rapid follow-up of time-domain events.  Data Release Products will be produced or updated annually, and will include well-calibrated single-epoch images, deep co-added images, and catalogs of objects, sources, and forced sources.
\looseness=-1

Prompt Products will include a substantial amount of data relevant to solar system science.  The planned LSST pipeline will identify detections of known and new solar system objects, and deliver calibrated photometry, point-spread-function (PSF) parameters, heliocentric orbital elements, phase function parameters, and certain observing geometry parameters.  It will also support transient observation alerts.  Some Data Release Products will also be relevant to solar system science, such as
%multi-aperture surface brightness measurements, which will be useful for comet studies, and 
cumulative orbit catalogs of solar system objects,
%(both previously known and newly discovered) 
based on detections made by LSST up to the time of each data release, allowing for survey efficiency and debiasing studies.
%However, these products will likely follow the release schedule of other Data Release Products, i.e., twice in the first year of the survey and then annually afterwards, as the LSST project specifications currently classify such measurements as non-time-sensitive.
\looseness=-1

\subsection{User-Generated Software and Data Products}\label{subsection:intro_ugproducts}

The LSST Science Platform will also facilitate creation of so-called User Generated Products by running user-developed tools that benefit from co-location with data within the LSST Archive Center.
%Possible User Generated Products relevant to solar system science include geometric and orbital parameters beyond those currently planned as Prompt Products (e.g., true anomalies, orbit plane angles, and barycentric orbital elements), custom measurements of image data (e.g., real-time multi-aperture, multi-azimuth photometry, and advanced activity detection parameters), and higher order data products (e.g., taxonomic classifications, rotation periods, and asteroid family associations).
Instead of individual scientists working completely independently, it would be beneficial for the community to minimize redundant effort by coordinating software development efforts, and providing data products of common interest as communally available User Generated Products.  Public availability of higher-order data products will also be useful for education and public outreach efforts, enabling educators, students, and citizen scientists to work on a broader array of scientific projects and topics than would be possible from LSST-provided data alone.
%To have maximum scientific and educational impact, these data products should be easily accessible by users who do not have an expert-level understanding of the LSST system.  
\looseness=-1

This white paper aims to lay out the expected software- and infrastructure-related needs of solar system scientists wishing to do science with LSST data and document ideas for addressing those needs. It follows on and expands upon the recently published SSSC software development roadmap \citep{schwamb2018_ssroadmap}. We intend it to be a living document.  Feedback from the community, or contributions to future updates of this work, are very welcome.
\looseness=-1

In this white paper, we focus on analysis of data from the LSST's Main Survey (also known as the Wide-Fast-Deep Survey), which will comprise the majority of LSST data. Many of the tools discussed will also be useful for analyzing data from other survey components.  
We omit discussion of the discovery of new objects from single exposures and orbit determination, which is expected to be handled by the LSST Project and the MPC,
%Our objective here is to attempt to lay out as all-inclusive a list as we can of software tools and infrastructure needs that we anticipate LSST solar system science will require.
%We note that not all of the tools listed here ultimately will be or should be developed by members of the SSSC on behalf of the community.
but retain discussion of some tools that overlap the scope of LSST Project and the MPC, to ensure that they are not overlooked if not provided as expected.
Software tools for which only narrow community interest exists, or for which individuals have highly specific preferences, are similarly listed here for completeness.
\looseness=-1

%%%%%%%%%%%%%%% SPECIFIC SOFTWARE DEVELOPMENT NEEDS %%%%%%%%%%%%%%%

\section{Specific Software and Infrastructure Development Needs}\label{section:swtools}

The SSSC has published a science roadmap \citep{schwamb2018_ssroadmap} listing ranked priorities of current SSSC members in four areas in solar system science: active objects, near-Earth objects, the inner solar system, and the outer solar system.
%Achieving most of these objectives will require large-scale analyses of LSST data, involving either the direct analysis of Prompt Products or Data Release Products, or the production and analysis of higher-order User-Generated data products.
%As discussed above, while software tools to perform such tasks could be developed by individual scientists for personal use, it would be worthwhile for the SSSC to minimize unnecessarily redundant effort by developing some of these tools of behalf of the community.
Reviewing the individual priorities listed in this roadmap, we have identified component tools and needs that will be required to achieve those objectives. Many tools and needs are relevant to multiple science priorities. We have also identified tools and needs relevant to facilitating solar system science with LSST data in general.  Key aspects of each are summarized in Table~\ref{table:swneeds_summary}, with detailed discussions in Appendix~\ref{section:appendix_detailed_sw_descriptions} and details of the mapping of software requirements to science roadmap priorities in Appendix~\ref{section:appendix_swneeds_sciencearea}.  

Software tasks and infrastructure needs are grouped here by approximate subject areas, and no prioritization is meant to be implied by their ordering.
%Prioritization of software requirements, as voted on by current members of the SSSC, is addressed in the SSSC Software Development Roadmap (Schwamb et al., in prep.).  
Some tools will require other higher-order data products as input, though, which affects their priority for development.  Highly time-sensitive tasks are also considered more urgent as they ideally should be ready by the start of the survey (cf.\ Section~\ref{subsection:swneeds_timescales}).
Dependencies are noted in  Appendix~\ref{section:appendix_detailed_sw_descriptions}, as applicable.

\begin{enumerate}

\item{{\bf Orbital object and detection parameter computation (Appendix~\ref{subsection:appendix_orbital_parameters}):} Computation of observational geometric parameters (e.g., antisolar vector position angles) and orbital properties (e.g., barycentric orbital elements) that will not be provided by the LSST Project.
%Needed for detection and characterization of activity and outbursts, and dynamical characterization of outer solar system objects.
}

\item{{\bf Orbital element and ephemeris uncertainty characterization (Appendix~\ref{subsection:appendix_uncertainty_characterization}):} Accurate characterization of orbital element and ephemeris uncertainties, to the extent (if any) that this is not adequately handled by the MPC.
%\hhh{specify interest is in detailed posteriors, etc.}
%Needed for faint object detection tools for LSST and priorities requiring follow-up observations.
}

\item{{\bf Extended object astrometry (Appendix~\ref{subsection:appendix_extended_object_astrometry}):} Determination of accurate astrometry for detections of extended sources to improve orbit determination for active objects.}

\item{{\bf Faint precovery and recovery identification (Appendix~\ref{subsection:appendix_lsst_precoveries_recoveries}):} Identification in LSST data of previously unidentified precovery detections and recovery detections fainter than LSST's 5-$\sigma$ detection threshold for newly discovered objects.
%Needed for orbit refinement for NEOs and other faint solar system objects.
}

\item{{\bf Extremely faint object detection (Appendix~\ref{subsection:appendix_faint_object_detection}):} Detection of objects fainter than the LSST's single-visit detection limit using methods such as shifting and stacking to search for objects with known orbits, or ``synthetic tracking'' to search for new objects.}

\item{{\bf Advanced moving object detection (Appendix~\ref{subsection:appendix_advanced_moving_object_detection}):} Detection of moving objects using algorithms other than that used by the current MOPS pipeline to improve detection rates of fast-moving near-Earth objects (NEOs) and extremely distant outer solar system objects.}

%\item{{\bf Fast- and slow-moving object detection (Appendix~\ref{subsection:appendix_slow_mover_detection}):}  Detection of small solar system bodies with non-sidereal rates slower than the detection limit of the default MOPS pipeline.  Needed for the detection of extremely distant outer solar system objects.}

\item{{\bf Phase function characterization (Appendix~\ref{subsection:appendix_phase_function_characterization}):} Determination of best-fit phase function parameters using a variety of phase function models, in addition to the filter-specific $H$,$G_{12}$ parameters expected to be provided by the LSST Project.
%Needed for physical characterization of small solar system bodies, and activity detection and characterization.
}

\item{{\bf Compositional characterization (Appendix~\ref{subsection:appendix_compositional_characterization}):} Determination of colors, likely taxonomic classifications, hydration states, and other compositional characteristics of solar system objects using phase function characterization results.}

\item{{\bf Rotational characterization (Appendix~\ref{subsection:appendix_rotational_characterization}):} Determination of properties related to an object's shape and spin state, e.g., rotational periods, axis ratios, and pole orientations.}

\item{{\bf Detection and characterization of multi-object systems (Appendix~\ref{subsection:appendix_multi_object_systems}):} Detection and characterization of resolved multi-object systems via detailed PSF analysis.}
% \dr{just clarifying what "resolved" means. Should clarify: does "characterization" mean solving for the mutual orbit parameters? Should Table 1 say "characterization" instead of "detection"}
% \hhh{changed Table 1 to "characterization"; characterization could mean solving for mutual orbit parameters, or could also just be easier stuff like angular separation or inferred [minimum] physical separation of components}

\item{{\bf Activity detection (Appendix~\ref{subsection:appendix_activity_detection}):} Detection of comet-like gas or dust emission activity using a variety of methods.}
%Needed for the detection of activity as well as the determination of physical properties requiring the exclusion of data where activity is present.}

\item{{\bf Activity characterization (Appendix~\ref{subsection:appendix_activity_characterization}):} Characterization of the strength, morphology, and other properties of visible comet-like features of active objects.
%Needed for the characterization of comet-like activity and its evolution.
}

\item{{\bf Outburst and disruption detection (Appendix~\ref{subsection:appendix_outburst_disruption_detection}):}  Detection of cometary outbursts or disruptions on asteroids via searches for activity or sudden brightening or fading events.\looseness=-1}

%\item{{\bf Asteroid pair identification (Appendix~\ref{subsection:appendix_asteroid_pairs}):}  }

\item{{\bf Advanced dynamical characterization (Appendix~\ref{subsection:appendix_advanced_dynamical_characterization}):}  Use of numerical integrations for computation of synthetic proper elements, and advanced dynamical characterization or classification of solar system bodies, including resonant objects, comets, and Centaurs.}
% \dr{Do we really need "outer". You don't want Mars Trojans identified, for example? What dynamical classification do inner solar system folks want? Since synthetic proper elements require some dynamical integration, I added them here.}
% \hhh{Resonant outer solar system object identification was specifically mentioned in the science roadmap and was discussed at the first readiness sprint; does identification of Mars Trojans require dynamical integrations?}
% \dr{Yes, resonances (including Trojans) typically require dynamical integrations of some kind.}

\item{{\bf Dynamical clustering identification (Appendix~\ref{subsection:appendix_dynamical_clustering_identification}):}  Identification of associations of small solar system bodies detected by LSST with known or new asteroid families, and identification of asteroid pairs or larger multi-object groupings.}
% \dr{I don't think "dynamical association identification" quite works here. Maybe "dynamical clustering analyses" since this is really about looking for dynamical groups. It's also not clear whether you mean a detailed study of asteroid families (including searching for new families) or a quick guess classification (doing better than a guess will be really hard in the LSST era). Note that this also requires proper elements, in general.}

%\item{{\bf Detection and characterization of non-gravitational perturbations (Appendix~\ref{subsection:appendix_nongrav_perturbations}):} Detection and chararacterization of the effects of non-gravitational perturbations on the orbits or spin states of small solar system bodies.}

%\item{{\bf Detection and characterization of mutual gravitational interactions (Appendix~\ref{subsection:appendix_mutual_interactions}):}  Detection and characterization of changes in the orbits of small bodies as a result of extremely close encounters with other small bodies to measure masses and bulk densities.}

%\item{{\bf Identification of resonant outer solar system objects (Appendix~\ref{subsection:appendix_resonant_oss_objects}):}  Identification of outer solar system objects in mean-motion resonances with the giant planets. Identified by the SSSC as an outer solar system science priority.}

\item{{\bf Occultation event prediction (Appendix~\ref{subsection:appendix_occultation_prediction}):}  Prediction of upcoming opportunities to observe occultations of background stars by small solar system bodies.}

\item{{\bf Detection of changes in physical properties (Appendix~\ref{subsection:appendix_physical_evolution}):} Detection and characterization of changes in physical properties (e.g., as a result of resurfacing events).}

\item{{\bf Detection of changes in dynamical properties (Appendix~\ref{subsection:appendix_dynamical_evolution}):} Detection and characterization of changes in dynamical properties as a result of non-gravitational perturbations or close encounters between small bodies.}

\item{{\bf System validation (Appendix~\ref{subsection:appendix_system_validation}):} Validation of satisfactory performance of the LSST system, to the extent (if any) that this is not within the scope of the LSST Project, including verification of satisfactory detection efficiencies for solar system populations.}

\item{{\bf Survey efficiency characterization and debiasing (Appendix~\ref{subsection:appendix_survey_debiasing}):}  Characterization of observational biases of the LSST survey to infer the true physical and dynamical properties of small solar system body populations.}

\item{{\bf Alert brokering and follow-up observation management (Appendix~\ref{subsection:appendix_alerts}):} Rapid identification of detections of interest, distribution of alerts for such detections to interested community members, and automated prioritization of targets for follow-up observations.}

\item{{\bf External data incorporation (Appendix~\ref{subsection:appendix_externaldata}):} Location, measurement and incorporation of both external archival data (e.g., serendipitous observations containing potential precoveries) and follow-up data, to facilitate orbit computation and data product generation. %Needed for priorities requiring use of previously obtained or follow-up observations by non-LSST facilities.
%\hhh{make sure we mention how much of this will be handled by MPC; maybe emphasize data mining of archival data since MPC will definitely not do this}
}

\item{{\bf Data access and visualization (Appendix~\ref{subsection:appendix_dataaccess}):} Infrastructure and tools to facilitate user interactions with LSST data products, such as browsing, application programming interface (API) interaction, image retrieval, data visualization, and follow-up observation planning.}

\end{enumerate}

\setlength{\tabcolsep}{4pt}
\renewcommand*{\arraystretch}{1.2}
\begin{table}[ht!]
\captionsetup{font=normalsize}
\centering
\caption{Summary of Software and Infrastructure Needs by Science Area}
\label{table:swneeds_summary}
\begin{tabular}{p{3.35in}ccccc}
\hline\hline
Software Task & Active$^{a}$ & NEOs$^{b}$ & ISS$^{c}$ & OSS$^{d}$ & Notes$^{e}$ \\
\hline\hline
\hangindent=.60cm\hangafter=1 1. Orbital parameter computation (\ref{subsection:appendix_orbital_parameters})
  & \checkmark & \checkmark & \checkmark & \checkmark & \faExclamationCircle\,---\,\faCopy \\
\hangindent=.60cm\hangafter=1 2. Orbital uncertainty characterization (\ref{subsection:appendix_uncertainty_characterization})
  & \checkmark & \checkmark & \checkmark & \checkmark & \faExclamationCircle\,---\,--- \\
\hangindent=.60cm\hangafter=1 3. Extended object astrometry (\ref{subsection:appendix_extended_object_astrometry})
  & \checkmark & --- & --- & --- & \faExclamationCircle\,\faCamera\,--- \\
\hangindent=.60cm\hangafter=1 4. Faint precovery/recovery identification (\ref{subsection:appendix_lsst_precoveries_recoveries})
  & \checkmark & \checkmark & \checkmark & \checkmark & \faExclamationCircle\,---\,--- \\
\hangindent=.60cm\hangafter=1 5. Extremely faint object detection (\ref{subsection:appendix_faint_object_detection})
  & \checkmark & \checkmark & \checkmark & \checkmark & \faClockO\,\faCamera\,--- \\
\hangindent=.80cm\hangafter=1 6. Advanced moving object detection (\ref{subsection:appendix_advanced_moving_object_detection})
  & --- & \checkmark & --- & \checkmark & \faExclamationCircle\,---\,--- \\
%\hangindent=.60cm\hangafter=1 7. Slow-moving object detection (\ref{subsection:appendix_slow_mover_detection})
%  & --- & --- & --- & \checkmark & \faClockO\,\faCamera\,--- \\
\hangindent=.60cm\hangafter=1 7. Phase function characterization (\ref{subsection:appendix_phase_function_characterization})
  & \checkmark & \checkmark & \checkmark & \checkmark & \faClockO\,---\,\faCopy \\
\hangindent=.60cm\hangafter=1 8. Compositional characterization (\ref{subsection:appendix_compositional_characterization})
  & \checkmark & \checkmark & \checkmark & \checkmark & \faClockO\,---\,\faCopy \\
\hangindent=.60cm\hangafter=1 9. Rotational characterization (\ref{subsection:appendix_rotational_characterization})
  & \checkmark & \checkmark & \checkmark & \checkmark & \faClockO\,---\,\faCopy \\
\hangindent=.60cm\hangafter=1 10. Multi-object system characterization (\ref{subsection:appendix_multi_object_systems})
  & \checkmark & \checkmark & \checkmark & \checkmark & \faClockO\,\faCamera\,--- \\
\hangindent=.80cm\hangafter=1 11. Activity detection (\ref{subsection:appendix_activity_detection})
  & \checkmark & \checkmark & \checkmark & --- & \faExclamationCircle\,\faCamera\,--- \\
\hangindent=.80cm\hangafter=1 12. Activity characterization (\ref{subsection:appendix_activity_characterization})
  & \checkmark & \checkmark & \checkmark & --- & \faExclamationCircle\,\faCamera\,--- \\
\hangindent=.80cm\hangafter=1 13. Outburst and disruption detection (\ref{subsection:appendix_outburst_disruption_detection})
  & \checkmark & \checkmark & \checkmark & --- & \faExclamationCircle\,\faCamera\,--- \\
%\hangindent=.80cm\hangafter=1 15. Dynamical association identification (\ref{subsection:appendix_asteroid_pairs})
%  & \checkmark & --- & \checkmark & --- & \faClockO\,---\,--- \\
\hangindent=.80cm\hangafter=1 14. Advanced dynamical characterization (\ref{subsection:appendix_advanced_dynamical_characterization})
  & \checkmark & --- & --- & \checkmark & \faClockO\,---\,--- \\
\hangindent=.80cm\hangafter=1 15. Dynamical clustering identification (\ref{subsection:appendix_dynamical_clustering_identification})
  & \checkmark & --- & \checkmark & --- & \faClockO\,---\,--- \\
%\hangindent=.80cm\hangafter=1 Non-grav.\ perturbation detection %(\ref{subsection:appendix_nongrav_perturbations})
%  & \checkmark & \checkmark & --- & --- & \faCalendar\,---\,\faCopy \\
%\hangindent=.80cm\hangafter=1 Mutual interaction characterization (\ref{subsection:appendix_mutual_interactions})
%  & --- & --- & \checkmark & --- & \faCalendar\,---\,\faCopy \\
%\hangindent=.80cm\hangafter=1 19. Resonant OSS object detection (\ref{subsection:appendix_resonant_oss_objects})
%  & --- & --- & --- & \checkmark & \faCalendar\,---\,--- \\
\hangindent=.80cm\hangafter=1 16. Occultation event prediction (\ref{subsection:appendix_occultation_prediction})
  & \checkmark & --- & \checkmark & \checkmark & \faClockO\,\faCamera\,--- \\
\hangindent=.80cm\hangafter=1 17. Physical property change detection (\ref{subsection:appendix_physical_evolution})
  & \checkmark & \checkmark & \checkmark & --- & \faCalendar\,---\,--- \\
\hangindent=.80cm\hangafter=1 18. Dynamical property change detection (\ref{subsection:appendix_dynamical_evolution})
  & \checkmark & \checkmark & \checkmark & --- & \faCalendar\,---\,--- \\
\hangindent=.80cm\hangafter=1 19. System validation (\ref{subsection:appendix_system_validation})
  & \checkmark & \checkmark & \checkmark & \checkmark & \faClockO\,---\,--- \\
\hangindent=.80cm\hangafter=1 20. Survey debiasing (\ref{subsection:appendix_survey_debiasing})
  & \checkmark & \checkmark & \checkmark & \checkmark & \faCalendar\,---\,--- \\
\hangindent=.80cm\hangafter=1 21. Alert and follow-up management (\ref{subsection:appendix_alerts})
  & \checkmark & \checkmark & --- & --- & \faExclamationCircle\,\faCamera\,--- \\
\hangindent=.80cm\hangafter=1 22. External data incorporation (\ref{subsection:appendix_externaldata})
  & \checkmark & \checkmark & \checkmark & \checkmark & \faExclamationCircle\,---\,--- \\
\hangindent=.80cm\hangafter=1 23. Data access tools (\ref{subsection:appendix_dataaccess})
  & \checkmark & \checkmark & \checkmark & \checkmark & \faExclamationCircle\,\faCamera\,--- \\
\hline\hline
\multicolumn{6}{l}{$^a$ Science roadmap priorities identified for active objects (Appendix~\ref{subsection:appendix_sciencepriorities_active})\vspace{-0.1cm}} \\
\multicolumn{6}{l}{$^b$ Science roadmap priorities identified for near-Earth objects (Appendix~\ref{subsection:appendix_sciencepriorities_neos})\vspace{-0.1cm}} \\
\multicolumn{6}{l}{$^c$ Science roadmap priorities identified for inner solar system objects (Appendix~\ref{subsection:appendix_sciencepriorities_innerss})\vspace{-0.1cm}} \\
\multicolumn{6}{l}{$^d$ Science roadmap priorities identified for outer solar system objects (Appendix~\ref{subsection:appendix_sciencepriorities_outerss})\vspace{-0.1cm}} \\
\multicolumn{6}{l}{$^e$ \faCalendar: low time sensitivity; \faClockO: moderate time sensitivity; \faExclamationCircle: high time sensitivity; \faCamera: requires\vspace{-0.1cm}} \\
\multicolumn{6}{l}{~~~ image data; \faCopy: requires ability to compute results from subsets of total LSST data set}
\end{tabular}
\end{table}

\section{General Software-Related Considerations}\label{section:swneeds_considerations}

\subsection{Practical Software Development Considerations}

Software tools for performing many of the tasks listed above already exist at various stages of development.  Some tools are essentially already LSST-ready (e.g., are fully automated and optimized for large data sets), while for other tools, e.g. those currently requiring significant human interaction, significant automation and optimization work is needed before they can be used on LSST-scale data sets.  We have attempted to list tools that we know to currently exist in the detailed task descriptions in Appendix~\ref{section:appendix_detailed_sw_descriptions}. We welcome additional information from members of the community about other tools that are currently available or are in advanced stages of development.  As tools are developed or adapated for LSST use prior to the start of the survey, we note that testing using existing large-scale data sets from facilities like NEOWISE\footnote{\tt https://irsa.ipac.caltech.edu/Missions/wise.html}, ATLAS\footnote{\tt http://atlas.fallingstar.com}, and the Zwicky Transient Facility\footnote{\tt https://www.ztf.caltech.edu/} (ZTF) could, and should, be an integral part of that development.

\subsection{Image Data Handling}

Some software tasks will require access to LSST images and not just catalog data (cf.\ Table~\ref{table:swneeds_summary}; Appendix~\ref{section:appendix_detailed_sw_descriptions}).  Such tasks include faint object detection, analysis of multi-object systems, activity detection and characterization, and forced photometry.  Due to the additional computational and network loads associated with using image data, it will be beneficial to determine image specifications for the various solar system-related tasks that require image data, and identify cases where the same image data can be used by multiple analyses.

For example, characterization of binary or other multi-object systems (Appendix~\ref{subsection:appendix_multi_object_systems}) via PSF analysis should only require small postage stamps of each detection (e.g., 100x100 pixels or smaller). The number of images to be retrieved could be further reduced, for example, by only examining detections of known multi-object systems or detections with significant residuals in initial PSF fit analyses.  Small postage stamps should also be sufficient for activity searches (Appendix~\ref{subsection:appendix_activity_detection}), although for this task, image data for {\it all} moving object detections will need to be analyzed to avoid biasing activity search results.  Meanwhile, other tasks such as characterization of large-scale cometary morphology features (Appendix~\ref{subsection:appendix_activity_characterization}), stacking single-visit images to search for faint known objects, or using synthetic tracking to search for faint unknown objects (Appendix~\ref{subsection:appendix_faint_object_detection}) will require much larger images.  Careful consideration will need to be given to the management of these image data to avoid overwhelming available system resources.

%We have indicated the software tasks that will require access to image data in Table~\ref{table:swneeds_summary}, with details about specific image data requirements for each task provided in Appendix~\ref{section:detailed_sw_descriptions}.

%\hhh{think about when postage stamps (30x30 pixels) will be okay and when larger images are needed}

%\msk{We'll generally need larger images for comet dust and ion tails, dust trails.}

%\mmi{If we want to look for faint recoveries (or arc extensions) on important objects via stacking, we will need the images. Ideally, as "raw" as possible apart for trivial bias/flat corrections: no resampling, no background subtraction, no "stitching together chips". Basically, we would need the full chips.}

%\dr{Some KBO binaries are wide enough (~4") that 30x30 would be too small (e.g., citep{Parker2011_wideTNObinaries}. Asteroid binaries are not this wide. Do we want to mention here the LSST plan that the alerts would contain a postage stamp (what size?) of all moving object detections? In practice, this means that alert brokers could provide much of the desired data.}

\subsection{Processing Timescales}\label{subsection:swneeds_timescales}

A key consideration in planning software development and implementation is the required processing timescale for each task.  Tasks pertaining to transient phenomena (e.g., activity), or objects requiring rapid follow-up by other telescopes, are considered highly time-sensitive.  These tasks need to be performed soon after LSST data are acquired to avoid loss of science.
Moderately time-sensitive tasks (e.g., physical property determination) should be performed in a relatively timely fashion (e.g., on a weekly or monthly basis), but extremely rapid or frequent execution is unnecessary.  These tasks often require a substantial amount of data to perform meaningful initial or updated analyses, minimizing the benefit of very rapid and frequent execution anyways.
Finally, tasks with low time sensitivity (e.g., non-gravitational perturbation detection) often require completion of a large fraction of the LSST survey, and so will be most effectively performed on a yearly basis, or even less frequently.
Our assessments of relative time sensitivities of software tasks are summarized in Table~\ref{table:swneeds_summary}, and are discussed further in the detailed task descriptions in Appendix~\ref{section:appendix_detailed_sw_descriptions}.
\looseness=-1

Processing timescales are important for prioritizing computing resources. Identification of low-urgency tasks that do not need to be performed particularly frequently or rapidly can free up resources for more time-sensitive tasks.  Required processing timescales will also be important for prioritizing tasks in the software development process. Tools for generating highly time-sensitive data products should be operational at the start of the survey, while lower-urgency tools could potentially still be in development at that time with minimal loss of science. Critically, the majority of LSST's small body discoveries are expected to occur within the first 1-2 years of the survey. As such, significant progress needs to be made, even on lower-urgency tasks, before the survey begins, to maximize our ability to capitalize on the massive number and rate of expected discoveries.
\looseness=-1

\section{Summary}\label{section:summary}

In this white paper discussing issues relevant to the development of community software tools and infrastructure for doing solar system science with LSST survey data, we identify and describe software and infrastructure needs relevant to the science objectives listed in the Solar System Science Collaboration Science Roadmap.  
We also discuss considerations related to efficient management of image data needs and the time sensitivity of different tasks.  It is our hope that this white paper will encourage further thought and discussion about specific software development needs of the LSST solar system science community. Given the expectation that most new objects will be discovered in the first two years of the survey and also that many time-sensitive science cases exist that require prompt or even real-time analysis of survey data as they are collected, our aim is for this white paper to also be a call to action for concerted efforts to meet as many of those needs as possible in advance of the expected start of survey operations in late 2022.

\section*{Acknowledgements}

We thank J.\ Camargo, C.\ Fuentes, M.\ Juri{\'c}, Z.\ Kne{\v z}evi\'c, N.\ Samarasinha, and M.\ Schwamb for helpful feedback on this work. HHH acknowledges support from NASA Early Career Fellowship grant 80NSSC18K0193.  
MTB appreciates support from STFC grant ST/P0003094/1 and travel support provided by STFC for UK participation in LSST through grant ST/N002512/1. JD acknowledges support from Czech Science Foundation Grant 18-04514J, and MG acknowledges support from the Academy of Finland grant \#299543.  This work also benefited from the hosting and financial support of an international team, led by C.\ Snodgrass and including HHH, MSPK, and MMK, by the International Space Science Institute to discuss main-belt comets, and hosting and support for SSSC Readiness Sprints provided by the Adler Planetarium, B612 Foundation, DIRAC Institute, LSST Corporation, Planetary Society, and the University of Washington.

%\hhh{review entire paper to make sure acronyms are defined once and at time of first usage}

%\acknowledgments

\renewcommand{\thesubsection}{\Alph{subsection}}

\clearpage
\appendix
\setcounter{table}{0}
\renewcommand{\thetable}{A\arabic{table}}

\section{Acronym Definitions\label{section:appendix_acronyms_summary}}

\begin{table}[ht]
\captionsetup{font=normalsize}
\centering
\caption{Acronym Definitions}
\label{table:abbreviations}
\begin{tabular}{ll}
\hline\hline
API     & Application Programming Interface (introduced in Section~\ref{section:swtools}) \\
BIM     & Backward Integration Method (introduced in Appendix~\ref{subsection:appendix_dynamical_clustering_identification}) \\
HCM     & Hierarchical Clustering Method (introduced in Appendix~\ref{subsection:appendix_dynamical_clustering_identification}) \\
LSST    & Large Synoptic Survey Telescope (introduced in Section~\ref{subsection:intro_lsst}) \\
MPC     & Minor Planet Center (introduced in Section~\ref{subsection:intro_lsst}) \\
MOID    & Minimum Orbit Intersection Distance (introduced in Appendix~\ref{subsection:appendix_orbital_parameters}) \\
MOPS    & Moving Object Processing System (introduced in Section~\ref{subsection:intro_lsst}) \\
NEO     & Near-Earth Object (introduced in Section~\ref{section:swtools}) \\
PSF     & Point-Spread Function (introduced in Section~\ref{subsection:intro_promptproducts_drproducts}) \\
SDSS    & Sloan Digital Sky Survey (introduced in Appendix~\ref{subsection:appendix_compositional_characterization}) \\
SSSC    & Solar System Science Collaboration (introduced in Section~\ref{subsection:intro_lsst}) \\
TCO     & Temporarily Captured Orbiter (introduced in Appendix~\ref{subsection:appendix_advanced_moving_object_detection}) \\
TNO     & Trans-Neptunian Object (introduced in Appendix~\ref{subsection:appendix_faint_object_detection}) \\
%AstDyS  & Asteroids Dynamic Site \\
%ATLAS   & Asteroid Terrestrial-impact Last Alert System \\
%CFHT    & Canada-France Hawaii Telescope \\
%DPDD & LSST Data Products Definition Document \\
%IAU  & International Astronomical Union \\
%ISO  & Interstellar Object \\
%ISS  & Inner Solar System \\
%JFC  & Jupiter-Family Comet \\
%KBO  & Kuiper Belt Object \\
%LPC  & Long-Period Comet \\
%MBA  & Main-Belt Asteroid \\
%MBC  & Main-Belt Comet \\
%MBO  & Main-Belt Object \\
%MOID & Minimum Orbit Intersection Distance \\
%NEA  & Near-Earth Asteroid \\
%NEOWISE & Near-Earth Object Wide-field Infrared Survey Explorer \\
%NES  & North Ecliptic Spur (survey) \\
%NOAO    & National Optical Astronomy Observatory \\
%OSS  & Outer Solar System \\
%PHA  & Potentially Hazardous Asteroid \\
%PHO  & Potentially Hazardous Object \\
%PS1  & Pan-STARRS1 \\
%PTF  & Palomar Transient Factory \\
%SDO  & Scattered Disk Object \\
%SSOIS   & Solar System Object Image Search \\
%TALCS   & Thousand Asteroid Lightcurve Survey \\
%TOM     & Target and Observation Manager \\
%UGP  & User Generated Product \\
%VO      & Virtual Observatory \\
%YORP    & Yarkovsky-O'Keefe-Radzievskii-Paddack [effect] \\
%ZTF     & Zwicky Transient Facility \\
\hline\hline
\end{tabular}
\end{table}

%\clearpage
\clearpage
\section{Detailed Software Task Considerations}\label{section:appendix_detailed_sw_descriptions}

%%%%%%%%%% ORBITAL OBJECT AND DETECTION PARAMETERS %%%%%%%%%%
\subsection{Orbital Object and Detection Parameter Computation}\label{subsection:appendix_orbital_parameters}

Observational geometric parameters and orbital properties of LSST's moving object detections will be relevant for a wide range of solar system science.
Examples of observational geometric parameters needed for various solar system science use cases include heliocentric and geocentric distances, true anomalies, and position angles of antisolar and heliocentric velocity vectors as projected on the sky. For orbital properties, these include heliocentric osculating orbital elements, barycentric orbital elements, and Minimum Orbit Intersection Distances, or MOIDs, for major planets.   Ideally, parameters needed by many solar system scientists should be readily available for general use by the community, rather than be left to individual scientists to derive on their own.
\looseness=-1

We list detection-level and object-level metadata parameters that are relevant to particular science tasks in Tables~\ref{table:geoparams} and \ref{table:objparams}.  At the current time, there are plans for MOPS to provide many of these parameters, as indicated in the aforementioned tables.  However, other parameters needed for key scientific tasks are not currently planned to be provided as Prompt Products or Data Release Products, and so may need to be provided as User Generated Products instead. For instance, projected antisolar vector and heliocentric velocity vector position angles, MOIDs for planets other than the Earth, and barycentric orbital elements will be needed, for tasks including activity detection and characterization, and dynamical classification of outer solar system objects. 
\looseness=-1

In cases where these metadata products are needed for urgent tasks, such as activity detection and characterization, this task should be considered highly time-sensitive.  Comet tails or trails are often oriented along projected antisolar or heliocentric velocity vector position angles, meaning that a specialized activity search algorithm might take advantage of this fact by searching for excess flux along each of these expected directions (cf.\ Appendix~\ref{subsection:appendix_activity_detection}(e)).  In this case, these parameters would be required for {\it every} moving object detection and need to be provided on a timescale similar to that required for other activity detection tasks (e.g., at least on a nightly basis).  Other parameters, on the other hand, such as MOIDs for planets other than the Earth, which could be useful for identifying possible close approaches that could induce surface color changes due to tidal disruptions (cf.\ Appendix~\ref{subsection:appendix_physical_evolution}), may not be needed as urgently and so can be considered only moderately time-sensitive.
\looseness=-1

We note that while analytic proper orbital elements \citep[cf.][]{milani1994_properelements} can be computed relatively quickly, they are of limited accuracy, especially for orbits with high eccentricities and inclinations.  By comparison, synthetic proper elements (cf.\ Appendix~\ref{subsection:appendix_dynamical_clustering_identification}) are significantly more reliable, and so if a synthetic proper element catalog can be kept reasonably up to date, either by the AstDyS group\footnote{\tt https://newton.spacedys.com/astdys/} or the SSSC, an analytic proper orbital element catalog could be considered a low priority for most purposes.
\looseness=-1

%With the exact list of data products to be provided by MOPS still to be determined, it is not currently known which of these data products will be provided by MOPS and which will need to be delivered by SSSC-developed UGP software tools.  From the perspective of scientists, of course, it is unimportant whether these parameters are provided by MOPS or SSSC-developed software tools, as long as they are available for use as end data products.  As such, for now, we simply list all basic data products that we anticipate will be needed for various science cases, so that once it becomes clearer what data products MOPS will provide, we can turn our attention to developing tools for providing the remaining data products.

\bigskip
{\centering
\begin{tcolorbox}[width=5.5in]
Key priorities
\begin{itemize}[leftmargin=*]
    \item{\vspace{-0.2cm}Computation of detection-level observing and orbital geometry parameters such as true anomalies, projected antisolar vector and heliocentric velocity vector position angles, and orbit plane angles}
    \item{\vspace{-0.2cm}Computation of object-level orbital parameters such as barycentric orbital elements and MOIDs for planets other than the Earth}
\end{itemize}
\end{tcolorbox}
}

\bigskip
\noindent\hangindent=.35cm\hangafter=1{\bf Processing requirements:} High time sensitivity.

\noindent\hangindent=.35cm\hangafter=1{\bf Software dependencies:} Does not require input from other software tools discussed here.  Output will be required for activity detection and characterization (Appendices~\ref{subsection:appendix_activity_detection} and \ref{subsection:appendix_activity_characterization}), outburst and disruption detection (Appendix~\ref{subsection:appendix_outburst_disruption_detection}), identification of resonant outer solar system objects (Appendix~\ref{subsection:appendix_advanced_dynamical_characterization}), and survey efficiency characterization and debiasing (Appendix~\ref{subsection:appendix_survey_debiasing}).

\noindent\hangindent=.35cm\hangafter=1{\bf Required input data:} 
Observation times, astrometric positions, and uncertainties for individual detections, and orbital elements for individual objects.
%\dr{Rewrote to be more specific and to include astrometric uncertainties since I assume that we'll want to propagate these. Ideally, we'd have orbital element uncertainties too}
%Observation dates and times of detections and orbital elements of corresponding solar system objects.

\noindent\hangindent=.35cm\hangafter=1{\bf Expected output data:} Object-level and detection-level metadata parameters beyond those provided by the LSST MOPS pipeline (cf.\ Tables~\ref{table:geoparams} and ~\ref{table:objparams}).

\setlength{\tabcolsep}{4pt}
\begin{table*}[ht]
\captionsetup{font=normalsize}
\def\arraystretch{1.3}
\centering
\caption{Desired detection-level geometric/observational parameters}
\smallskip
%\large
\begin{tabular}{p{2.4in}cp{3.10in}}
\hline\hline
\multicolumn{1}{l}{Metadata product}
 & PP/DR?$^a$
 & \multicolumn{1}{l}{Examples of expected uses} \\
\hline
\hangindent=.25cm\hangafter=1 Heliocentric and geocentric\newline distances & Y
  & \hangindent=.25cm\hangafter=1 Normalizing measured object brightnesses; studying cometary activity strength as a function of heliocentric distance \\
\hangindent=.25cm\hangafter=1 Solar phase angle & Y
  & \hangindent=.25cm\hangafter=1 Normalizing measured object brightnesses; studying surface properties via phase function analyses \\
\hangindent=.25cm\hangafter=1 Heliocentric ecliptic latitude and\newline longitude & Y
  & \hangindent=.25cm\hangafter=1 Identifying observational biases or real non-uniform distributions in discoveries of cometary activity or faint distant objects\\
%One-way light travel time$^*$
%  & \hangindent=.25cm\hangafter=1 Improving time precision for fast rotator lightcurve studies and astrometric deflection studies \\
Residuals from best-fit orbit solution & Y
  & \hangindent=.25cm\hangafter=1 Identifying non-gravitational perturbations e.g. from low-level activity, or perturbations due to mutual asteroid interactions \\
Predicted ephemeris uncertainty & Y
  & \hangindent=.25cm\hangafter=1 Assessing likelihood of correct object identification \\
%\hangindent=.25cm\hangafter=1 Solar and lunar elongation angle, and\newline lunar illumination percentage$^\dagger$
%  & \hangindent=.25cm\hangafter=1 Assessing background brightness suitability for cometary activity searches (could also be addressed with a general data quality parameter like surface brightness magnitude limit)\\
Galactic latitude and longitude & Y
  & \hangindent=.25cm\hangafter=1 Assessing reliability of detection characterization measurements \\
\hangindent=.25cm\hangafter=1 Angular distance to and magnitude of nearest known background \newline source in field & Y
  & \hangindent=.25cm\hangafter=1 Assessing reliability of detection characterization measurements \\
\hangindent=.25cm\hangafter=1 Angular distance to and magnitude of nearest identified saturated source in field  & N
  & \hangindent=.25cm\hangafter=1 Assessing reliability of detection characterization measurements \\
\hangindent=.25cm\hangafter=1 Projected antisolar and heliocentric\newline velocity vector sky-plane angles & N
  & Searching for activity and interpreting comet morphology \\
Orbit plane angle & N
  & \hangindent=.25cm\hangafter=1 Identifying observations of small solar system bodies with favorable dust detection opportunities \\
True anomaly & N
  & \hangindent=.25cm\hangafter=1 Characterizing cometary activity strength as a function of orbit position \\
\hline
\hline
\multicolumn{3}{l}{\vspace{-0.1cm}$^a$ Expected to be provided as a Prompt Product or Data Release Product? (Y: yes; N: no)} \\
%\multicolumn{2}{l}{$^\dagger$ Not expected to be provided as a Prompt Product} \\
\end{tabular}
%{$^a$ Expected to be provided as a Prompt Product?.} \\
%{$^\dagger$ Desirable for assessment of likely data/image quality} \\
\label{table:geoparams}
\end{table*}

\setlength{\tabcolsep}{4pt}
\begin{table*}[ht]
\captionsetup{font=normalsize}
\def\arraystretch{1.3}
\centering
\caption{Desired object-level orbit-related parameters}
\smallskip
%\large
\begin{tabular}{p{2.4in}cp{3.10in}}
\hline\hline
\multicolumn{1}{l}{Metadata product}
 & PP/DR?$^a$
 & \multicolumn{1}{l}{Examples of expected uses}
 \\
\hline
Orbit period & N
  & \hangindent=.25cm\hangafter=1 Enabling straightforward computation of number of orbits completed within a specified time period to facilitate sample selection for studies requiring observations of objects over multiple orbits; potentially facilitating dynamical classifications \\
\hangindent=.25cm\hangafter=1 Heliocentric osculating orbital\newline elements & Y
  & \hangindent=.25cm\hangafter=1 Preferred for dynamical analyses of inner solar system objects \\
\hangindent=.25cm\hangafter=1 Barycentric osculating orbital\newline elements & N
  & \hangindent=.25cm\hangafter=1 Preferred for dynamical analyses of outer solar system objects \\
\hangindent=.25cm\hangafter=1 Tisserand parameters with respect to Jupiter and Neptune & N
  & \hangindent=.25cm\hangafter=1 Dynamically characterizing various small bodies (e.g., asteroids, comets, Centaurs) [relatively simple to compute, however, so may be just as easily and potentially more efficiently computed at the time of any query, rather than being pre-computed and stored] \\
\hangindent=.25cm\hangafter=1 MOIDs for Earth & Y
  & \hangindent=.25cm\hangafter=1 Identifying potentially hazardous NEOs or NEOs with close observation opportunities \\
\hangindent=.25cm\hangafter=1 MOIDs for planets other than Earth & N
  & \hangindent=.25cm\hangafter=1 Identifying objects with potential close planetary encounters, facilitating searches for changes in surface properties during such encounters; only necessary for some combinations of small bodies and planets (e.g., not needed for NEOs and Neptune) \\
Analytic proper orbital elements & N
  & \hangindent=.25cm\hangafter=1  Identifying preliminary associations of small bodies with known or new asteroid families \\
Synthetic proper orbital elements & N
  & \hangindent=.25cm\hangafter=1 Identifying associations of small bodies with known or new asteroid families; see Appendix~\ref{subsection:appendix_advanced_dynamical_characterization} \\
Asteroid family associations & N
  & \hangindent=.25cm\hangafter=1 Associating small bodies with others of probable similar compositions; see Appendix~\ref{subsection:appendix_dynamical_clustering_identification} \\
\hangindent=.25cm\hangafter=1 Pre-perihelion unperturbed orbital\newline elements of long-period comets & N
  & \hangindent=.25cm\hangafter=1 Dynamically classifying long-period comets; see Appendix~\ref{subsection:appendix_advanced_dynamical_characterization} \\
%\hangindent=.25cm\hangafter=1 Tisserand parameter (for various\newline planets, not just Jupiter)
%  & \hangindent=.25cm\hangafter=1 Identification of objects with strong dynamical interactions with given planets \\
\hline
\hline
\multicolumn{3}{l}{\vspace{-0.1cm}$^a$ Expected to be provided as a Prompt Product or Data Release Product? (Y: yes; N: no)} \\
%\multicolumn{2}{l}{\vspace{-0.1cm}$^*$ Desirable for scientific purposes} \\
%\multicolumn{2}{l}{$^\dagger$ Desirable for data quality assessment purposes} \\
\end{tabular}
%{$^*$ Desirable for scientific purposes.} \\
%{$^\dagger$ Desirable for assessment of likely data/image quality} \\
\label{table:objparams}
\end{table*}

\clearpage
%%%%%%%%%% ORBITAL ELEMENT AND EPHEMERIS UNCERTAINTY COMPUTATION %%%%%%%%%%

\subsection{Orbital element and ephemeris uncertainty determination}\label{subsection:appendix_uncertainty_characterization}

Appropriate propagation of astrometric uncertainties, both for LSST and external data, through to orbital element uncertainties, will be important for assessment of the quality of computed orbits and the reliability of ephemeris predictions: particularly for high-priority follow-up targets, impact hazard assessment, and dynamical classifications.  We expect that the Minor Planet Center will bear primary responsibility for orbit determination for LSST-discovered objects, and along with that task, should also provide realistic orbital element uncertainties, including covariances.  Presently, they do not provide these uncertainties.
Realistic orbital element uncertainties are required to compute realistic ephemeris uncertainties, which in turn are needed for planning observations of high-priority follow-up targets. This task should be considered highly time-sensitive.

To the extent (if any) that orbital element uncertainties are not adequately handled by the MPC in the LSST era, it will be of interest to the community to develop tools to perform these computations using various approaches that have been described in the literature \citep[e.g.,][]{chernitsov2017_orbitconfidenceregions,veres2017_astrometricerrors}.  The probabilistic orbital element plots generated by NASA's Scout NEO hazard assessment tool\footnote{\tt https://cneos.jpl.nasa.gov/scout/} provide another example of how orbital element uncertainties could be characterized and visualized.   %Well-characterized orbital element uncertainties are also the foundation of many other analyses like dynamical classification.

%\dr{ZZZ Thought this paragraph should be made stronger; we REALLY need these uncertainties. ZZZ}

For objects only observed over a short arc, such as newly discovered NEOs, traditional least-square orbital fitting procedures may fail, because discovery tracklets contain insufficient information to fully estimate all six orbital elements and their covariances. In these cases, alternative orbital determination methods using a ranging approach should be considered \citep[e.g.,][]{granvik2009,farnocchia2015_ranging}.
The ranging approach produces sets of orbital elements that are compatible with observations. These same sets of orbits can be used to derive realistic ephemeris uncertainties by simply determining the shapes and sizes of the regions corresponding to the ephemeris points computed from each orbit. Such regions will often have highly irregular shapes, due to the non-linearity of orbit determination in the short-arc regime.

\bigskip
{\centering
\begin{tcolorbox}[width=5.5in]
Key priority
\begin{itemize}[leftmargin=*]
    \item{\vspace{-0.2cm}Computation of appropriate uncertainties for orbital elements and ephemeris predictions based on detection-level astrometric uncertainties, to the extent (if any) that this is not adequately handled by the MPC}
\end{itemize}
\end{tcolorbox}
}

\bigskip
\noindent\hangindent=.35cm\hangafter=1{\bf Processing requirements:} High time sensitivity.

\noindent\hangindent=.35cm\hangafter=1{\bf Software dependencies:} Will require external data incorporation for some input (Appendix~\ref{subsection:appendix_externaldata}).  Output will be required for alert brokering and follow-up observation management (Appendix \ref{subsection:appendix_alerts}).

\noindent\hangindent=.35cm\hangafter=1{\bf Required input data:} Astrometric positions, times, and uncertainties from LSST and other facilities.

\noindent\hangindent=.35cm\hangafter=1{\bf Expected output data:} Realistic orbital element and ephemeris uncertainties.

\clearpage
%%%%%%%%%% ACTIVE OBJECT ASTROMETRY %%%%%%%%%%

\subsection{Extended object astrometry}\label{subsection:appendix_extended_object_astrometry}

Techniques for measuring accurate astrometry of extended objects (i.e., active comets) will be needed to maximize the accuracy of orbit determination for active objects. Comet comae tend to peak at the nucleus, but asymmetries in the dust distribution as projected in the plane of the sky will affect centroiding.  The offset will be within the seeing disk, so for LSST, it is expected to be on the order of $\sim$100~mas.  It is a systematic bias that affects orbital and non-gravitational parameter retrieval of the nucleus.  This can be mitigated with a comet-specific method, such as centroiding with progressively smaller box sizes, using the coordinate variation with size to predict the 0-pixel aperture centroid, and therefore the comet nucleus location.

\bigskip
{\centering
\begin{tcolorbox}[width=5.5in]
Key priority
\begin{itemize}[leftmargin=*]
    \item{\vspace{-0.2cm}Accurate centroiding of extended objects (i.e., comets and active asteroids) to obtain accurate astrometry for orbit determination purposes}
\end{itemize}
\end{tcolorbox}
}

\bigskip
\noindent\hangindent=.35cm\hangafter=1{\bf Processing requirements:} High time sensitivity.  Image data required.

\noindent\hangindent=.35cm\hangafter=1{\bf Software dependencies:} Does not require input from other software tools discussed here.
%Output will be required for alert brokering and follow-up observation management (Appendix \ref{subsection:appendix_alerts}).

\noindent\hangindent=.35cm\hangafter=1{\bf Required input data:} Image data for detections of extended objects

\noindent\hangindent=.35cm\hangafter=1{\bf Expected output data:} Accurate astrometric measurements of extended objects.

\clearpage
%%%%%%%%%% FAINT PRECOVERY AND RECOVERY IDENTIFICATION %%%%%%%%%%

\subsection{Faint precovery and recovery identification}\label{subsection:appendix_lsst_precoveries_recoveries}

For certain objects that are discovered by LSST, orbital arcs covered by available 5-$\sigma$ astrometry from LSST will be inadequate for determining orbits to sufficient precision to ensure reliable future recovery of the objects.  
These include both fast-moving NEOs and other small and faint objects that are just barely detectable by LSST in the first place.
To extend the orbital arcs of such observationally challenging objects, it will be useful to search for both precoveries in previously acquired LSST data, and recoveries in subsequently acquired LSST data that are below the 5-$\sigma$ detection threshold used by the baseline LSST pipeline. (We consider the extension of this task to non-LSST data in Appendix~\ref{subsection:appendix_externaldata}).
In many cases, astrometry of 3-$\sigma$ or even $<$3-$\sigma$ detections may be able to provide valuable constraints on poorly constrained orbits. When added into an orbit solution, their residuals can provide additional information about the probability that such detections are real, and likely to be associated with the object in question.

Software to extend arcs should first identify objects that are in need of orbit refinement (e.g., by selecting objects with orbital element uncertainties larger than certain thresholds). Then, for newly discovered objects, it should identify which past LSST images coincide with expected positions of each target of interest at the time of observation. Finally, the software should perform a search for $<$5-$\sigma$ sources within the expected uncertainty of the object.  For existing objects in need of orbit refinement, the software could also run on all new LSST images as they are obtained, identifying images with the possibility of containing detections of targets of interest and then performing a search for $<$5-$\sigma$ sources near the expected position of each object.  The software could search until the orbital element uncertainties of a given object fall below a certain threshold, indicating that continued orbit refinement is no longer urgently required. 

Orbit refinement of observationally challenging objects is needed to allow ephemeris predictions to be as accurate as possible, for highly successful follow-up observations. As such, this task should be considered highly time-sensitive.

\bigskip
{\centering
\begin{tcolorbox}[width=5.5in]
Key priority
\begin{itemize}[leftmargin=*]
    \item{\vspace{-0.2cm}Searches of both archived and new LSST data for $<$5$\sigma$ detections corresponding to expected positions of newly-discovered solar system objects with poorly constrained orbits for possible incorporation into orbit solutions}
\end{itemize}
\end{tcolorbox}
}

\bigskip
\noindent\hangindent=.35cm\hangafter=1{\bf Processing specifications:} High time sensitivity.  Image data required.

\noindent\hangindent=.35cm\hangafter=1{\bf Software dependencies:} Does not require input from other software tools discussed here.  Output will be required for software tasks related to challenging orbit determination situations, e.g., orbital element and ephemeris uncertainty determination (Appendix~\ref{subsection:appendix_uncertainty_characterization}), advanced moving object detection (Appendix~\ref{subsection:appendix_advanced_moving_object_detection}), and detection and characterization of non-gravitational perturbations (Appendix~\ref{subsection:appendix_dynamical_evolution}).

\noindent\hangindent=.35cm\hangafter=1{\bf Required input data:} Existing LSST astrometry, existing orbit solution parameters and uncertainties, past LSST images coinciding with expected positions of objects of interest.

\noindent\hangindent=.35cm\hangafter=1{\bf Expected output data:} Low S/N astrometry for objects with low-precision orbit solutions that can in turn be used to compute higher-precision orbit solution parameters.

\clearpage
%%%%%%%%%% EXTREMELY FAINT OBJECT DETECTION %%%%%%%%%%

\subsection{Extremely faint object detection}\label{subsection:appendix_faint_object_detection}

%\hhh{mention Gerdes work on DES and Juric and Connolly work on ZTF}
 
The ability to detect objects fainter than the single-visit detection limit of LSST will be important for science priorities related to the discovery of faint outer solar system objects. It may also be useful for orbit refinement for faint inner solar system objects, and even NEOs.  Stacking analyses will also be beneficial for characterizing the inactive nuclei of active objects. Finally, it is also key for the physical characterization of faint inactive objects that are only directly observable for short periods of time; particularly NEOs and possible impactors, but also faint inner or outer solar system objects that are just barely above the LSST detection limit at perihelion or close-Earth approach.

%\hhh{astrometric recoveries of faint ISS objects or even NEOs?} %  (B-1, B-4, B-5, B-6, B-8, B-9, C-1, C-2)

Image stacking is the primary method by which we anticipate that scientists will search for extremely faint objects (i.e., objects fainter than the nominal single-visit detection limit for the standard LSST pipeline) in LSST data.  The addition of multiple images of the same portion of the sky, ideally obtained relatively closely in time, creates composite images with effective image depths deeper than that of a single LSST exposure. This permits the detection of objects too faint to be detected in a single LSST image \citep[e.g.,][]{juric2019_lsstdeepOSSOsearches}.  For LSST, this type of analysis will be applicable for both known and unknown objects, and it is under consideration by a number of the LSST Science Collaborations.

%For known objects for which orbits are sufficiently well-constrained that ephemeris positions can be reasonably predicted (at least for data relatively close in time to previous detections), stacking analyses can assist with further confirming and constraining orbit solutions.  By extending available orbit arcs, such analyses can enhance our ability to derive reliable orbits for objects like NEOs that often recede quickly from the Earth following their discoveries, with corresponding rapid decreases in brightnesses, or newly discovered active objects that are bright when active but become much fainter as activity weakens and disappears.

For known solar system objects for which orbits are sufficiently well-constrained that ephemeris positions can be reasonably predicted (e.g., for observations obtained relatively close in time to previous detections before ephemeris uncertainties become too large), stacking analyses can assist with constraining the sizes of inactive nuclei of active objects.  Applied to data sets for which geometric circumstances do not change significantly (e.g., images obtained on the same night or potentially within the same week or so, depending on the type of orbit an object occupies), stacking analyses can yield object detections unavailable in single exposures. Photometry on these stacked detections can potentially add key constraints to phase function fitting analyses, improving absolute magnitude estimates.

Also, for known NEOs with poorly determined orbits, it is often possible to accurately predict the motion of the object at a given time, even when its position on the plane of the sky is highly uncertain. Stacking a set of images with the known motion rates may show a faint detection inside the uncertainy region: the corresponding astrometry may result in a dramatic improvement of the precision of the object's orbit, improving the predictability of possible future impacts.

For unknown objects, an established method known by several names, including `shift-and-stack', synthetic tracking, or digital tracking, is a powerful detection technique. Digital images of the same area of sky are shifted, assuming different non-sidereal velocities, and then added together to search for unknown faint objects \citep[e.g.,][]{gladman1997_kbsearches,gladman1998_pencilbeamtnosurvey}. This technique should be well suited for searching for unknown distant trans-Neptunian objects (TNOs) in LSST data, since the slow non-sidereal velocities of these objects on the sky mean that their positions change only a small amount between survey observations, even for relatively widely temporally spaced visits of several days or even weeks. This minimizes the parameter space of non-sidereal rates and directions that need to be explored.  Recently, \citet{whidden2019_tnosearchalgorithm} introduced a similar type of shift-and-stack approach, which starts with a maximum likelihood estimate for the probability of the detection of a source within a series of images, and then uses a massively parallel algorithm to rapidly search large numbers of candidate moving object trajectories using Graphics Processing Units (GPUs).  Code for conducting such searches is publicly available\footnote{\tt https://github.com/dirac-institute/kbmod} and could be used to search LSST data for distant TNOs (see also Appendix~\ref{subsection:appendix_advanced_moving_object_detection} for related discussion).
\looseness=-1

With available computing power increasing in recent years, synthetic tracking has begun to be used on more nearby (and therefore faster-moving) objects such as main-belt asteroids or even near-Earth objects \citep{zhai2014_synthetictracking,heinze2015_digitaltracking}, for which the parameter space of possible rates and directions of non-sidereal motion was previously too large to efficiently search, especially for images widely separated in time.  The relatively large interval of several days expected between return visits to a given region of sky by LSST means that the viability of synthetic tracking for detecting NEOs or inner solar system objects is questionable at the moment. Given current uncertainty about future computing capabilities, though, future re-consideration of application of this technique to NEOs and inner solar system objects could be worthwhile.
\looseness=-1

Weekly or monthly searches for known and unknown faint objects should be sufficient, although more frequent or targeted searches may be desired for objects for which astrometry is required to enable follow-up observations by other facilities.

\bigskip
{\centering
\begin{tcolorbox}[width=5.5in]
Key priorities
\begin{itemize}[leftmargin=*]
    \item{\vspace{-0.2cm}Stacking of image data at the positions of known solar system objects to increase image depth for extending orbital arcs, thus improving orbit solutions, or obtaining more data for measuring physical properties}
    \item{\vspace{-0.2cm}``Synthetic'' or ``digital'' tracking of image data to search for faint unknown solar system objects.}
\end{itemize}
\end{tcolorbox}
}

\bigskip
\noindent\hangindent=.35cm\hangafter=1{\bf Processing requirements:} Moderate time sensitivity. Full-frame image data required.

\noindent\hangindent=.35cm\hangafter=1{\bf Software dependencies:} Will require orbital element and ephemeris uncertainty determination (Appendix~\ref{subsection:appendix_uncertainty_characterization}) and external data incorporation for some input (Appendix~\ref{subsection:appendix_externaldata}).  Output will be required for slow-moving object detection (Appendix~\ref{subsection:appendix_advanced_moving_object_detection}).

\noindent\hangindent=.35cm\hangafter=1{\bf Required input data:} LSST images coinciding with expected positions of objects of interest as well as in general regions of the sky in which objects of interest may be located (e.g., along the ecliptic plane).

\noindent\hangindent=.35cm\hangafter=1{\bf Expected output data:} Detections of solar system objects fainter than the single-visit detection limit of LSST.

\clearpage
%%%%%%%%%% ADVANCED MOVING OBJECT DETECTION %%%%%%%%%%

\subsection{Advanced moving object detection}\label{subsection:appendix_advanced_moving_object_detection}

%By the end of LSST's ten year survey it will have discovered millions of new minor planets. LSST's contribution to Solar System science will enable a high resolution study of the many different populations of minor planets in our Solar System, allowing a detailed look into its formation and destruction processes. 

As discussed in Section~\ref{subsection:intro_lsst}, the solar system community will primarily rely on the LSST MOPS pipeline to identify known solar system objects and make new discoveries. However, the pipeline has some limitations that result in lower detection efficiencies for certain solar system object populations. Additional software will augment and improve detection efficiencies for solar system object populations such as fast-moving NEOs and slow-moving outer solar system objects. It will also be useful for objects with extreme colors or high-amplitude lightcurves, where there is a high potential for only obtaining a single detection in a night, instead of the sets of at least two detections that MOPS requires to function.
Given these considerations, it will be useful to consider developing alternative algorithms to boost discovery rates of certain types of solar system objects.

The default LSST approach to moving object detection will consist of taking two or more images of the same area of the sky spaced closely enough (currently expected to be $\sim$30~minutes) that any moving object in the first image can be relatively unambiguously identified with its re-observations in subsequent images. Series of such closely separated observations (i.e., obtained on the same night) are called ``tracklets''. Having constructed tracklets, it is then relatively straightforward (though still computationally expensive) to link these into tracks and then into orbits \citep{Kubica2007_kdtreelinking,Denneau2013_panstarrsmops, Jones2018_discoverymachine}.  While effective for a large fraction of solar system objects, this approach nonetheless is sub-optimal for certain types of objects. Such an approach typically requires the adoption of an upper limit to an object's non-sidereal velocity (which essentially defines the radius around an initial detection within which the linking algorithm searches for candidate subsequent detections to form tracklets; the larger the velocity limit is, the more potential tracklets must be considered).  Slow-moving objects are also difficult to identify with this method if a survey's cadence and data processing/flagging mean that their motion between two consecutive visits to the same sky region in a night is not detectable. This is often the case for surveys optimized to search for faster-moving objects like NEOs, and for exceptionally distant TNOs. Surveying at distances $\gtrsim300$ au requires the LSST transient detection system together with MOPS to identify motions of $\lesssim 0.5$ arcsec/hr, and this may be better suited to dedicated software (see also Appendix~\ref{subsection:appendix_faint_object_detection}).

The need by MOPS for tracklets means that visits to the same sky fields multiple times per night are required, placing strict constraints on the LSST survey cadence. Over periods of several months, the Earth's motion creates non-linear apparent motion of solar system objects on the sky. This makes linking tracklets over long periods of time challenging and computationally expensive, which can limit discovery yields.
For very nearby objects, a similar issue with non-linearity of the object's motion in the sky may also become apparent on a much shorter timescale of just a few hours. In this case, the non-linearity of the observer motion during the daily axial rotation of the Earth produces a diurnal parallax effect, changing the apparent direction and speed of an NEO's motion in the sky.
This issue becomes extremely relevant for objects in an imminent impact trajectory, i.e., those that are discovered just hours to a few days before a collision with the Earth. In these cases, the diurnal parallax combines with the quickly approaching trajectory to result in a dramatic change in the apparent velocity of the incoming impactor, which may make the identification of the tracklet much more complex.
\looseness=-1

%by linking LSST detections over longer time periods or of objects with rates of non-sidereal motion slower or faster than the velocity limits of the default MOPS pipeline.  

For NEOs, tests of an algorithm that can link six single detections over a 60 night window have demonstrated the ability to increase LSST discoveries of potentially hazardous asteroids by $\sim$25\%, from the survey's predicted baseline completeness of 65.6\% to 82.5\% at $H$$\,\leq\,$22 \citep{Jones2018_discoverymachine}. The ephemeris-space Multiple-Address-Comparison \citep[eMAC;][]{granvik2005} and the orbital-element-space Multiple-Address-Comparison \citep[oMAC;][]{granvik2008} algorithms implemented in the OpenOrb package \citep{granvik2009} have been shown to scale log-linearly with the number of tracklets involved, thereby showing that a rigorous solution to the linking problem can be computationally tractable also in the era of large surveys. The linking process starts with the computation of an orbital ranging solution for each tracklet. Candidate linkages are then sought by comparing either orbital elements (oMAC) or, typically geocentric, ephemerides at multiple common epochs (eMAC) using an efficient tree-like data structure. Candidate linkages are then discarded if an orbital fit that reproduces all tracklets involved cannot be located. \citet{fedorets2019} have recently shown that using these algorithms in the linking process will increase the discovery rate of Earth's temporarily-captured natural satellites by an order of magnitude, compared to the baseline MOPS processing. Lastly, recent work by \citet[][HelioLinC]{Holman2018_heliolinc} has not only shown significant promise in making the linking problem less computationally intensive, but also enabling the linking of tracklets over longer periods of time. HelioLinC shifts the frame of linking to the heliocenter. By assuming a heliocentric distance and its rate of change, tracklets belonging to a unique minor planet can be propagated to a common epoch and clustered at very little computational cost. 

Linking single-night detections across long temporal arcs of at least weeks is necessary for dealing with the long baselines needed for the detection of objects in the distant outer solar system \citep[e.g.][]{larsen2007_spacewatch, schwamb2010_palomar, Bannister:2013, rabinowitz2012_lasilla, brown2015_catalina, ashton2019_ossos}).
Such surveys have made substantial past use of the Orbfit algorithm of \citet{bernstein2000_orbfit}. HelioLinC offers a promising alternative algorithm that appears to work well for TNO detection \citep{Holman2018_heliolinc,Holman:2018_dwarf}. 
Preliminary work with the ``Tracklet-less Heliocentric Orbit Recovery'' algorithm (Moeyens et al., in prep) has  successfully linked known non-NEO solar system object detections in two-week windows of Zwicky Transient Facility data, without imposing any cadence restrictions or using tracklets. If this or similar algorithms can be extended to LSST, they may allow for the discovery of additional small solar system objects.
Furthermore, if scalable to LSST's performance requirements, and if made to discover all populations of solar system objects at equal or higher completeness than the currently planned MOPS, these algorithms would remove the tracklet cadence constraint and potentially allow LSST to observe twice the area of sky in a single night.  Given these potential benefits, some of these alternative approaches (e.g., HelioLinC) are under active consideration by the MOPS development team for detection linking.

% \hhh{For very fast-moving objects (e.g., very close NEOs), detections will be trailed due to the appreciable motion of the object during the typical LSST exposure.  Very fast moving objects present challenges to traditional linking algorithms since they increase the search radius that must be considered when looking for corresponding detections to form tracklets, which would normally increase the likelihood of confusion with detections of other unrelated objects.  However, their trailing properties (i.e., length and direction) provide additional useful information beyond what is available for untrailed detections.  As such, it will be useful to consider alternative orbit determination algorithms that incorporate trail information for very fast-moving objects (which the MOPS development team is also doing).}

For very fast-moving objects such as very close NEOs or Temporarily Captured Orbiters \citep[TCOs; also sometimes known as ``mini-moons''; cf.][]{jedicke2018_minimoons}, detections will be trailed due to the appreciable motion of the object during the typical LSST exposure. Very fast moving objects present challenges to traditional linking algorithms since they increase the search radius that must be considered when looking for corresponding detections to form tracklets, which would normally increase the likelihood of confusion with detections of other unrelated objects.  However, their trailing properties (i.e., length and direction) provide additional useful information beyond what is available for untrailed detections. As such, alternative linking algorithms that can make use of information on trailed pairs for very fast-moving objects will be desirable, and are also being considered for implementation by the MOPS development team.

%The current implementation of the LSST Moving Object Processing System (MOPS) relies on building ``tracklets'', which are sky-plane motion vectors that constrain the direction and velocity of moving objects. Tracklets require at least two detections of a moving object to be made within a night, and in the case of the LSST, within 20-30 minutes. This cadence requirement means that fast moving objects such as NEOs and PHAs may not have a second detection that falls within the angular velocity range with which tracklets are built. Without a corresponding tracklet, MOPS will not discover such an object. An algorithm that can link six single detections over a 60 night window can increase PHA completeness by a little over 16\%, from the baseline completeness of 65.6\% to 82.5\% at $H \leq 22$. It is therefore prudent to devise such a method. Early prototype codes show promising results and suggest that this may be possible.  [references?]

%Implementation of advanced fast-moving object detection algorithms will likely involve weekly to monthly processing on unlinked transient sources from the LSST data stream. Nightly processing of the alert stream is also a possibility. 

\bigskip
{\centering
\begin{tcolorbox}[width=5.5in]
Key priorities
\begin{itemize}[leftmargin=*]
    \item{\vspace{-0.2cm}Linking detections over long periods of time than MOPS currently allows, allowing for improved detection rates of slow-moving outer solar system objects, using software like HelioLinC}
    \item{\vspace{-0.2cm}Potential removal of the need for tracklets for moving object discovery, allowing for improved discovery rates of solar system objects from isolated detections}
\end{itemize}
\end{tcolorbox}
}

\bigskip
\noindent\hangindent=.35cm\hangafter=1{\bf Processing requirements:} Moderate to high time sensitivity.

\noindent\hangindent=.35cm\hangafter=1{\bf Software dependencies:} Does not require input from other software tools discussed here.

\noindent\hangindent=.35cm\hangafter=1{\bf Required input data:} Unlinked transient sources detected by the LSST pipeline, either via the LSST Alert Stream or periodic data releases.

\noindent\hangindent=.35cm\hangafter=1{\bf Expected output data:} Newly discovered small solar system objects beyond those discovered by the baseline LSST MOPS pipeline.

%%%%%%%%%% SLOW-MOVING OBJECT DETECTION %%%%%%%%%%

%\bigskip
%\subsection{Slow-moving object detection}\label{subsection:appendix_slow_mover_detection}
%The baseline LSST MOPS pipeline will be unable to detect motion of solar system objects below a certain sky-plane velocity limit, limiting its ability to discover extremely distant and therefore extremely slow-moving outer solar system objects.  Discovery of such objects will require development of dedicated software to link and derive orbits from transient detections that may only exhibit appreciable motion over several days or weeks.
%\bigskip
%\noindent\hangindent=.35cm\hangafter=1{\bf Processing requirements:} Moderate time sensitivity.
%\noindent\hangindent=.35cm\hangafter=1{\bf Software dependencies:} Does not require input from other software tools discussed here.
%\noindent\hangindent=.35cm\hangafter=1{\bf Required input data:} Unlinked transient sources detected by the LSST pipeline, either via the LSST Alert Stream or periodic data releases.
%\noindent\hangindent=.35cm\hangafter=1{\bf Expected output data:} Newly discovered slow-moving objects beyond those discovered by the baseline LSST MOPS pipeline.

\clearpage
%%%%%%%%%% PHASE FUNCTIONS %%%%%%%%%%

\subsection{Phase function characterization}\label{subsection:appendix_phase_function_characterization}

Determination of best-fit phase function parameters for inactive objects using a variety of phase function models will be needed for priorities in all science areas related to characterization of sizes, composition, and other physical properties (e.g., densities, which require size estimations) of small solar system bodies, as well as activity detection and characterization (Appendices~\ref{subsection:appendix_activity_detection} and \ref{subsection:appendix_outburst_disruption_detection}); which will rely on accurate knowledge of the expected brightnesses of cometary nuclei at given viewing geometries).  Additionally, compositional characterization of solar system objects with LSST data will rely on comparisons of absolute magnitudes of objects measured in different filters (Appendix~\ref{subsection:appendix_compositional_characterization}), while phase function determination will also be needed for improving the accuracy of rotational property determination efforts (Appendix~\ref{subsection:appendix_rotational_characterization}).
Eventually, joint fits that simultaneously characterize phase function, color, and rotational variations may allow for precise and comprehensive characterization for the majority of small bodies (cf.\ Appendix~\ref{subsection:appendix_rotational_characterization}).
\looseness=-1

$H$,$G_{12}$ parameters \citep{muinonen2010_threeparamphasefunction} in each individual LSST filter are expected to be provided for all solar system objects by MOPS, but it may be of interest to some researchers for best-fit parameters for other phase function models to also be available.  The $H$,$G_{12}$ phase function model was designed to improve upon the predictive power of the previous IAU standard photometric model, the $H$,$G$ phase function model \citep{bowell1989_astphotmodels_ast2} for sparse data sets.
For observationally challenging objects for which very little data may be able to be obtained (e.g., extremely small or distant objects, or fast-moving objects with limited observing windows), though, it will be useful to consider how best to characterize as many of their basic properties (e.g., size and color) as possible given the limited data that LSST is expected to be able to collect.
Meanwhile, more complex models such as the Hapke photometric phase function model \citep{hapke1981_bidirectionalreflspec1_jgr,hapke1984_bidirectionalreflspec3,hapke1986_bidirectionalreflspec4,hapke1981_bidirectionalreflspec2,hapke2002_bidirectionalreflspec5} would allow for more detailed probing of likely material properties for objects for which sufficient and appropriate data (e.g., dense photometry at phase angles of $\alpha<2^{\circ}$) are available, from parameters characterizing such phase function properties as the width of the opposition surge, for example \citep[cf.][]{waszczak2015_ptflcs}.

For size estimation alone in cases where insufficient data are available for deriving $H$,$G_{12}$ fits in all filters, a ``single-filter'' phase function fit (where assumptions about an object's color are used to transform photometry obtained across all LSST filters to a single common filter, such as $r$) could be useful for at least obtaining a reasonable absolute magnitude estimate.
Simpler phase functions (e.g., linear functions, or an $H$,$G$ phase function model where the frequently used $G$=0.15 is assumed, reducing the set of free parameters to just the absolute magnitude, $H$) may also be able to provide useful information.
For example, if reasonable magnitude estimates can be derived at non-near-opposition phase angles using fewer data points than are needed for a full phase function fit, these could be used to derive color estimates earlier in the survey than would otherwise be possible relying only on full $H$,$G_{12}$ fits.  If reasonable magnitude estimates can be derived from these simpler phase function models, at least over certain phase angle ranges, this approach may also be useful for performing photometric searches for activity early in the survey.  Studies will be required, though, to assess whether such simplified approaches can actually deliver superior information (e.g., accurate absolute magnitudes, and accurate magnitude predictions at non-near-opposition phase angles) with fewer data points than $H,G_{12}$ fitting.
We note that given the much smaller range of phase angles over which TNOs can typically be observed ($0^{\circ}\leq\alpha\lesssim2^{\circ}$), their phase functions are already typically approximated as linear \citep[at least down to $\alpha\sim0.2^{\circ}$, at which point an opposition surge is typically observed;][] {sheppard2002_kbophotometry,belskaya2006_tnooppositioneffect}.

\looseness=-1

%Meanwhile, parameters for more complicated phase function models, such as the Hapke model, will require much larger sets of photometry data to compute, and so will only be derivable for a much smaller subset of objects than the set for which $H$,$G_1$,$G_2$ values are available.  However, for that smaller set of objects (which for LSST, should still be substantial) for which Hapke modeling is possible, far more detailed compositional characterization will be possible.
%\looseness=-1

Especially in the cases of inactive comet nuclei, care must be taken to only include photometry that is uncontaminated by ejected dust or gas when computing phase functions. Given the discoveries of unexpected activity even for dynamically asteroidal objects in recent years \citep{jewitt2015_actvasts_ast4}, evaluation of {\it every} detection to be used for phase function computation to check for the presence of activity would be a highly desirable component of that process.  As more data are acquired for individual objects, activity detection sensitivity may change as phase function and rotational parameter solutions are refined, and as such, periodic recomputation of phase function parameters to ensure the omission of detections determined later to contain activity will be desirable.
\looseness=-1

Initial computation of $H$,$G_{12}$ parameters or parameters for other simpler phase functions, if they are found to be useful, should be performed relatively quickly for making apparent magnitude predictions and color estimations.  Computation of more complex phase function parameters, such as Hapke model parameters, is less urgent. % Given the amount of data needed to compute new phase functions or make substantive improvements to previously calculated phase functions, extremely frequent analyses (e.g., daily) are unlikely to be useful. \mtb{I disagree with this ;) }

We note that tools developed for computing best-fit phase function parameters should have the ability to use data sets restricted by certain parameters, e.g., specified time periods or viewing geometries.  Characterization of surface property changes over time due to activity or disruptive processes, which will rely on measurements of phase function parameters and colors (for which phase function solutions are required to compute), is a priority for active object and inner solar system object science (Appendix~\ref{subsection:appendix_physical_evolution}).  It has also been noted that computed absolute magnitudes can actually vary at different viewing aspect angles \citep[e.g.][]{muinonen2010_threeparamphasefunction}, and so the ability to compute independent phase function solutions for objects at different viewing geometries may be an important capability to have for activity detection via photometric enhancement detection (Appendix~\ref{subsection:appendix_activity_detection}), which relies on accurate magnitude predictions for inactive objects.

%\dr{Needs to be edited, some paragraphs repeat.}

\bigskip
{\centering
\begin{tcolorbox}[width=5.5in]
Key priorities
\begin{itemize}[leftmargin=*]
    \item{\vspace{-0.2cm}Computation of phase function parameters in individual filters for phase function models other than $H$,$G_{12}$, including both less complex (e.g., linear, $H$,$G$ assuming $G$=0.15, and $H$,$G$ with $G$ as a free parameter) and more complex ($H$,$G_1$,$G_2$ and Hapke) phase function models}
    \item{\vspace{-0.2cm}Computation of equivalent phase function parameters for a single filter for various phase function models, making assumptions for object colors}
    \item{\vspace{-0.2cm}Procedures to ensure omission of detections where activity is present from phase function parameter computations}
\end{itemize}
\end{tcolorbox}
}

\bigskip
\noindent\hangindent=.35cm\hangafter=1{\bf Processing specifications:} Moderate time sensitivity.  Ability to compute parameters only using data from specified time periods needed (cf.\ Appendix~\ref{subsection:appendix_physical_evolution}).
%Monthly computation of new or updated phase function parameters for objects with adequate data sets for the phase function model being used should be sufficient.

\noindent\hangindent=.35cm\hangafter=1{\bf Software dependencies:} Will require activity detection (Appendix~\ref{subsection:appendix_activity_detection}; for removal of activity-contaminated detections) for input.

\noindent\hangindent=.35cm\hangafter=1{\bf Required input data:} Photometry data, observing geometry metadata (heliocentric and geocentric distances, phase angles), and activity detection parameters.

\noindent\hangindent=.35cm\hangafter=1{\bf Expected output data:} Parameter values for various phase function models in different filters.

\clearpage
%%%%%%%%%% COMPOSITIONAL CHARACTERIZATION %%%%%%%%%%

\subsection{Compositional characterization}\label{subsection:appendix_compositional_characterization}

Compositional characterization of small solar system bodies will be needed for science priorities related to general physical characterization of small bodies, investigations into how compositional properties are correlated with others such as the presence of activity or dynamical properties, identification of objects with probable cometary origins, investigations requiring size determinations, which will require assumptions for albedos, which in turn can be estimated from taxonomic classifications, and studies of compositional similarities between potentially related objects.  Needed compositional characterization functions will include determinations of colors, spectral slopes, taxonomic types, and hydration states of primitive asteroids.
\looseness=-1

Determining colors (i.e., the differences in magnitudes measured in different filters for an object, assuming simultaneous observations) or spectral slopes (i.e., the slope in wavelength-magnitude space, usually expressed in percent of reddening per 1000~\AA\ in wavelength) is complicated by the effects of an object's phase function and rotation on its observed brightness at any given time, which must be removed before the wavelength dependence of an object's observed brightness (i.e., its colors) can be reliably determined.  Normally, these effects can be minimized and colors can be measured by obtaining multi-filter observations on the same night (minimizing brightness variations due to changes in viewing geometry) and ideally as close in time as possible (e.g., by conducting observations that are actually simultaneous or by bracketing observations in certain filters by others, thus minimizing or at least accounting for the effects of brightness variations due to rotation).

Neither of these approaches is likely to work with LSST data, given its expected sparse cadence.  Colors should instead be derivable from LSST data by phase function fitting to single-filter datasets (Appendix~\ref{subsection:appendix_phase_function_characterization}), i.e., by comparing $H$ magnitudes computed from data in different filters, which has the effect of averaging out rotational brightness variations over time. The accuracy of these colors will depend on both the phase angle coverage achieved in each filter and the amplitude and period of each object's rotational brightness variations. These in turn will depend on the orbital parameters and spin and shape characteristics of individual objects. However, over a 10-year survey, LSST should provide enough measurements of main-belt asteroids and TNOs for reasonably accurate color determination. It is possible that the colors of many NEOs will be less secure, unless they are large and bright enough to be observed over a significant time period by LSST.  Independent follow-up observations may be required to obtain more secure colors for smaller and fainter NEOs.
Alternatively, given the large amount of LSST data that should eventually be available for many objects, simultaneous fits for the phase functions, colors, and rotational properties of objects observed by LSST may be possible (cf.\ Appendix~\ref{subsection:appendix_rotational_characterization}).

%\dr{This is NOT how I think it will work in practice. In practice, we'll do a joint phase-color-rotation fit; there will be plenty of data to do so and the results will be better than ignoring rotation.}

%We also note that while computing colors by comparing $H$ magnitudes is probably the most reasonable default approach since $H$ magnitudes will already be computed for other purposes, another approach for computing colors, especially for objects with limited available data, is to compare magnitudes at non-zero phase angles, e.g., $\alpha\sim20^{\circ}$.  This approach would allow colors to be computed from linear phase function solutions, which generally require less data to derive than $H$,$G_{12}$ phase functions.
\looseness=-1

Likely taxonomic classifications for solar system objects will likely be determined using the nomenclature of the Bus \& Binzel \citep{bus2002_smasstaxonomy} and Bus-DeMeo \citep{demeo2009_taxonomy} systems and, possibly, new systems informed from LSST observations themselves.  The precise method by which objects will be classified is not yet clear, however.  The classification method will likely rely mostly on $ugrizy$ colors, similar to how \citet{carvano2010_sdsstaxonomy} used Sloan Digital Sky Survey (SDSS) colors to derive likely taxonomic types for asteroids detected by the SDSS, but may also incorporate other input such as phase function slope parameters.  Consideration will also need to be given to whether and how taxonomic classifications can be assigned to objects without full $ugrizy$ colors, since it is probable that there will be objects of interest, such as fast-moving NEOs with limited observing windows, for which these will not be available.

%\hhh{Besides being diagnostic of surface composition, taxonomic classifications can also be used to estimate albedos, which are in turn needed for estimates of size, a key piece of data for evaluating NEO impact hazards.}
%\msk{Taxonomic classification could (should) also be done on subsets of ugrizy, especially for NEAs that might not get the full ugrizy filter set.  Albedo has some correlation with class, and getting albedo is important for estimating size, and I presume NEA size is an important quantity to know.}
\looseness=-1

The determination of hydration states from multi-band broadband photometry was discussed by \citet{rivkin2012_hydratedasts}, who used SDSS data to identify asteroids likely to have spectra with a 0.7~$\mu$m absorption feature \citep{vilas1989_phyllosilicatefeatures,vilas1994_hydrationfeatures}, which in turn has been found to be correlated with the presence of 3~$\mu$m absorption attributed to hydrated minerals 
\citep{feierberg1985_3micronfeature,jones1990_cpdasteroids,rivkin2003_ctypeasteroids}. It will presumably be possible to utilize a similar technique to determine hydration states of primitive asteroids using broadband colors determined from LSST data.
\looseness=-1

For very faint objects or those with limited observing windows (leading to issues like poor sampling at small phase angles or over a sufficient range of phase angles), it may be difficult to derive complete $H$,$G_{12}$ or even $H$,$G$ phase functions in all filters.
One approach for dealing with such situations could be to simply fit linear phase functions to data in different filters obtained over phase angle ranges of $10^{\circ}\lesssim\alpha\lesssim30^{\circ}$ (over which many asteroids exhibit approximately linear phase function behavior).  Colors could then be computed from reduced magnitudes at non-zero phase angles (e.g., $\alpha=20^{\circ}$), which will likely be easier to constrain for a larger number of objects, rather than requiring colors to be derived from absolute $H$ magnitudes at $\alpha=0^{\circ}$.

As is the case for phase function determinations, care must be taken to only include photometry that is uncontaminated by ejected dust or gas when computing rotational properties. Otherwise, there is a risk of miscalculation of an object's lightcurve amplitude or shape.  As discussed above, evaluation of detections of all types of objects for the presence of activity is highly recommended for rotational lightcurve and lightcurve inversion studies. \looseness=-1

Monthly computation of new or updated compositional properties for objects with adequate data sets should be sufficient.  Reasonably timely determination of compositional properties is needed as they may be used to identify particular targets of interest for further analysis using LSST data or follow-up observations with other facilities.  Given the amount of data needed to make new determinations of compositional properties or make substantive improvements to previously calculated properties, though, much more frequent analyses are unlikely to be useful.

\bigskip
{\centering
\begin{tcolorbox}[width=5.5in]
Key priorities
\begin{itemize}[leftmargin=*]
    \item{\vspace{-0.2cm}Computation of colors by comparing $H$ magnitudes computed for different filters, or via alternative methods such as comparing predicted magnitudes at non-zero phase angles using linear phase function fits}
    \item{\vspace{-0.2cm}Determination of likely taxonomic classifications and hydration states of solar system objects from broadband colors}
    \item{\vspace{-0.2cm}Procedures to ensure omission of detections where activity is present from compositional property determinations}
\end{itemize}
\end{tcolorbox}
}

\bigskip
\noindent\hangindent=.35cm\hangafter=1{\bf Processing specifications:} Moderate time sensitivity.  Ability to compute parameters only using data from specified time periods needed (cf.\ Appendix~\ref{subsection:appendix_physical_evolution}).

\noindent\hangindent=.35cm\hangafter=1{\bf Software dependencies:} Will require activity detection (Appendix~\ref{subsection:appendix_activity_detection}; for removal of activity-contaminated detections) for input.

\noindent\hangindent=.35cm\hangafter=1{\bf Required input data:} Phase function parameter values in different filters, observing geometry metadata (heliocentric and geocentric distances, phase angles), and activity detection parameters.

\noindent\hangindent=.35cm\hangafter=1{\bf Expected output data:} Parameter values related to compositional characterization of small solar system objects (e.g., colors, spectral slopes, taxonomic classifications, hydration states).

\clearpage
%%%%%%%%%% ROTATIONAL CHARACTERIZATION %%%%%%%%%%

\subsection{Rotational characterization}\label{subsection:appendix_rotational_characterization}

Determination of rotational properties for inactive solar system objects will be required for science priorities related to studies of rotational periods, shapes, and spin angular momentum distributions of objects in different populations, identification and characterization of binary and other multi-object systems, and investigations of rotational fission events.

Examples of rotational properties of interest include rotation periods, lightcurve amplitudes, lightcurve shapes, and spin-axis orientations \citep{kaasalainen2004_sparsephotometricmodels,durech2009_sparsephotometry}, and even full shape models for objects for which a sufficiently large amount of photometric data is available \citep{kaasalainen2001_lightcurveinversion1,kaasalainen2001_lightcurveinversion2,hanus2012_sparsephotometry}.  Detection of binarity or other higher-order multiplicity \citep{pravec2006_binarysurvey,scheirich2009_binarylightcures,thirouin2014_tnobinaries,scheirich2015_binaryNEO}, as well as detection and characterization of non-principal axis rotation \citep{pravec2005_tumblingasteroids,samarasinha2015_nparotation} will also be important tasks related to analyzing the rotational lightcurves of inactive objects.

Phase dispersion minimization (PDM) is a technique that is commonly used for period determination \citep{stellingwerf1978_pdm}.  This method makes no a priori assumptions about the shape of a lightcurve and simply seeks to find periods that minimize the scatter in lightcurves phased to those periods.  Other period-finding methods that could also be utilized include Fourier and Lomb-Scargle analysis \citep[e.g.,][]{szabo2016_keplerasteroidlightcurves}.  These techniques will likely be most easily applied to data from mini-surveys or Deep Drilling Fields where a number of closely-spaced detections of an object are obtained over a period of several hours or so, as these data sets most closely mimic the kind of data sets that are normally obtained to determine an object's rotational lightcurve using targeted classical observations. The sparse nature of most of the data sets for individual asteroids that are expected to be obtained by LSST's Wide-Fast-Deep survey will mean that additional steps (e.g., corrections for changes in observational geometry and observations in different filters) will be needed before they can analyzed using standard period-finding algorithms.  That said, substantial work has been done in recent years for rotation period determination of asteroids from sparse data \citep{kaasalainen2004_sparsephotometricmodels,durech2009_sparsephotometry}, and preliminary analyses have suggested that $\sim$50 photometric points per asteroid (where many asteroids should have hundreds of detections each) may be sufficient for deriving reasonable simultaneous fits to rotational lightcurves, phase functions, and colors with minimal assumptions (see next paragraph).

Derivation of a full shape model requires finding the global best-fit solution for the inverse problem, i.e., a model that gives the best $\chi^2$ fit to the data. The model's parameters include the sidereal rotation period, direction of the spin  axis, the shape (approximated by a convex polyhedron), and parameters of the phase function. In practice, finding a best-fit solution can be accomplished by using the gradient-based lightcurve inversion algorithm of \citet{kaasalainen2001_lightcurveinversion2} many times from different starting points in the available parameter space to map all local minima in $\chi^2$ and select the global minimum. This scanning of the available parameter space is a computationally demanding process but can be split into smaller tasks that are solved separately. This approach is currently used in the distributed computing project Asteroids@home \citep{durech2015_asteroidsathome} and can also be used for LSST data. Photometric data in different filters are treated as separate lightcurves with their offsets being free parameters of the optimization. In this way, colors as shifts between filters are optimized and determined together with other parameters. Phase function parameters are assumed to be the same for all filters or, if sufficient data are available, they can be independently determined for different filters to account for phase-reddening effects.

Inversion algorithms and computational resources (via the Asteroids@home network of volunteers) are ready to process LSST asteroid photometry. However, a crucial step of deciding whether a model that mathematically fits available data represents physical reality or not is far from being solved optimally. The current approach to this problem is based on $\chi^2$ significance levels and various reliability checks \citep{durech2018_asteroidmodels}, but more modern methods using regression trees \citep{waszczak2015_ptflcs} or neural networks might be a better option. This is the most important part of the inversion pipeline that still needs to be developed before the start of LSST operations.

For the majority of objects observed by LSST, it will likely be helpful to use measured or assumed colors to convert LSST photometry of a given object to a single filter to maximize the number of data points available for rotational lightcurve analysis.  That said, in cases where sufficient data are available, independent construction of rotational lightcurves in individual filters for a given object would be worthwhile, as it would allow for studies such as searches for rotational compositional heterogeneity.

Even in the absence of sufficient data to construct full lightcurves, it will still be possible to identify extreme objects such as contact binaries from sparse photometry by searching for objects with large inferred photometric ranges \citep[e.g.,][]{mann2007_contactbinarytrojans,lacerda2008_contactbinaries,sonnett2015_binaries}.  In the cases of super-fast rotators (i.e., with rotation periods comparable to or shorter than a single LSST exposure), it may be possible to estimate rotation periods from analyzing brightness changes along the trailed detections of objects with large sky-plane velocities.  In the case of highly active comets, rotation properties can sometimes be determined from the analysis of morphology changes (e.g., from the motion of jets) or simple aperture photometry \citep[e.g.,][]{millis1986_halleyrotation,li2017_252p}, but it is unclear at the present time whether either type of analysis will be possible with LSST's expected cadence.

A complementary approach to modeling spin and shape properties of individual objects is to model distribution of these properties on a population level \citep{mcneill2016_mbabrightnessvariations,cibulkova2018_mbashapes,mommert2018_gaiaasteroidshapes}. Theoretical background and practical algorithms have already been developed \citep{nortunen2017_asteroidshapeelongation,nortunen2017_asteroidshapespins}, but the performance of these algorithms has yet to be optimized for LSST-scale data sets.

Monthly computation of new or updated rotational properties for objects with adequate data sets should be sufficient.  Reasonably timely determination of rotational properties is needed as they may be used to identify particular targets of interest for further analysis using LSST data or follow-up observations with other facilities.  Given the amount of data needed to make new determinations of rotational properties or make substantive improvements to previously calculated properties, though, much more frequent analyses are unlikely to be useful.

\bigskip
{\centering
\begin{tcolorbox}[width=5.5in]
Key priorities
\begin{itemize}[leftmargin=*]
    \item{\vspace{-0.2cm}Determination of rotation rates and axis ratios for large numbers of solar system objects from sparse photometric data in a range of filters}
    \item{\vspace{-0.2cm}Determination of rotational lightcurve characteristics for solar system objects in individual filters to search for compositional heterogeneity}
    \item{\vspace{-0.2cm}Determination of advanced spin state properties (e.g., spin axis orientation, shape models) for solar system objects for which sufficient photometric data are available}
    \item{\vspace{-0.2cm}Identification of objects with large inferred photometric ranges, even in cases where insufficient data are available to construct full rotational lightcurves}
\end{itemize}
\end{tcolorbox}
}

\bigskip
\noindent\hangindent=.35cm\hangafter=1{\bf Processing requirements:} Moderate time sensitivity.  Image data potentially required in certain specific cases.  Ability to compute parameters only using data from specified time periods needed (cf.\ Appendix~\ref{subsection:appendix_physical_evolution}).

\noindent\hangindent=.35cm\hangafter=1{\bf Software dependencies:} Will require activity detection (Appendix~\ref{subsection:appendix_activity_detection}; for removal of activity-contaminated detections) for input.

\noindent\hangindent=.35cm\hangafter=1{\bf Required input data:} Photometry data, phase function parameters, observing geometry metadata (heliocentric and geocentric distances, phase angles), color data, and activity characterization parameters.

\noindent\hangindent=.35cm\hangafter=1{\bf Expected output data:} Parameter values related to rotational properties of solar system objects.

\clearpage
%%%%%%%%%% MULTI-OBJECT SYSTEM DETECTION %%%%%%%%%%

\subsection{Detection and characterization of multi-object systems}\label{subsection:appendix_multi_object_systems}

The ability to detect resolved multi-object systems via PSF analyses will be important for science priorities related to studies of fission events %(A-3, C-9)
and the dynamics and other characteristics of binary and other multi-object systems. %(A-5, D-6).
This capability will also be relevant to science priorities related to the detection of binary or higher order multi-object systems via lightcurve analyses.% (B-7).
\looseness=-1

The general procedure by which we envision that this task could be carried out is by first identifying detections of known solar system objects that exhibit significant deviations from simple PSF fits (or trailed PSF fits, as applicable) as determined by the baseline LSST pipeline.  If these surface brightness profile deviations cannot be attributed to contamination from nearby sources or activity, image data for flagged detections could then be analyzed in more detail for indications of binarity or higher-order multiplicity \citep[e.g.,][]{agarwal2016_288p,agarwal2017_288p}.  Further characterization could also conceivably be performed, such as performing a double or multiple PSF fit in order to estimate relative astrometry. In many cases, sufficient orbital motion will be detectable over the course of the LSST survey for full orbital fits. More consideration is needed, however, to assess how best to automate these procedures, or determine which portions of this procedure lend themselves best to automation and for which portions manual analysis may be preferred.

Semi-resolved or unresolved multi-object systems, especially those where one object is significantly brighter than the other(s), might also be detectable from astrometric residuals due to the photocenters of such systems being frequently offset from their barycenters.  We also note that some outer solar system binaries will be so well separated ($\gtrsim1''$) in LSST data that they are tracked as separate objects, which is currently not well flagged by the MPC. Both of these cases could potentially lead to challenges (albeit, rarely) with MOPS linking and so likely require some consideration beyond simple interest in detecting multi-object systems.

Monthly searches for multi-object systems should be sufficient for enabling follow-up observations for characterization of those systems.  Processing will potentially be more urgent for objects with short available observing windows (e.g., NEOs), but less urgent for objects that remain visible for longer periods of time.

\bigskip
{\centering
\begin{tcolorbox}[width=5.5in]
Key priority
\begin{itemize}[leftmargin=*]
    \item{\vspace{-0.2cm} Identification of objects exhibiting significant deviations from simple PSF fits that are potentially indicative of binarity or higher order multiplicity, and determination of parameters relevant to the characterization of multi-object systems}
\end{itemize}
\end{tcolorbox}
}

\bigskip
\noindent\hangindent=.35cm\hangafter=1{\bf Processing requirements:} Moderate to high time sensitivity.  Images required.

\noindent\hangindent=.35cm\hangafter=1{\bf Software dependencies:} Will require activity detection (Appendix~\ref{subsection:appendix_activity_detection}; for removal of activity-contaminated detections) for input.

\noindent\hangindent=.35cm\hangafter=1{\bf Required input data:} PSF and trail-fitting residual data from the Prompt Product data stream to flag candidates for further analysis. Image data required for detections flagged by PSF and trail-fitting residual data.

\noindent\hangindent=.35cm\hangafter=1{\bf Expected output data:} Parameter values related to rotational properties of solar system objects

\clearpage
%%%%%%%%%% ACTIVITY DETECTION %%%%%%%%%%

\subsection{Activity detection}\label{subsection:appendix_activity_detection}

At the intersection of the study of inactive and active objects is the determination of whether activity exists in any given observation of a solar system object.  The simplest case where activity detection is relevant is the discovery of a new comet, i.e., the detection of activity for a newly discovered solar system object.  It is also likely that with the greater image depth and sensitivity of LSST relative to current surveys, many currently known solar system objects that have been hitherto classified as asteroids will be discovered to exhibit comet-like activity, as has already happened with several active asteroids, asteroids on cometary orbits, and Centaurs.  
\looseness=-1

We broadly define activity here as the deviation of an object's appearance or brightness from that of a point-source-like object with a known absolute magnitude.  We note here that activity may not necessarily be ``cometary'', i.e., produced as a result of the sublimation of volatile ices, but could instead also be produced by disruptive processes such as impacts, rotational destabilization, or surface avalanches \citep[e.g.,][]{ishiguro2011_scheila2,hirabayashi2015_rotationalshedding,jewitt2015_actvasts_ast4,steckloff2016_hartley2avalanches,hofmann2017_impacttriggeredlandslides}.
\looseness=-1

There is a wide variety of potential methods that can be used to search for activity in solar system objects.  For the LSST survey, it will be highly desirable to implement multiple activity detection methods, as all methods tend to be preferentially sensitive to certain types of active morphologies and not others.
%For example, a nucleus-dominated object with a faint coma that does not appreciably change the object's PSF width will likely not be identified as active by an algorithm that relies on PSF measurements for activity detection.  Similarly, an algorithm that assumes radially symmetric activity will not be particularly sensitive to objects with faint dust tails oriented in a particular direction. 
Activity detection methods that could be used for the LSST survey include the following:
\looseness=-1

\bigskip
{\it (a) Visual inspection:} The most straightforward method of activity detection is simple visual inspection to search for objects with comet-like appearances \citep[e.g.,][]{hsieh2009_htp,gilbert2009_cfhtmbcs,gilbert2010_cfhtmbcs,chandler2018_safari}.  While requiring substantially more human effort compared to automated methods, visual inspection has the advantage of being sensitive to a wide range of active morphologies, from strong comae and tails to weak comae or faint dust trails, and is not as prone (depending on the experience of the individual performing the visual inspection) to many of the more obvious types of false positives that some automated search algorithms may produce \citep[cf.][]{hsieh2015_ps1mbcs}.
\looseness=-1

At the scale of the LSST survey, visual inspection of all moving objects by professional astronomers will be an overwhelming task.  Citizen science \citep[cf.][]{lintott2011_galaxyzoo,hsieh2016_comethunters,schwamb2017_comethunters} may provide a means for making human inspection of all solar system object detections from LSST to search for activity a more realistic possibility. However, careful design of user training materials, classification tasks, and methods of interpreting submitted results, as well as ongoing engagement with citizen scientists by professional astronomers will be required to ensure reliable results derived from citizen scientist input.  Strong outreach and recruitment efforts via traditional and social media will also be needed to achieve sufficient effective processing capacity (in the form of user numbers and involvement levels) to keep pace with the LSST data stream.

Whether performed by professional scientists or citizen scientists, however, we note that visual inspection will not be as sensitive as automated activity detection algorithms for certain active morphologies.  Examples of such morphologies are a compact coma that slightly broadens an object's PSF but is otherwise undetectable by eye \citep[e.g.,][]{hsieh2012_288p}, or completely unresolved dust emission contained entirely within an object's seeing disk \citep[e.g.,][]{hsieh2015_324p}.

\medskip
{\it (b) PSF width comparison:} Perhaps the most straightforward and common automated method of activity detection is the comparison of the PSF width of the detection of a solar system object with the average PSF width of nearby reference stars \citep[e.g., as done for the Pan-STARRS1 survey;][]{hsieh2015_ps1mbcs}.  This approach is best-suited for detecting active objects with strong near-nucleus comae that comprise a significant fraction of an object's total scattering surface area, but is less well-suited for detecting objects with bright nuclei and faint coma (such that the coma does not appreciably affect the nucleus's PSF profile).

\medskip
{\it (b) Photometric enhancement detection:} Another method for detecting extremely low-level activity is the use of photometric analysis to investigate whether an object's brightness deviates from what is expected based on prior observations of the object when it was believed to be inactive.  An increase in brightness for an object that is otherwise stellar in appearance could indicate the presence of unresolved dust emission associated with the object.  This technique was used to discover activity in 95P/(2060) Chiron \citep{bus1988_chiron,tholen1988_chiron,meech1989_chiron,hartmann1990_chiron}, detect the start of activity of 324P/La Sagra in 2015 \citep{hsieh2015_324p}, and search for active asteroids in the MPC's archive of submitted observations \citep{cikota2014_activeasts}.\looseness=-1

This method is useful for detecting activity that would otherwise be undetectable by the human eye (and can also detect activity that {\it is} visible of course).  However, it requires reliable predictions for how bright an object is expected to be at any given time, assuming that it is inactive, as well as accurate photometry at the time of observation, especially in cases including overlapping or nearby field stars or galaxies.  Information about the level of expected rotational variation in the object's brightness would also be useful for assessing the significance of minor enhancements in measured brightnesses.
\looseness=-1

\medskip
{\it (d) Multi-aperture photometry analysis:} \citet{waszczak2013_ptfmbcs} used a method for searching for activity associated with moving object detections by the Palomar Transient Factory where the ratio of the fluxes of an object measured using photometry apertures of two sizes was compared to that of nearby stars to obtain a measure of how ``condensed'' an object was.  The larger aperture used to measure photometry for active objects would be expected to capture more flux than it would for inactive objects, resulting in a larger ratio of the flux measured using the larger aperture to the flux measured using the smaller aperture.

\medskip
{\it (e) Multi-azimuth photometry analysis:} \citet{sonnett2011_cfhtmbcs} used a method for searching for activity associated with asteroid detections by the Canada-France-Hawaii Telescope as part of the Thousand Asteroid Lightcurve Survey \citep[][]{masiero2009_talcs} where an annulus of sky around each detection was divided into pie segments to search for anomalous brightness in one of the segments that would correspond to azimuthally directed cometary features.  Given that previously discussed detection methods are most effective at detecting azimuthally symmetric activity (e.g., spherical coma), a detection method such as this one designed to find azimuthally asymmetric activity (e.g., tails or trails) would be very useful to implement for the LSST survey.

\medskip
{\it (f) NoiseChisel:} {\tt NoiseChisel} is a noise-based non-parametric technique optimized for detecting extremely faint diffuse objects and fainter parts of brighter diffuse objects developed by \citet{akhlaghi2015_noisechisel} for detecting and studying irregular or clumpy galaxies.  Unlike traditional ``signal-based'' source detection algorithms, which typically perform best when targets conform to expected morphologies (e.g., Gaussian PSFs), {\tt NoiseChisel} makes no a priori assumptions about the morphological properties of the diffuse targets that it is designed to detect and instead identifies sources by first characterizing the noise characteristics of the background sky and identifying areas exhibiting deviations from that background.

%An initial test has indicated that this technique could show promise for detecting extremely diffuse activity from small solar system objects (Figure~\ref{figure:noisechisel}) and would be worth considering for use on LSST data.

%\begin{center}
%\begin{figure}[ht]
%\centering
%%	\includegraphics[width=\columnwidth]{example}
%\includegraphics[width=3.5in]{fig_noisechisel_p2012t1_example_horizontal.pdf}
%\caption{Left: Image of Comet 358P while active; Right: {\tt NoiseChisel} results showing extent of detected activity above sky noise for the image on the left at the same spatial scale. From M.\ Akhlaghi (private communication)}
%\label{figure:noisechisel}
%\end{figure}
%\end{center}

\medskip
{\it (g) Forced photometry:} Forced photometry refers to photometric measurements performed at the expected position of an object in the absence of an actual detection at that position.  For very faint objects that would otherwise be below LSST's detection limit, brightness increases expected from activity or disruptions could temporarily raise their brightnesses above LSST's detection limit.  As such, for each newly discovered object, it would be desirable to search earlier images to identify times when no detection was reported when the expected position of the object was observed, and then to perform forced photometry in order to determine the upper limits of the object's brightnesses in those images.

If a newly discovered object should have been detectable in earlier images given an absolute magnitude inferred from its current measured brightness (e.g., by assuming $G=0.15$; see Appendix~\ref{subsection:appendix_phase_function_characterization}), it would suggest that the object currently has a brighter intrinsic magnitude than in the past, indicating the possible presence of activity.  Prompt identification and follow-up of newly detected objects suspected of having elevated brightnesses at the time of detection could enable the study of asteroid disruptions in real time, while later analysis of the LSST archive for ``orphan'' detections of objects that should have been detected at other times, but have no other recorded detections, could constrain disruption rates in various small-body populations in the solar system \citep[cf.][]{denneau2015_asteroiddisruptions}.

Similarly, catastrophically disrupted asteroids may be detectable from decreases in brightnesses below LSST's detection limit associated with their near or complete disintegrations following their discoveries.  As such, for each known object, it will also be desirable to search newly obtained images in which the object should appear and identify times when the position of the object is observed but the object is not detected, and then to perform forced photometry at the expected position of the object in order to determine the upper limit of the object's brightness in that image.  If an object should have been detectable given its absolute magnitude computed using previous detections, a lack of detection could indicate that the object has disrupted into smaller, undetectable fragments.  In the absence of direct detections of disruptions, perhaps due to the unobservability of objects at the time of disruption (e.g., for NEOs or comets disrupting at small heliocentric distances, and therefore small solar elongations), this procedure would allow for the indirect detection of disrupted objects.

\medskip
{\it (h) Non-gravitational perturbation detection:}  Active comets are known to experience non-Keplerian dynamical evolution due to non-gravitational forces associated with directed mass loss \citep[e.g.,][]{krolikowska2004_lpcnongraveffects,krolikowska2006_81p,krolikowska2006_lpcnongraveffects,hui2017_activeastsnongrav}.  In principle, comet-like mass loss could therefore be identified via the detection of non-gravitational perturbations, for which software is desired for other purposes as well (Appendix~\ref{subsection:appendix_dynamical_evolution}).

\bigskip
Another useful processing step would be co-addition of separate LSST images to allow deeper searches for material around bodies.  Although this may not be applicable to some impact-related activity, where impact ejecta would sometimes display a fast evolution over days, detection of steady low-level activity would be enhanced with this. Coaddition of two images once per 4 nights over 16 days would increase the signal to noise by $\sim$3. This would allow searches for activity to lower brightness levels from potentially active objects (e.g. Damocloids and other high-eccentricity inert objects).  It should be possible to re-purpose software developed for stacking images for faint object detection (Appendices~\ref{subsection:appendix_lsst_precoveries_recoveries} and \ref{subsection:appendix_faint_object_detection}) for this task.

As a practical matter, we note that visual inspection (perhaps accompanied by more detailed data analysis in certain marginal cases) by professional astronomers with experience in comet identification will likely continue to be a final required step in the verification of the presence of activity for a given object, regardless of initial screening methods (including citizen science), as it has been for many other previous comet search efforts, including those employing automated screening techniques \citep[e.g.,][]{solontoi2010_sdsscomets,waszczak2013_ptfmbcs,hsieh2015_ps1mbcs}.  Confirmation that specific objects are active may also only be available upon acquistion of deeper follow-up observations (cf.\ Appendix~\ref{subsection:appendix_alerts}).  Regardless of the eventual method of confirmation of activity, diligent record-keeping in terms of noting which active object candidates identified by a particular detection algorithm were eventually confirmed to be active and which were not will be extremely useful for refining those activity detection algorithms to improve their reliability, as well as providing training sets for developing new activity detection algorithms using machine learning.

Daily, if not immediate, searches for activity using several of these methods (as some are better suited than others for detecting particular types of activity) would be preferred to enable rapid follow-up and characterization of evolving activity.  As such, this task should be considered highly time-sensitive.

\bigskip
{\centering
\begin{tcolorbox}[width=5.5in]
Key priorities
\begin{itemize}[leftmargin=*]
    \item{\vspace{-0.2cm} Searches for activity using multiple techniques including visual inspection (either by professional astronomers or citizen scientists), PSF width comparison with stellar sources, photometric enhancement detection, multi-aperture photometry analysis, multi-azimuth photometry analysis, {\tt NoiseChisel}, forced photometry, and non-gravitational perturbation detection}
    \item{\vspace{-0.2cm} Stacking of image data to search for faint activity below the detection limit of single-visit exposures}
    \item{\vspace{-0.2cm} Mechanism(s) to enable the visual screening of flagged active object candidates by professional astronomers and keep records of the results in a systematic fashion to enable later studies of detection efficiencies and build training sets for machine learning}
\end{itemize}
\end{tcolorbox}
}

\bigskip
\noindent\hangindent=.35cm\hangafter=1{\bf Processing requirements:}  High time sensitivity.  Image data required.

\noindent\hangindent=.35cm\hangafter=1{\bf Software dependencies:} Will require phase function determination (Appendix~\ref{subsection:appendix_phase_function_characterization}) and rotational property determination (Appendix~\ref{subsection:appendix_rotational_characterization}) for input when available.  Some overlap expected with software development for faint object detection (Appendices~\ref{subsection:appendix_lsst_precoveries_recoveries} and \ref{subsection:appendix_faint_object_detection}), activity characterization (Appendix~\ref{subsection:appendix_activity_characterization}), outburst and disruption detection (Appendix~\ref{subsection:appendix_outburst_disruption_detection}), and non-gravitational perturbation detection (Appendix~\ref{subsection:appendix_dynamical_evolution}).  Output will be required for tasks related to physical characterization of objects where the removal of activity-contaminated detections is required, such as phase function determination (Appendix~\ref{subsection:appendix_phase_function_characterization}), compositional characterization (Appendix~\ref{subsection:appendix_compositional_characterization}), and rotational property determination (Appendix~\ref{subsection:appendix_rotational_characterization}), as well as alert brokering and follow-up observation management (Appendix~\ref{subsection:appendix_alerts}).

\noindent\hangindent=.35cm\hangafter=1{\bf Required input data:} Image data for all solar system object detections, photometry data, detection PSF data, and orbit solution residual data.

\noindent\hangindent=.35cm\hangafter=1{\bf Expected output data:} Activity detection parameter values.

\clearpage
%%%%%%%%%% ACTIVITY CHARACTERIZATION %%%%%%%%%%

\subsection{Activity characterization}\label{subsection:appendix_activity_characterization}

The characterization of the strength, morphology, and other properties of visible comet-like features will be required for several active object science priorities. This information will be crucial for such science tasks as characterizing the evolution of an object's activity and determining the most likely cause of observed activity (e.g., sublimation, or impact or rotational disruption).

A basic measure of activity strength is the observed flux of an active object.  The flux of an active object can be measured or parameterized in a variety of ways.  One of the most commonly used method of parameterization is the use of the $Af\rho$ parameter \citep{ahearn1984_bowell}. $Af\rho$ is a proxy for dust production that is constant under steady-state conditions. In reality, most comets exhibit a change in $Af\rho$ with aperture size, wavelength, and heliocentric distance, and the amount of variation can be used to infer properties of the dust grains and activity. $Af\rho$ can be calculated from the apparent magnitude in variously sized apertures for each filter and requires knowledge of an active object's orbit. 

More advanced versions of $Af\rho$ may also be desirable, such as a nucleus-removed $Af\rho$ or $Af\rho$ normalized to a constant phase angle (e.g., $A({\theta}=0^{\circ})f\rho$). The former case is useful for weakly active comets where the nucleus contributes significantly to the total flux, and can be calculated for objects for which the pre-activity nucleus size is reasonably well known. The latter can be achieved either by using a fixed correction for the dust phase angle dependence (the common standard is the Halley-Marcus phase function compiled by D.~Schleicher\footnote{\tt http://asteroid.lowell.edu/comet/dustphase.html}) or by an iterative fit to long-term brightness behavior corrected for heliocentric distance effects.  Nucleus-removed or phase-corrected $Af\rho$ values need only be calculated after a sufficient quantity of data have been acquired. For high activity or nearby comets, $Af\rho$ should be calculated to very large apertures, at least tens of arcseconds. Use of $Af\rho$ in conjunction with a proximity indicator for the nearest bright stars will allow a quick assessment of optimal aperture size for comparison with neighboring nights.  For comparative studies of the same comet over time or of different comets, it will be desirable to have the ability to compute $Af\rho$ values for fixed physical aperture sizes (e.g., in km) at the distance of each active object at the time of observation.  If surface brightness profiles of all active objects are characterized using average radial surface brightness profiles or multi-aperture photometry (discussed below), this should be relatively straightforward to accomplish (e.g., by computing the comet's flux at the desired aperture size from a measured surface brightness profile, or if multi-aperture data is available, simply selecting the aperture size closest to the desired value or interpolating to find the comet's flux at intermediate aperture sizes).
\looseness=-1

If the nucleus size of the underlying object is known, the total excess flux due to dust for an active object can also be computed, and given assumptions about average grain sizes and grain densities, the total excess mass of visible dust can be computed \citep[e.g.,][]{hsieh2015_324p}.  This information could then potentially be used to estimate net dust production rates for comets observed multiple times during the same active periods.

Characterization of the surface brightness profiles of active objects will also be desired for a variety of science tasks such as characterizing the morphological evolution of an object's activity and inferring the physical properties of ejected material.  In the simplest case of spherically symmetric activity, parameterization of coma morphology can simply consist of determination of a particular power law index describing the average radial surface brightness profile of the coma in different filters.  For objects that exhibit asymmetric activity, simple analyses could include some type of quantification of the degree of circular symmetry for a given active object, and estimation of the position angles and lengths of any cometary tails or trails that can be identified.  Software developed for performing multi-aperture and multi-azimuth photometry for activity detection (Appendix~\ref{subsection:appendix_activity_detection}) should prove useful for this task by providing a means for numerically parameterizing cometary surface brightness profiles.  Performing such analyses on data obtained using different filters would also enable simple searches to be conducted for comets with detectable morphological variations in different filters \citep[e.g.,][]{bauer2003_c2001t4} that could indicate the presence of compositional gradients or gas emission.

For sufficiently bright comets, more advanced morphology characterization could include automated basic Finson-Probstein dust modeling \citep{finson1968_cometdustmodeling1,kramer2017_c2010l5}, which is commonly used to infer ejection times, ejection velocities, and grain size distributions for cometary dust emission from observed morphology.  Simple automated fitting of cometary surface brightness profiles to a syndyne-synchrone network characteristic of this type of dust modeling could produce approximate estimates of parameters such as grain sizes or ejection start times for a large number of comets observed by LSST, allowing for searches for trends with such parameters as comet brightness, nucleus size, dynamical characteristics, and heliocentric distance. Searches for significant morphological deviations from a simple syndyne-synchrone network could reveal instances of collimated dust emission, i.e., the presence of jets.

As active objects fade in brightness, either due to intrinsic weakening of their activity or simply due to geometric effects such as increasing distance from the Sun and Earth, continued monitoring of the evolution of their activity can provide valuable constraints on maximum particle sizes, which in some cases can dominate total ejected masses.  Software developed for stacking images for faint object detection (Appendices~\ref{subsection:appendix_lsst_precoveries_recoveries} and \ref{subsection:appendix_faint_object_detection}) and activity detection (Appendix~\ref{subsection:appendix_activity_detection}) should prove useful for enhancing the visibility of faint features such as residual dust trails.

%\msk{Identification of dust trail candidates?}

Daily calculations of activity strength parameters (e.g., $Af\rho$) will be needed to enable prompt detection of outbursts (Appendix~\ref{subsection:appendix_outburst_disruption_detection}), and in turn, enable follow-up observations.  Monthly computation of other new or updated activity characterization parameters should be sufficient for other science use cases.

\bigskip
{\centering
\begin{tcolorbox}[width=5.5in]
Key priorities
\begin{itemize}[leftmargin=*]
    \item{\vspace{-0.2cm} Computation of parameters characterizing activity strength (e.g., $Af\rho$)}
    \item{\vspace{-0.2cm} Characterization of surface brightness profiles using parameters such as power law indices describing the average radial surface brightness profile, quantification of the degree of circular symmetry for a given active object, and estimation of position angles and lengths of any cometary tails or trails}
    \item{\vspace{-0.2cm} Mechanisms for comparing morphologies of active objects observed with different filters that could indicate the presence of compositional gradients or gas emission}
    \item{\vspace{-0.2cm} Automated Finson-Probstein dust modeling to make approximate estimates of grain sizes and ejection times, and to search for significant morphological deviations from simple syndyne-synchrone networks that could reveal instances of collimated dust emission (i.e., jets)}
\end{itemize}
\end{tcolorbox}
}

\bigskip
\noindent\hangindent=.35cm\hangafter=1{\bf Processing requirements:} Moderate to high time sensitivity.  Image data required.

\noindent\hangindent=.35cm\hangafter=1{\bf Software dependencies:} Will require phase function determination (Appendix~\ref{subsection:appendix_phase_function_characterization}) and rotational property determination (Appendix~\ref{subsection:appendix_rotational_characterization}) for input when available.  Some overlap expected with software development for faint object detection (Appendices~\ref{subsection:appendix_lsst_precoveries_recoveries} and \ref{subsection:appendix_faint_object_detection}), and activity detection (Appendix~\ref{subsection:appendix_activity_detection}).

\noindent\hangindent=.35cm\hangafter=1{\bf Required input data:} Multi-aperture, multi-azimuth photometry, and LSST images of active object detections.

\noindent\hangindent=.35cm\hangafter=1{\bf Expected output data:} Parameter values related to activity characterization.

\clearpage
%%%%%%%%%% OUTBURST AND DISRUPTION DETECTION %%%%%%%%%%

\subsection{Outburst and disruption detection}\label{subsection:appendix_outburst_disruption_detection}

The ability to detect outburst and disruption events for both asteroidal and cometary objects will be needed for active object, NEO, and inner solar system science priorities related to the detection and characterization of cometary outbursts or asteroid disruption events.

Cometary outbursts are known to occur, but their frequency and strength (or even the range of strengths) are not well constrained. Missions to comets have shown that small outbursts are relatively common \citep{ahearn05,vincent16} but generally below the detection threshold for most Earth-based observatories. Small (few 0.1 mag) outbursts are occasionally detected from Earth but require high cadence observations \citep[e.g.,][]{boehnhardt16,knight17} to detect unless they are accompanied by significant morphological changes \citep[e.g.,][]{eisner17}. Large ($\sim$1 mag) outbursts are relatively infrequent and, at present, most often discovered by amateurs (review by \citealt{ishiguro16}).  Meanwhile, outbursts from asteroids are also possible, either due to impact or rotational disruptions \citep[e.g.,][]{jewitt2010_p2010a2,jewitt2011_scheila,bodewits2011_scheila,ishiguro2011_scheila2}, or in the case of main-belt comets \citep[MBCs;][]{hsieh2006_mbcs}, the triggered sublimation of subsurface ices.

LSST will be ideal for constraining the frequency and magnitude of both cometary and asteroidal outbursts. Detections of outbursts require regular brightness monitoring with high photometric precision, ideally with all observations in the same filter and measured in the same aperture size.  In the case of LSST Wide-Fast-Deep observations, which will utilize several different filters, measured or assumed colors could potentially be used to compute equivalent magnitudes in a single reference filter for the purposes of outburst detection.

An outburst can be identified by a significant brightening (detected either directly or via a parameter like $Af\rho$) above that expected from normal activity variations and changing geometric conditions  \citep[cf.][]{mcloughlin2015_outburstphotometry}. In some cases, an outburst may have occured long enough prior to the observations that it cannot be identified photometrically in small apertures, but it might be detected in larger apertures (e.g., older material), or via its morphology. Thus, many of the tools described in this document will be critical to the prompt identification of outbursts. 

Thresholds for triggering alerts and/or automatic follow-up observations should be variable, with conservative thresholds required before triggering additional observations or community-wide alerts, but lower thresholds possible to be set by individual data-rights holders who request notification of potential outbursts and are willing to accept a higher rate of false positives. It is advisable that experienced professional astronomers visually inspect images of potential outbursts before issuing a significant alert to minimize false positives.

Forced photometry was discussed earlier in the context of activity detection (Appendix~\ref{subsection:appendix_activity_detection}), but it can also be applied similarly in the context of outburst and disruption detection.  In addition to applying forced photometry on previously obtained LSST images where newly discovered objects should have been in the field of view (as discussed in Appendix~\ref{subsection:appendix_activity_detection}(g)), performing forced photometry on {\it new} LSST data in which {\it known} objects should be in the field of view but are not detected could also reveal disruption events.  The aim of this technique would be to search for objects that have been detectable up to the present time but undergo a catastrophic disruption, creating fragments that are all individually too faint to be detected by LSST.  If a comparison of the 5-$\sigma$ limiting magnitude at the expected location of a given object with the object's expected brightness based on previous observations indicates that the object should have been detectable but was not seen, it could indicate that the object has been catastrophically disrupted.  In this case, prompt follow-up observations in the form of deep imaging would be highly desirable in order to search for evidence of the disruption, e.g., detections of the resulting fragments of the disruption event.

Daily searches for outbursts or disruption events will be needed to enable prompt follow-up observations for characterization and monitoring of both cometary and asteroid outburst and disruption events.  As such, this task should be considered highly time-sensitive.

\bigskip
{\centering
\begin{tcolorbox}[width=5.5in]
Key priorities
\begin{itemize}[leftmargin=*]
    \item{\vspace{-0.2cm} Identification of cometary outbursts (i.e., significant brightening events above ``normal'' expected brightness evolution) for active comets.}
    \item{\vspace{-0.2cm} Development of a framework for characterizing the nature of different outbursts and prioritizing follow-up observations}
    \item{\vspace{-0.2cm} Use of forced photometry to identify possible disruption events where a newly discovered object should have been detected earlier but was not (i.e., where ejected material may have caused a sudden and substantial increase in an object's scattering surface area, causing a previously undetectable object to be detectable) or a known object is not detected when it should have been (i.e., where a catastrophic disruption may have destroyed the object, leaving behind fragments too small to be detected)}
\end{itemize}
\end{tcolorbox}
}

\bigskip
\noindent\hangindent=.35cm\hangafter=1{\bf Processing requirements:} High time sensitivity.  Image data required.
 
\noindent\hangindent=.35cm\hangafter=1{\bf Software dependencies:} Will require phase function determination (Appendix~\ref{subsection:appendix_phase_function_characterization}) and rotational property determination (Appendix~\ref{subsection:appendix_rotational_characterization}).  Some overlap expected with software development for activity detection and characterization (Appendices~\ref{subsection:appendix_activity_detection} and \ref{subsection:appendix_activity_characterization}).  Output will be required for alert brokering and follow-up observation management (Appendix~\ref{subsection:appendix_alerts}).

\noindent\hangindent=.35cm\hangafter=1{\bf Required input data:} Multi-aperture photometry, previously obtained LSST image data for which newly discovered objects are expected to be in the field of view, and newly obtained LSST image data for which previously known objects are expected to be in the field of view.

\noindent\hangindent=.35cm\hangafter=1{\bf Expected output data:} Detections of likely outburst and disruption events.

%%%%%%%%%% ASTEROID PAIR IDENTIFICATION %%%%%%%%%%

%\bigskip
%\subsection{Asteroid pair identification}\label{subsection:appendix_asteroid_pairs}
%\bigskip
%\noindent\hangindent=.35cm\hangafter=1{\bf Processing requirements:} Moderate time-sensitivity. 
%\looseness=-1
%\noindent\hangindent=.35cm\hangafter=1{\bf Software dependencies:} Does not require input from other software tools discussed here.  Some overlap expected with software development for family association identification (Appendix~\ref{subsection:appendix_families}).
%\noindent\hangindent=.35cm\hangafter=1{\bf Required input data:} Osculating orbital elements.
%\noindent\hangindent=.35cm\hangafter=1{\bf Expected output data:} Identifications of pairs (or larger groupings) of dynamically similar solar system objects

\clearpage
%%%%%%%%%% ADVANCED CHARACTERIZATION %%%%%%%%%%

\subsection{Advanced dynamical characterization}\label{subsection:appendix_advanced_dynamical_characterization}

Use of numerical integrations will be required for the advanced dynamical characterization of certain types of solar system objects, including the identification of candidate resonant outer solar system objects, classification of resonant inner solar system objects (e.g., Earth and Mars Trojans), determination of original unperturbed orbital elements for long-period comets, and characterization of the recent dynamical histories of Centaurs.  Computation of synthetic proper elements using numerical integrations will also be desirable for dynamical clustering analyses (Appendix~\ref{subsection:appendix_dynamical_clustering_identification}).

Development of automated tools for performing dynamical classifications for LSST discoveries will necessarily require defining classification schemes with specific criteria for different types of recent dynamical histories of interest for these objects.  The full output of backward integrations, and not just the resulting classifications, will also be of interest to some scientists, for instance for the dynamical history of Centaurs: this forms a data product of interest itself.

%Identification of outer solar system objects in mean-motion resonances with the giant planets is a listed priority of outer solar system scientists in the SSSC.  
The detailed structure of the population of TNOs provides important constraints on early Solar System dynamical history and, in particular, on Neptune's early orbit evolution \citep[e.g.,][]{malhotra1995_plutoorigin,murrayclay2005_neptuneresonance,levison2008_kborigin,nesvorny2016_neptunemigration}. Identification of resonant outer solar system objects is typically achieved through the use of numerical integrations to characterize their dynamical evolution over $\sim$10~Myr timescales \citep[e.g.,][]{gladman2008_ossnomenclature,gladman2012_resonanttnos,volk2018_neptuneresonantobjects}. Accurate classifications will require well characterized and low orbital element uncertainties (Appendix~\ref{subsection:appendix_uncertainty_characterization}), since uncertainties in semimajor axis of $<0.1$\% are preferable for successful high-quality resonance classification \citep{Bannister:2016ossosdesign}.  
These classifications should be version-controlled and their history retained as its own data product, since near-resonant objects can have their dynamical classification change over time, as their orbital parameters become better refined with additional oppositions of observed arc.

Similarly, numerical integrations will also be desirable for dynamically classifying Earth and Mars Trojan candidates \citep{brasser2002_terrestrialplanettrojans,scholl2005_marstrojans,marzari2013_earthtrojanstability} to determine their stability timescales, and in the case of Earth Trojans, fill in a significant gap in our understanding of a potentially significant population of near-Earth objects \citep{malhotra2019_earthtrojans}.

Analyses involving dynamical integrations are also required to identify the source regions of long-period comets \citep{krolikowska2014_warsawcatalogue,krolikowska2014_oortspikecomets,dybczynski2016_lpcorigin}, an important task for the orbital classification of such comets.  The dynamical classification of Centaurs based on their recent dynamical histories \citep[e.g., whether an object has occupied a JFC orbit in the recent past or appears to still be undergoing its initial migration into the inner solar system;][]{wood2018_chironrings} may also be valuable for developing a better understanding of Centaur activity \citep[cf.][]{jewitt2009_actvcentaurs}.  

%\dr{What about synthetic proper elements? What about inner solar system objects, no dynamical classification?}

Searches for upcoming collisions of small bodies with major planets, like Shoemaker-Levy 9's impact on Jupiter, will ensure these rare events can be observed \citep[e.g.,][]{orton1995_sl9irtf,weaver1995_sl9hubble,sanchezlavega2010_jupiterimpact}.  Analyses can be restricted to objects with calculated MOIDs with one of the major planets (cf.\ Appendix~\ref{subsection:appendix_orbital_parameters}) below a certain threshold (perhaps parameterized by the Hill radius of the relevant planet) to minimize the computational resources needed for this task.

Weekly to monthly dynamical classification of resonant inner and outer solar system objects, Centaurs, and long-period comets should be sufficient for relevant science investigations.  While the primary scientific interest in the identification of resonant outer solar system objects, Centaurs with specific dynamical histories, and long-period comets is in the statistical properties of those populations, timely identification of objects of interest would still be useful for enabling the prioritization of follow-up observations for additional compositional characterization efforts.  Searches for possible upcoming planetary impact events can be conducted on similar timescales, and can be limited to newly discovered objects that have calculated MOIDs with one of the major planets below a certain threshold to minimize the computational burden associated with these searches.

%Characterization of the evolution of comet orbits will be needed for the dynamical classification of those comets.  Comet orbits are known to evolve as they pass through the solar system, making characterization of this evolution important for the identification of, for example, dynamically new comets, potential interstellar objects, and dynamically anomalous Jupiter-family comets.

%Computation of probable orbital elements of high-eccentricity long-period comets from the Oort Cloud prior to significant interaction with the major planets will be of particular interest for identifying comets entering the solar system for the first time, identifying what parts of the Oort Cloud they may sample, and the identification of potentially interstellar objects.

%\hhh{Also need integrations for Centaur classifications...anything else?}

\bigskip
{\centering
\begin{tcolorbox}[width=5.5in]
Key priorities
\begin{itemize}[leftmargin=*]
    \item{\vspace{-0.2cm} Use of dynamical integrations to perform various dynamical characterization tasks such as identifying TNOs in mean-motion resonances with Neptune, dynamically classifying Earth and Mars Trojan candidates, determining possible or likely source regions of long-period comets, and characterizing the recent dynamical histories of Centaurs.}
\end{itemize}
\end{tcolorbox}
}

\bigskip
\noindent\hangindent=.35cm\hangafter=1{\bf Processing requirements:} Moderate time sensitivity.

\noindent\hangindent=.35cm\hangafter=1{\bf Software dependencies:} Object and detection metadata computation (specifically barycentric osculating orbital elements; Appendix~\ref{subsection:appendix_orbital_parameters}).

\noindent\hangindent=.35cm\hangafter=1{\bf Required input data:} Current heliocentric or barycentric osculating orbital elements of solar system objects of interest and their uncertainties

\noindent\hangindent=.35cm\hangafter=1{\bf Expected output data:} Synthetic proper elements, identification of resonant and non-resonant outer solar system objects, any alteration in resonant classification for TNOs, unperturbed orbital elements for long-period comets, and parameters related to the dynamical histories of Centaurs

\clearpage
%%%%%%%%%% DYNAMICAL CLUSTERING IDENTIFICATION %%%%%%%%%%

\subsection{Dynamical clustering identification}\label{subsection:appendix_dynamical_clustering_identification}

Identification of dynamically associated objects will be important for studies of correlations between taxonomy of small solar system objects and their dynamical properties, and compositional relationships among members of families, clusters, and pairs.  Families are groups of small bodies clustered in orbital element space that are believed to have formed from the catastrophic fragmentation of parent bodies \citep{hirayama1918_astfam}.  Over 120 asteroid families are currently known, ranging in estimated ages from several Gyrs to less than 1 Myr, encompassing about $\sim$40\% of all known asteroids \citep[cf.][]{nesvorny2015_astfam_ast4,rosaev2017_hobson}.  We expect that many of the new asteroids discovered by LSST will be found to be members of already known families, but many new families \citep[including sub-families within other families; e.g.,][]{nesvorny2002_karin,nesvorny2006_karin,nesvorny2008_beagle} will also likely be discovered among the large number of new asteroids expected to be discovered by LSST. Typical asteroid clustering algorithms are not effective for outer solar system objects \citep[one outer solar system family associated with the Haumea system has been identified using other means, though;][]{brown2007_haumeafamily}, but the wealth of LSST data is expected to motivate the development of alternative methods for searching for outer solar system families \citep{Marcus2011_kbofamilies}. 

Perhaps the most straightforward automated family-related analysis that could be performed on LSST data is the attribution of LSST-discovered asteroids to known families.  This type of analysis is typically accomplished using the Hierarchical Clustering Method \citep[HCM;][]{zappala1990_hcm,zappala1994_hcm} to find dynamically linked asteroids\footnote{e.g., {\tt https://www.boulder.swri.edu/$\sim$davidn/family/family.html}}, where compositional information can also be incorporated into such analyses to help with interloper removal \citep[e.g.,][]{masiero2013_astfams_neowise,radovic2017_interlopers}.  Estimated asteroid diameters can also be correlated with proper semimajor axes, $a$, to identify possible associations with diffuse, old families \citep[][]{Spoto2015, Bolin2017}.  For very young families, automated application of the Backward Integration Method \citep[BIM;][]{nesvorny2002_karin,Radovic2017b} would be useful for confirming membership of newly linked asteroids.  We note that as the LSST survey extends asteroid discoveries to much smaller sizes, where smaller asteroids will also be more dispersed, the ability of HCM to provide reliable asteroid family identifications may be challenged and should be monitored as the survey progresses.

Whatever clustering technique is used, family discovery and characterization will require computation of proper elements \citep[either synthetic or analytically computed; Appendices~\ref{subsection:appendix_orbital_parameters} and \ref{subsection:appendix_advanced_dynamical_characterization};][]{knezevic2000_synthelements} for all asteroids which have sufficiently well-determined orbits for proper element computations to be considered reliable.  Synthetic proper elements have been found to be more reliable for family determination at high eccentricities and inclinations \citep[e.g.,][]{gilhutton2006_highifamilies}, and so as mentioned in Appendix~\ref{subsection:appendix_orbital_parameters}, will generally be preferred over analytic proper elements for asteroid family analyses.

Currently, updated online catalogs of both synthetic and analytically computed proper elements are maintained by the AstDyS website\footnote{\tt https://newton.spacedys.com/astdys/}, and preparations are currently being made to continue maintaining these catalogs during the LSST era (Kne{\v z}evi\'c 2019, personal communication).  Automated updates are also already currently being made on a regular basis to the membership lists of known families maintained by AstDyS using well-tested software \citep{knezevic2014_automatedfamilyclassification,milani2014_astfamilies}, where the AstDyS group also expects to continue providing this service in the LSST era.  As such, at the present time, it is probably unnecessary to expend additional SSSC resources on these tasks.

%but it has not yet been determined whether the AstDyS group will have the resources to continue to maintain these catalogs in the LSST era.  If it cannot, the SSSC may need to consider maintaining its own proper element database.
%\bb{This issue of large amounts of compuatational time being needed to determine proper elements of LSST-discovered Main Belt asteroids is exacerbated by the fact that LSST will be discovering smaller, more numerous Main Belt asteroids. The computation of proper elements for family association of these object must be considered a priority due to their importance in studying young asteroids families and the effects of stochastic YORP and size-dependent thermal interia on the dynamics of family asteroids \citep[][]{nesvorny2015_astfam_ast4,Bottke2015a,Bolin2017b}.} 
%We note that synthetic proper element determination can be a natural outcome of advanced dynamical classification (Appendix~\ref{subsection:appendix_advanced_dynamical_characterization}), so these efforts may be combined.

%\citet{milani2014_astfamilies} described a method for doing this using large data sets

Automated identification of new families from the massive amount of new asteroids expected to be discovered by LSST represents a more challenging problem, and one that will require more thought on how best to proceed.  We note here that SSSC-developed tools for performing HCM and BIM analyses could potentially make use of automated implementations of those analyses that are already available to the public at the online Asteroid Families Portal\footnote{\tt http://asteroids.matf.bg.ac.rs/fam/}.

The ability to identify extremely dynamically similar solar system objects will be important for science priorities related to the studies of fission events %(A-3, C-9)
and of genetic relationships between asteroid pairs. %(C-7).
Young asteroid pairs (i.e., resulting from fission events within the last $\sim$1~Myr) were first identified by \citet{vokrouhlicky2008_asteroidpairs}, who searched for asteroids with unusually similar osculating orbital elements ($a$, $e$, $i$, $\varpi$, $\Omega$). This approach was used instead of the usual method of searching for members of asteroid families with similar orbits using proper orbital elements because fragments produced in very recent disruption events will typically not become significantly dispersed by planetary perturbations and radiation forces until $\gtrsim$1~Myr after the disruption event \citep{vokrouhlicky2008_asteroidpairs}.  Asteroid pairs are scientifically interesting because their young ages make them ideal test beds for space weathering studies.

Given the large number of new solar system objects expected to be discovered by LSST, periodic searches of the latest catalog of osculating orbital elements (including both asteroids discovered by LSST and those discovered by other facilities) for dynamically similar objects will uncover many new young asteroid pairs, and perhaps even larger groupings of recently separated fragments.
%Further characterization of asteroid pairs, such as age determination \citep[e.g.,][]{galad2012_asteroidpairages}, will likely be left to individual scientists.  
It may also be of interest at some point to develop community software for deriving approximate age estimates (e.g., without considering non-gravitational forces or perturbations from other minor planets).

%\dr{Disagree with the "further charaterization... will be left to future scientists; recommend we remove that sentence.}

Monthly computation or updating of proper orbital elements and new identifications of family associations should be sufficiently frequent for most science cases.  Given the amount of data needed to identify new asteroid families or to derive orbits of sufficient quality for newly discovered objects for determining proper elements, much more frequent analyses are unlikely to be useful.  That said, tools for making preliminary determinations of possible family associations of newly discovered asteroids based on their osculating orbital elements or analytical proper elements soon after their discoveries could be potentially useful for providing context for those discoveries and assessing the desirability of follow-up observations (e.g., especially for active objects or potential members of very young asteroid families).

Monthly searches for dynamically similar solar system objects should be sufficiently frequent for most science cases.  Timely identification of young asteroid pairs is desirable as it would allow for spectroscopic follow-up of objects to study their compositional properties, but except for small and fast-moving NEOs with short available observing windows, follow-up will typically not be needed extremely urgently.

\bigskip
{\centering
\begin{tcolorbox}[width=5.5in]
Key priorities
\begin{itemize}[leftmargin=*]
    \item{\vspace{-0.2cm} Computation of synthetic proper orbital elements for main-belt asteroids (to the extent, if any, that this is not adequately handled by the AstDyS group in the LSST era).}
    \item{\vspace{-0.2cm} Use of HCM analyses (and possibly others) to associate new asteroids with known asteroid families based on previously determined properties of those families}
    \item{\vspace{-0.2cm} Identification of potential pairs of small asteroids and young asteroid family members based on osculating orbital elements.}
\end{itemize}
\end{tcolorbox}
}

\bigskip
\noindent\hangindent=.35cm\hangafter=1{\bf Processing requirements:} Moderately time-sensitive.  

\noindent\hangindent=.35cm\hangafter=1{\bf Software dependencies:} Will require compositional characterization (Appendix~\ref{subsection:appendix_compositional_characterization}) for input for some types of analyses.

\noindent\hangindent=.35cm\hangafter=1{\bf Required input data:} Osculating orbital elements and compositional characterization parameters.

\noindent\hangindent=.35cm\hangafter=1{\bf Expected output data:} Analytically computed and synthetic proper orbital elements, identifications of family associations for known asteroids, identifications of new asteroid families, and identification of pairs (or larger groupings) of dynamically similar solar system objects

\clearpage
%%%%%%%%%% OCCULTATION EVENT PREDICTION %%%%%%%%%%

\subsection{Occultation event prediction}\label{subsection:appendix_occultation_prediction}

The prediction of upcoming occultation events by small solar system objects is an identified science priority for inner and outer solar system scientists in the SSSC.  Occultation observations are a powerful means for characterizing solar system objects in ways that would otherwise not be possible via direct observations with ground-based telescopes \citep[e.g., setting tight constraints on shapes and sizes, detecting rings and jets, and deriving more precise orbits;][]{berard2017_chariklorings,betzler2017_haleboppoccultations,ortiz2017_haumeaoccultation,ortiz2015_chironrings,kammer2018_MU69occultation,desmars2019_plutooccultations,bragaribas2013_quaoar,bragaribas2014_chariklo,leiva2017_chariklo}.

Occultation predictions require a high-precision stellar position catalog and high-precision ephemeris predictions for solar system objects.  The Gaia catalog \citep{gaiacollaboration2016_gaia,gaiacollaboration2018_gaiadr2} is widely considered to be the best high-precision stellar position catalog currently available.  For the latter, various techniques have been developed to produce occultation-prediction-quality ephemeris predictions from existing astrometric observations and orbital solutions \citep[e.g.,][]{fraser2013_kboccultations,desmars2015_occultationprediction}.  From these, prediction maps can then be generated showing where and when occultation events are predicted to be observable \citep[cf.][]{assafin2010_occultations,bandahuarca2019_occultations}.

As predicted occultations approach, continued refinement of ephemeris predictions using newly acquired astrometry will be desirable for verifying and refining the expected circumstances of the occultation as close up to the occultation observations as possible. As such, automated tools for performing this continuous refinement for objects identified as presenting occultation opportunities, or otherwise tools that can be used on demand by users to perform this task using all LSST observations available up to the current time (and potentially external data as well; cf.\ Appendix~\ref{subsection:appendix_externaldata}), are needed.  \citet{bandahuarca2019_occultations} describe a system for refining orbits of TNOs discovered in Dark Energy Survey data, and generating, updating, and publishing occultation predictions on a publicly available website as part of the Lucky Star project\footnote{\tt http://lesia.obspm.fr/lucky-star/}.  This system has been developed in anticipation of LSST and so could be a viable option for generating and curating occultation predictions and making them available to the community in the LSST era.

Another model for observing occultations in the LSST era is provided by the Research and Education Collaborative Occultation Network\footnote{\tt http://tnorecon.net/} (RECON) project \citep{buie2016_recon}, which aims to observe a larger number of occultations than would otherwise be observable using traditional methods (i.e., obtaining high-precision astrometry of every candidate object and star and deploying observers to predicted observing sites).  This is accomplished by employing a large network of fixed telescopes arranged along a rough North-South line that can take advantage of lower-precision occultation predictions by covering an area much larger than the uncertainty area of each prediction, essentially passively intercepting occultations as they pass through the network, rather than actively ``chasing'' specific occultations.

For any system designed to make occultation predictions, a major bottleneck is the retrieval of new astrometry data to identify new potential occultation observation opportunities and update existing ones. This will likely be an even greater issue in the LSST era given the survey's extremely high expected data rate.  As such, it will likely be particularly beneficial for occultation prediction software to be run within a LSST Data Access Center (cf.\ Appendix~\ref{subsection:appendix_dataaccess}) in close proximity to raw astrometric data, rather than on remote systems which would require those raw astrometric data to be downloaded first.

Weekly (or even more frequent) searches for upcoming occultation opportunities will be needed for relevant science investigations.  Frequent searches for upcoming occultation events will be desirable in the event that a particular event is predicted to occur in the very near future, but many upcoming events will likely be farther in the future, allowing more time for observation planning.  

\bigskip
{\centering
\begin{tcolorbox}[width=5.5in]
Key priorities
\begin{itemize}[leftmargin=*]
    \item{\vspace{-0.2cm} Timely predictions of upcoming likely occultation event observation opportunities.}
    \item{\vspace{-0.2cm} Tools for facilitating automated or manual refinement of occultation event predictions close to the time of those events.}
\end{itemize}
\end{tcolorbox}
}

\bigskip
\noindent\hangindent=.35cm\hangafter=1{\bf Processing requirements:} Moderate time sensitivity.

\noindent\hangindent=.35cm\hangafter=1{\bf Software dependencies:} Does not require input from other software tools discussed here.

\noindent\hangindent=.35cm\hangafter=1{\bf Required input data:} Osculating elements for small solar system objects.

\noindent\hangindent=.35cm\hangafter=1{\bf Expected output data:} Identification of likely upcoming occultation events and relevant details of those events.

\clearpage
%%%%%%%%%% PHYSICAL PROPERTY EVOLUTION %%%%%%%%%%

\subsection{Detection of changes in physical properties}\label{subsection:appendix_physical_evolution}

Certain science priorities % A-4 and C-3 
will require that users are able to determine certain physical properties from subsets of the total LSST data set (e.g., over specified time periods) in order to search for changes in those properties over time (cf.\ Table~\ref{table:swneeds_summary}).  Such cases include searches for changes in rotation periods of comets due to outgassing torques \citep[e.g.,][]{belton2011_tempel1,bodewits2018_41Pspinslowdown,schleicher2019_41protation,knight2011_tempel2,samarasinha2011_103p}, changes in the rotation rates of asteroids due to the Yarkovsky-O'Keefe-Radzievskii-Paddack (YORP) effect \citep{rubincam2000_yorp}, and changes in surface colors of comets or asteroids due to cometary activity, close planetary encounters, or impacts \citep[e.g.,][]{binzel2010_neoearthencounters,demeo2014_marsresurfacing,bodewits2014_scheila}.
While the LSST Project is expected to provide data products like absolute magnitudes (from which surface colors can be derived) and orbital elements based on the entire LSST data set, it is unlikely that it will also provide data products derived from subsets of that data set.  As such, User Generated tools will need to be developed in order to address science priorities related to changes in these and other physical properties of objects over time.
\looseness=-1

Two conceivable approaches to performing this type of analysis are (1) to allow users to recompute a given quantity at any time using only data obtained during manually specified time periods as input, or (2) to systematically compute and store various quantities at regular intervals during the survey, using only data obtained within each interval as input (where optimum interval lengths will vary for each parameter depending on the minimum amount of data needed to compute reliable values for the parameter in question).
The first approach might be preferable for searches for changes in properties associated with specific events whose timing is known, e.g., close planetary encounters, perihelion passages for comets, or impacts detected in real time.  Performing analyses using this approach will likely require infrastructure to allow users to essentially conduct on-demand processing of customized data sets, but could otherwise use the same automated data analysis tools developed to process the entire LSST data set.

Meanwhile, the second approach would facilitate systematic searches for unpredicted changes in properties of interest by allowing scientists to perform untargeted searches for changes over time in these properties for a large number of objects.
While the first approach is mainly useful for analyses of individual objects and events, this second approach actually offers more flexibility, allowing for both targeted studies of specific objects and events (where users could simply select pre-computed properties derived for time intervals before and after an event of interest) and untargeted studies.  As such, while it may still be useful to implement the capability to use software tools in the manner described in the first approach as part of the general data access infrastructure provided via the LSST Science Platform (Appendix~\ref{subsection:appendix_dataaccess}), we suggest that the second approach may be a preferable way to address the specific issue of studying the time evolution of physical properties.

On a related note, some physical properties of objects have also shown potential to vary with other parameters other than time.  For instance, inferred taxonomic classifications of asteroids observed by SDSS have been shown to have a dependency on the phase angle of the observations \citep[cf.][]{carvano2015_phaseangletaxonomy}.  As such, the ability to compute various physical properties using subsets of data selected using criteria other than time, such as viewing geometry parameters, may also be valuable to develop in the future.

\bigskip
{\centering
\begin{tcolorbox}[width=5.5in]
Key priorities
\begin{itemize}[leftmargin=*]
    \item{\vspace{-0.2cm} Systematic computation of physical properties at regular time intervals using only data obtained within those intervals as input}
    \item{\vspace{-0.2cm} Mechanisms for computing physical properties using manually specified subsets (e.g., selected by time or viewing geometry circumstances) of the total LSST data set}
\end{itemize}
\end{tcolorbox}
}

\bigskip
\noindent\hangindent=.35cm\hangafter=1{\bf Processing requirements:} Moderate to low time sensitivity.

\noindent\hangindent=.35cm\hangafter=1{\bf Software dependencies:} Requires use of other software tools developed to determine various physical and dynamical properties of solar system objects. %Does not require input from other software tools discussed here.

\noindent\hangindent=.35cm\hangafter=1{\bf Required input data:} As required by individual software tools.

\noindent\hangindent=.35cm\hangafter=1{\bf Expected output data:} Physical and dynamical properties computed using data limited to specified time intervals.

\clearpage
%%%%%%%%%% DYNAMICAL PROPERTY EVOLUTION %%%%%%%%%%

\subsection{Detection of changes in dynamical properties}\label{subsection:appendix_dynamical_evolution}

The detection and characterization of the effect of non-gravitational perturbations on the dynamical properties of small solar system objects will be important for science priorities related to studies of the effects of radiative effects (i.e., the Yarkovsky effect) and outgassing effects on active and inactive solar system objects. %(A-6,B-8). 
In the cases of small NEOs and TCOs, characterization of non-gravitational perturbations may enable determination of whether an object has a natural or artificial density \citep[e.g.,][]{Micheli2012,Bolin2014}, while non-gravitational perturbations may also enable the indirect detection of comet-like activity in objects for which no visual indication of activity is present \citep[e.g.,][]{micheli2018_oumuamua}.
 
The majority of small bodies will not show detectable non-gravitational orbital perturbations, but automated analysis of astrometric residuals from orbit fitting would be one way to select potentially perturbed objects for more detailed analysis.
%In order to obtain a significant detection of a non-gravitational force acting on the orbit of a solar system object, it is essential to extract astrometry from the images with the highest possible precision, and provide accurate uncertainties for each positional measurement. 
Astrometric positional measurements and uncertainties of candidate objects will then need to be fitted with a variety of non-gravitational models: in particular, for small objects it will be necessary to fit a radial acceleration component (due to solar radiation pressure), while for larger ones a tangential component (due to the Yarkovsky effect) will also be needed. For cometary objects, and for any object in a comet-like orbit, it will be useful to test for a possible non-gravitational signature due to outgassing, which can manifest in any direction (radial, tangential and/or out-of-plane).
Furthermore, in case a marginal detection is achieved, it is essential to provide the ability to visually inspect the raw images corresponding to some key astrometric points, in order to verify that the supposed signal is not an artifact of one (or a few) individual detections being affected by local issues (e.g. image defects, edge effects, tracking irregularities, and so on).

%Detection of non-gravitational evolution of an object's spin state will require the ability to determine the rotational properties of an object over specified periods so that searches can be conducted for changes in those properties.  As described in Section~\ref{subsection:swneeds_timeevolution}, these searches could be conducted in at least two conceivable ways.  First, tools could be developed to allow users to recompute rotation rates using only data obtained during manually specified time periods as input, such as before and after a disruption event on a particular object. Alternatively, rotation rates could be systematically computed at regular intervals using only data obtained during those intervals, which would then allow searches to be conducted for changes in those rates over time for a large number of objects.  While the first approach would be useful for individual analyses of objects and events, the second approach would likely be more flexible, permitting both targeted studies of specific objects and events (where users could simply select rotation rates derived from data obtained around the time of the event of interest), while also allowing untargeted searches for rotation rate changes in a large number of objects.

The detection and characterization of changes in the orbits of small solar system bodies as a result of extremely close encounters with other small bodies will be needed for the determination of asteroid masses and bulk densities from mutual gravitational interactions.  This task will require software tools for identifying close encounter events and then for analysis of identified events \citep[e.g.,][]{siltala2017_asteroidmassestimation}.

Monthly or even annual searches for objects exhibiting evidence of non-gravitational perturbations due to the Yarkovsky effect should be sufficient.  Given the long time baselines that are typically required to detect such perturbations, more frequent searches are unlikely to be useful.  Meanwhile, it may be of interest to conduct searches for non-gravitational perturbations due to comet-like mass loss on much shorter timescales for objects likely to exhibit such behavior, especially for specific objects of high interest like 1I/2017 U1 `Oumuamua \citep{micheli2018_oumuamua}.

Annual searches for potential close-encounter events of interest for further analysis should be sufficiently frequent for relevant science investigations.  Given that we anticipate the main interest in studies of this nature will be the determination of masses of large numbers of asteroids, and not the characteristics of individual deflection events, we expect that more frequent searches will not be of high scientific interest.

\bigskip
{\centering
\begin{tcolorbox}[width=5.5in]
Key priorities
\begin{itemize}[leftmargin=*]
    \item{\vspace{-0.2cm} Identification of long-term evolution of orbital elements of solar system objects that could indicate the influence of non-gravitational effects such as the Yarkovsky effect or comet-like mass loss activity.}
    \item{\vspace{-0.2cm} Identification of significant impulsive changes in orbital elements of solar system objects that could indicate deflection events due to close mutual encounters.}
\end{itemize}
\end{tcolorbox}
}

\bigskip
\noindent\hangindent=.35cm\hangafter=1{\bf Processing requirements:} Moderate to low time sensitivity.

\noindent\hangindent=.35cm\hangafter=1{\bf Software dependencies:} Requires use of other software tools developed to determine various physical and dynamical properties of solar system objects. %Does not require input from other software tools discussed here.

\noindent\hangindent=.35cm\hangafter=1{\bf Required input data:} As required by individual software tools.

\noindent\hangindent=.35cm\hangafter=1{\bf Expected output data:} Dynamical properties computed using data limited to specified time intervals.  identification of objects likely to have experienced non-gravitational perturbations.

\clearpage
%%%%%%%%%% SYSTEM VALIDATION %%%%%%%%%%

\subsection{System validation}\label{subsection:appendix_system_validation}

While we expect the LSST Project, and specifically the software team responsible for developing the MOPS pipeline, will conduct extensive tests of the performance of pipeline processing software using internally defined metrics, it may be useful for the broader solar system community to define our own science-driven metrics and perform our own assessments to the extent (if any) that this process is not adequately handled by the LSST Project.  Examples of assessments that may be desirable include verification of satisfactory detection efficiencies for various solar system object populations of interest and verification of the detection of individual known solar system objects.

Verification of satisfactory detection efficiency by the survey could be accomplished by survey simulators similar or identical to that described in Appendix~\ref{subsection:appendix_survey_debiasing}, where detections corresponding to synthetic populations of objects are injected into LSST data and rates of recovery of these objects by the full LSST pipeline are measured.  As with survey simulators used for population debiasing, realistic solar system population models will also be needed (cf.\ Appendix~\ref{subsection:appendix_survey_debiasing}). While the value of such a simulator for survey debiasing would be in determining the detection efficiency for various objects by the LSST survey, allowing scientists to infer the unbiased intrinsic properties of a population from the detected sample of objects, the detection efficiencies themselves could be evaluated for acceptability by scientists interested in specific populations.  Early identification of unsatisfactory detection rates for unusual objects of interest such as interstellar objects, Earth Trojans, or slow-moving outer solar system objects could be very valuable, flagging a need for urgent calls for improvements to software algorithms for extracting such objects from the LSST data stream (cf.\ Appendix~\ref{subsection:appendix_advanced_moving_object_detection}), changes to follow-up strategies (cf.\ Appendix~\ref{subsection:appendix_alerts}), or in extreme cases, potential changes to the LSST survey cadence itself.

Verification of satisfactory detection rates for known solar system objects will similarly be desirable to identify other software or procedures in need of improvement.  This task has some overlap with forced photometry tools for activity and outburst detection (Appendices~\ref{subsection:appendix_activity_detection}(g) and \ref{subsection:appendix_outburst_disruption_detection}), in that it will involve the identification of newly-acquired LSST data in which known objects are expected to be detectable (i.e., are within the field of view and are expected to be bright enough to be detected) and checking to see whether those objects are indeed detected.  If detections are present but not successfully linked as known solar system objects, it could point to deficiencies and needs for improvement in linking algorithms (cf.\ Appendix~\ref{subsection:appendix_advanced_moving_object_detection}).

\bigskip
{\centering
\begin{tcolorbox}[width=5.5in]
Key priorities
\begin{itemize}[leftmargin=*]
    \item{\vspace{-0.2cm}Creation and implentation of tools to determine the detection efficiency of LSST for different populations of synthetic solar system objects, taking into account the LSST survey's real-world cadence and weather conditions, as well as object sizes, colors, rotational properties, potential activity, and other properties}
\end{itemize}
\end{tcolorbox}
}

\bigskip
\noindent\hangindent=.35cm\hangafter=1{\bf Processing requirements:} High time sensitivity.

\noindent\hangindent=.35cm\hangafter=1{\bf Software dependencies:} Does not require input from other software tools discussed here.

\noindent\hangindent=.35cm\hangafter=1{\bf Required input data:} Survey operation and performance parameters.

\noindent\hangindent=.35cm\hangafter=1{\bf Expected output data:} Detection efficiencies for various solar system populations or individual solar system objects with certain physical or dynamical properties.

\clearpage
%%%%%%%%%% SURVEY EFFICIENCY CHARACTERIZATION AND DEBIASING %%%%%%%%%%

\subsection{Survey efficiency characterization and debiasing}\label{subsection:appendix_survey_debiasing}

Characterization of the efficiency of the LSST survey and its biases will be important for all science priorities related to the determination of the physical and dynamical properties of solar system body populations.

Accomplishing this task will require detailed information about survey characteristics and instrument performance (both on a per-image basis). Given the large field of view of the LSST camera, it may be necessary to characterize detection efficiencies at a resolution per raft or per chip, since there may be different likelihoods of detecting an object depending on where it falls on the focal plane.  One straightforward way of accomplishing this is by creating and running survey simulators that generate large numbers of synthetic objects, and determines if and when each one of them would have been detected, linked and discovered by LSST, given real-world observing cadence, weather, instrument, pipeline, and other considerations.  Such a survey simulator and assessment framework have already been developed for evaluating proposed cadences for LSST\footnote{\tt https://www.lsst.org/scientists/simulations/opsim;\newline $~~~~~$https://www.lsst.org/scientists/simulations/maf}, using modeled pointing and weather history for pre-survey simulations. The software to perform detection efficiency characterization and debiasing could be an adaptation of this existing software to the survey's real-world observing history.  

Survey simulators can test the detectability of model objects (sometimes known as ``grid objects'') across the full range of orbital element phase space, providing an assessment of a well-characterized survey's ability to detect unexpected types of objects, and providing significant non-detection constraints on the absence of populations \citep[e.g.][]{lawler2018_simulator}. 
Mapping out the phase space of orbital elements that can appear in a field of view \citep[][]{Jedicke2016} has the advantage of being more computationally efficient compared to the synthetic object generation method, allowing survey efficiency to be computed at a higher resolution without requiring correspondingly more computational time. 
Simulators can also test the plausibility of realistic model solar system populations \citep[e.g.,][]{grav2011_solarsystemmodel,lawler2018_simulator}.
These model populations should ideally take into account such parameters and conditions as object sizes, colors, rotation periods and amplitudes, non-sidereal rates, and activity strengths.

%In addition, the field of view method can also easily characterize detection efficiency variation within the field of view of LSST.

Determination of survey efficiency characteristics and debiasing of survey results is a high priority task. It will typically require a large amount of data to be both collected and incorporated in order to be performed properly, since it is the overall performance of the survey that is of primary interest, not its performance on any particular night or even in any particular month.  Annual analyses, perhaps coinciding with annual data releases, will be sufficient for relevant science investigations. 

That said, especially in the early portion of the survey, assessment of the effectiveness and efficiency of the survey cadence and data processing pipelines for detecting certain types of solar system objects may be very valuable for determining whether any populations are particularly poorly sampled so that changes to the cadence or processing software can at least be requested, and hopefully implemented.  This particular application of survey efficiency characterization tools is addressed separately in Appendix~\ref{subsection:appendix_system_validation}.

\bigskip
{\centering
\begin{tcolorbox}[width=5.5in]
Key priorities
\begin{itemize}[leftmargin=*]
    \item{\vspace{-0.2cm} Creation of a survey simulator that determines the detection efficiency of LSST for different populations of synthetic solar system objects, taking into account factors such as cadence, weather, and instrument and pipeline performance, as well as object sizes, colors, rotational properties, potential activity, and other physical properties.}
    \item{\vspace{-0.2cm} Implementation of the field-of-view approach to characterizing survey efficiency for LSST data}
    \item{\vspace{-0.2cm} Creation of tools for correcting selection effects for different solar system populations.}
\end{itemize}
\end{tcolorbox}
}

\bigskip
\noindent\hangindent=.35cm\hangafter=1{\bf Processing requirements:} Low time sensitivity.

\noindent\hangindent=.35cm\hangafter=1{\bf Software dependencies:} Does not explicitly require input from other software tools discussed here, but properties measured for real-world solar system objects (e.g., phase functions, colors, rotation periods and amplitudes, and activity levels; Appendices~\ref{subsection:appendix_phase_function_characterization}, \ref{subsection:appendix_compositional_characterization}, \ref{subsection:appendix_rotational_characterization}, and \ref{subsection:appendix_activity_characterization}) may be useful for improving realism of underlying model populations.

\noindent\hangindent=.35cm\hangafter=1{\bf Required input data:} Survey operation and performance parameters.

\noindent\hangindent=.35cm\hangafter=1{\bf Expected output data:} Information for debiasing population-level properties determined by other software tools.

\clearpage
%%%%%%%%%% ALERT BROKERING AND FOLLOW-UP OBSERVATION MANAGEMENT %%%%%%%%%%

\subsection{Alert brokering and follow-up observation management}\label{subsection:appendix_alerts}

Alert distribution and follow-up observation management will be a crucial part of doing science in the LSST era.  Not all objects are expected to be adequately characterized by LSST data alone, necessitating the use of follow-up observations by other facilities that are triggered by the identification of LSST detections or events of interest. LSST will produce its own alert stream containing all transient sources identified by the LSST pipeline, where all alerts will be immediately publicly available to all members of the worldwide community.  Community alert brokers (yet to be identified) will be responsible for filtering this alert stream and potentially appending additional relevant metadata to alerts of interest.

Currently, the baseline LSST alert system\footnote{\tt http://ls.st/LDM-612} is expected to take into account data that is produced by the main data reduction pipeline (including MOPS) and will send out alerts within 60 seconds of an observation being taken (according to the LSST Data Products Definition Document).  These alert packets will mainly contain LSST-generated Prompt Product data.
%, but could conceivably also include some previously computed Data Release or User Generated data products (e.g., taxonomic classifications or asteroid family associations) that are already available in the LSST database system.  
They will not contain the results of User Generated analyses (such as advanced activity, outburst, and disruption search parameters) performed after the release of the initial alert.

LSST alerts will likely be distributed to the community via community alert brokers, where the Arizona-NOAO Temporal Analysis and Response to Events System\footnote{\tt https://www.noao.edu/ANTARES/} \citep[ANTARES;][]{saha2014_antares,saha2016_antares,narayan2018_alertbrokers}, Las Cumbres Observatory's Make Alerts Really Simple (MARS) system\footnote{\tt https://mars.lco.global/}, and the Lasair system\footnote{\tt https://lasair.roe.ac.uk/} \citep[][]{smith2019_lasair} being developed by the University of Edinburgh and Queen's University, Belfast, represent examples of systems that could be employed or adapted for solar system alert distribution.  A community alert broker catering specifically to the solar system community could be designed to receive LSST alerts in real time while providing the option of delaying their distribution to the community until certain time-sensitive User Generated data products can be produced for inclusion in those alert packets and also for use in the selection by users of which alerts they wish to receive.

These solar-system-specific alerts could consist of both ``pre-set'' alerts whose conditions are set to trigger on events or detections of broad interest (e.g., activity or outburst detections, or potentially hazardous objects or interstellar object candidates), and custom alerts whose conditions can be set by individual users.
In addition to being ``pushed out'' to users via an alert broker, detections triggering pre-set alerts (especially those comprising high priorities for follow-up) could also simply listed online on web pages similar to the Minor Planet Center's NEO Confirmation Page\footnote{\tt https://www.minorplanetcenter.net/iau/NEO/toconfirm\_tabular.html} (NEOCP) and Potential Comet Confirmation Page\footnote{\tt https://minorplanetcenter.net/iau/NEO/pccp\_tabular.html} (PCCP).
For users who require alerts as quickly as possible (e.g., those seeking to trigger follow-up observations of newly discovered NEO or interstellar object candidates who are not particularly interested in activity in those objects), such a system should also offer the option to receive alerts without waiting for additional User Generated data products to become available.

Given the large number of alerts that will be generated by LSST that will be of interest to not only the solar system science community, but other fields as well, and the limited number of ground-based facilities capable of performing meaningful follow-up of LSST discoveries, a system for coordinating community efforts to obtain follow-up observations of high-priority targets will likely be highly desirable.  Design of such a system will not just be a technical challenge, but a sociological one as well, where issues that need to be addressed will include how data sharing, or not sharing, would be handled, and how to equitably serve both members of the community who only have access to publicly available facilities and those who also have access to private facilities in addition to those same public facilities. As such, any community-wide follow-up management system will likely require the engagement and buy-in of a large fraction of the interested community to be successful.

Once candidates for follow-up have been identified, observation managers may be a useful tool for facilitating actual observations at other telescopes.  Managers, such as those built with Las Cumbres Observatory's Target and Observation Manager (TOM) Toolkit\footnote{\tt https://lco.global/tomtoolkit/}, can submit requests for new observations at participating observatories and have them manually or even automatically scheduled for execution.  Such a manager or scheduling system that considers moving targets, can be integrated with a community coordination system (ideally including as broad and diverse a range of participating facilities as possible), and could potentially even assign targets to different sites based on some sort of optimization scheme (e.g., taking into account geographic location, aperture size, observing capabilities, and so on) would be extremely useful.

Observation coordination websites\footnote{e.g., {\tt http://wirtanen.astro.umd.edu/obs\_campaigns.shtml}} have been developed for particular targets of interest and so something similar could conceivably be developed for LSST-discovered targets or events of high interest to follow-up observers.  Whether such websites could be generated for indefinite numbers of targets or would only be created for specific rare objects will likely depend on availability of computational and personnel resources, and interest.  Such websites could potentially also implement some version of the functionality provided by the Talk feature\footnote{\tt https://www.zooniverse.org/talk} on Zooniverse citizen science projects to enable discussion among interested scientists or even amateur astronomers, again with the level of implementation of this feature depending on the level of interest among potential follow-up observers.

For objects with relatively simple follow-up requirements (e.g., astrometric follow-up or simple monitoring that does not require going significantly deeper than the LSST's single-visit image depth), sufficient self-follow-up may actually be provided by the LSST itself in the course of its normal survey operations.  Upcoming LSST pointings (or at least the likelihood of those pointings) are likely to be published ahead of time, meaning that it may be possible to calculate the probabilities of LSST being able to perform its own recoveries of various follow-up targets within a given time frame.  This information could then be taken into account by observation management software when prioritizing objects for follow-up by external facilities.  We note that this specific functionality is not expected to be provided by the LSST Project, and so software for computing self-recovery likelihoods for follow-up targets based on anticipated upcoming LSST pointings would need be developed by members of the community.

%\hhh{need to write software to determine likelihood of self-followup by LSST of NEOs based on likelihood of upcoming LSST observation within a certain length of time}

\bigskip
{\centering
\begin{tcolorbox}[width=5.5in]
Key priorities
\begin{itemize}[leftmargin=*]
    \item{\vspace{-0.2cm}A system to distribute alerts of LSST detections of interest to solar system scientists that include solar system-relevant data products (both previously computed object properties and real-time measurements of individual detections) in addition to standard LSST alert packet data, and can also be filtered on the basis of those solar system-specific data products}
    \item{\vspace{-0.2cm}A system (or systems) to manage follow-up observations, potentially allowing for automatic triggering of such observations for high-priority targets, and to facilitate community coordination of follow-up campaigns}
\end{itemize}
\end{tcolorbox}
}

\bigskip
\noindent\hangindent=.35cm\hangafter=1{\bf Processing requirements:} High time sensitivity.

\noindent\hangindent=.35cm\hangafter=1{\bf Software dependencies:} Requires output from almost all software tools described here.
%Does not require input from other software tools discussed here.  

\noindent\hangindent=.35cm\hangafter=1{\bf Required input data:} Output from other software tools.

\noindent\hangindent=.35cm\hangafter=1{\bf Expected output data:} General and customized alerts of interest to both the community at large and individual scientists with both baseline LSST data and relevant higher-order SSSC-generated data products, prioritization of potential follow-up targets, and a mechanism for the management of follow-up observations.

\clearpage
%%%%%%%%%% EXTERNAL DATA INCORPORATION %%%%%%%%%%

\subsection{External data incorporation}\label{subsection:appendix_externaldata}

As software tools are developed to analyze LSST data, it will be worth considering how sub-optimal data sets will be handled, i.e., where LSST is unable to collect sufficient data to complete a particular analysis task for a given object.  Some of the smallest objects (particularly NEOs and interstellar objects, but also main-belt asteroids) that LSST will be able to detect will only be observable for short periods of time, limiting the amount of data that LSST will be able to collect for those objects.  As many of the most observationally challenging objects will likely also be some of the most interesting, it will be worth considering how to maximize the scientific return on the limited data that will be collected for these objects (e.g., discussions about addressing this issue for phase function determination and compositional characterization in Appendices~\ref{subsection:appendix_phase_function_characterization} and \ref{subsection:appendix_compositional_characterization}), as well as options for supplementing these data with observations from other telescopes for deriving higher-order data products.\looseness=-1

Data of interest will include both data from targeted follow-up observations and serendipitous detections of LSST-discovered objects in archival data from other facilities.  Identification of archival data of interest could be accomplished with a tool similar to the Canadian Astronomy Data Centre's (CADC) Solar System Object Image Search\footnote{\tt http://www.cadc-ccda.hia-iha.nrc-cnrc.gc.ca/en/ssois/} (SSOIS), although tools for the reduction and analysis of that data would need to be developed for it to be of use at large scales.  Incorporation of follow-up data could be implemented as part of the general management of follow-up observations (Appendix~\ref{subsection:appendix_alerts}), at least for non-proprietary observations.

The supplementing of LSST data with external data could potentially be part of a broader effort to link outside data sources (e.g., the Lowell Observatory AstorbDB database\footnote{\tt https://asteroid.lowell.edu/main/}, NEOWISE\footnote{\tt https://irsa.ipac.caltech.edu/Missions/wise.html}, AstDyS proper element catalogs\footnote{\tt http://hamilton.dm.unipi.it/astdys/}, and other future complementary catalogs) to the LSST solar system database to allow for the generation of more higher-order data products or to enable more sophisticated science investigations than would be possible using LSST data alone.  Any such effort would need to include quality assurance procedures for external data sources, and depending on the nature of specific external data sets (e.g., whether they contain proprietary data, or whether they have been formally published and are therefore citeable or not), may also require development of policies governing the data and authorship rights of contributors of external data.

The time sensitivity of tasks involving incorporation of external data will vary depending on specific tasks.  For example, compositional characterization and detection of non-gravitational perturbations are not considered particularly urgent tasks, and so searches for and analyses of archival or follow-up data to aid in these tasks should likewise not be considered particularly urgent.  On the other hand, refinement of orbital elements (to allow for more precise ephemeris predictions to enable more efficient and effective follow-up efforts) for active objects, NEOs, and possible interstellar objects is a considerably more time-sensitive task, and so in such cases, automated searches of archival data for additional astrometric points to aid with this task (as needed) might be considered more urgent.

%On the other hand, incorporation of external data will have considerably higher urgency for other tasks, such as incorporation of external astrometric data for orbit refinement for active objects and NEOs (cf.\ Appendices~\ref{subsection:appendix_activity_detection} and \ref{subsection:appendix_alerts}).
%As such, clearly-defined methods for incorporating external data will be extremely useful to have by the start of the LSST survey for some (but not all) applications.

\bigskip
{\centering
\begin{tcolorbox}[width=5.5in]
Key priorities
\begin{itemize}[leftmargin=*]
    \item{\vspace{-0.2cm} Mechanisms for incorporating external data (either from targeted follow-up observations of LSST-discovered objects, or from serendipitous observations of those objects in archival data from other facilities) into computations using LSST data to produce higher-order data products}
    \item{\vspace{-0.2cm} Mechanisms for linking LSST data products with external data sources and catalogs}
\end{itemize}
\end{tcolorbox}
}

\bigskip
\noindent\hangindent=.35cm\hangafter=1{\bf Processing requirements:} Dependent on specific applications.

\noindent\hangindent=.35cm\hangafter=1{\bf Software dependencies:} Dependent on specific applications.

\noindent\hangindent=.35cm\hangafter=1{\bf Required input data:} Dependent on specific applications.

\noindent\hangindent=.35cm\hangafter=1{\bf Expected output data:} Dependent on specific applications.

\clearpage
%%%%%%%%%% DATA ACCESS TOOLS %%%%%%%%%%

\subsection{Data access and visualization}\label{subsection:appendix_dataaccess}

In terms of specific mechanisms for access to user-generated data products, other than a solar-system-specific alert broker system which will actively ``push'' data out to users and is discussed in more detail in Appendix~\ref{subsection:appendix_alerts}, significant functionality should be provided by the LSST Science Platform\footnote{\tt https://ldm-542.lsst.io/}.  The Science Platform is expected to provide access to catalog, image, and ancillary data via a number of different interfaces, including Python Application Programming Interfaces (APIs), web APIs, and a graphical user interface.  These interfaces are intended to provide access to both Project-developed data products (i.e., Prompt Products and Data Release Products) and User Generated data products.  In particular, raw images, processed visit images, difference images, coadded images, and templates are all expected to be accessible via the Science Platform.

Current plans for the LSST Science Platform include three ``Aspects'': a Portal Aspect, Notebook Aspect, and API Aspect.  The objective of the Portal Aspect is to provide web-based query and visualization tools for LSST data products and facilitate exploratory data analysis of available catalog and image data products.  It is planned for this Aspect to be able to include additional analysis and visualization tools developed by individual users and collaborations.  The Notebook Aspect will provide access to a Python-oriented computational environment, which will be hosted at LSST Data Access Centers, using Jupyter notebooks\footnote{\tt https://jupyter.org/}.  Users of this aspect will be able to run Python code in close proximity to LSST databases, will have access to a range of LSST processing and analysis software, and will also be encouraged to develop and run custom processing, analysis, and visualization Python code in this environment.  Finally, the API Aspect will provide remote access to LSST data (both Project-developed and User-Generated) through web APIs, many of which will likely be based on Virtual Observatory\footnote{\tt http://www.ivoa.net/} (VO) standards.

While this environment will be provided by the LSST Project, it will be up to individual users and collaborations to develop tools specific to their respective fields.  While not necessarily mandated by LSST or SSSC policy, release of such tools or resulting data products for community use will help to reduce redundant effort, freeing up resources for more advanced analyses, and will also encourage wider use of those data across the scientific community as well as by the public as part of education and public outreach efforts \citep[e.g.,][]{olsen2017_datalab,olsen2018_datalab_outreach}.

Examples of tools that could be useful to develop include follow-up observation planning tools for identifying times when objects of interest are passing over or nearby bright background sources or dense background fields (this could also be a component of follow-up observation management; Appendix~\ref{subsection:appendix_alerts}), tools for visually screening candidate active objects (Appendix~\ref{subsection:appendix_activity_detection}), and visualization tools for reviewing rotational lightcurve solutions (Appendix~\ref{subsection:appendix_rotational_characterization}) or plotting the evolution of physical properties over time (Appendix~\ref{subsection:appendix_physical_evolution}).

At the current time, we do not anticipate that additional development of data access infrastructure beyond what will be provided by the LSST Science Platform is needed, although development of specific user generated software and tools (many of which are described elsewhere in this white paper) to be used on the Science Platform will be required of course.

%Ideally, the majority of the data access tools described here should be developed well in advance of the start of the LSST survey, so that they can be used to facilitate science from the very beginning of the survey.  In practice, however, we expect data access needs to evolve as the survey proceeds, as usage patterns and community demands become clearer once data is actually available, rather than simply being anticipated. As such, while we consider development of certain data access tools (e.g., particularly those needed for alert generation and enhancement) to be an urgent need, we anticipate that data access tool development will likely continue well into the survey mission.

%XXX visualization tools, user interface, browsing, website, DataLab, image retrieval XXX

%Observation planning, e.g., predicting when small bodies will pass over background stars or galaxies or near particularly bright background sources

\bigskip
{\centering
\begin{tcolorbox}[width=5.5in]
Key priorities
\begin{itemize}[leftmargin=*]
    \item{\vspace{-0.2cm} Identification and development of user generated solar system science-specific software and tools for use on the LSST Science Platform}
\end{itemize}
\end{tcolorbox}
}

\bigskip
\noindent\hangindent=.35cm\hangafter=1{\bf Processing requirements:} Will depend on specific applications.

\noindent\hangindent=.35cm\hangafter=1{\bf Software dependencies:} Not applicable. %Does not require input from other software tools discussed here.

\noindent\hangindent=.35cm\hangafter=1{\bf Required input data:} Not applicable.

\noindent\hangindent=.35cm\hangafter=1{\bf Expected output data:} Not applicable.

%\subsubsection{Mass Determination from Non-Gravitational Force Analysis}\label{subsubsection:massdeflections_active}
%
%See \citet{sosa2009_cometmasses} and \citet{sosa2011_lpcmasses}

%%%%%%%%%% SOFTWARE NEEDS BY SCIENCE AREA %%%%%%%%%%

\clearpage
\section{Software Needs by Science Area\label{section:appendix_swneeds_sciencearea}}

\subsection{Introduction\label{subsection:appendix_sciencepriorities_intro}}

In this section, we list the specific software tools and needs that will be required to achieve the science priorities listed in the Solar System Science Roadmap developed by the LSST Solar System Science Collaboration.

%%%%%%%%%% SOFTWARE REQUIREMENTS FOR ACTIVE OBJECTS %%%%%%%%%%

\bigskip
\subsection{Science Priorities and Software Requirements for Active Objects\label{subsection:appendix_sciencepriorities_active}}

\hangindent=.80cm\hangafter=1
\noindent 1. {\bf Discovery, orbital classification, and monitoring of active objects to explore physical and orbital characteristics:} Will require tools for
orbital object and detection parameter computation (for activity detection and evolution characterization),
orbital element and ephemeris uncertainty characterization (for dynamical refinement),
extended object astrometry (for orbit refinement),
faint precovery and recovery identification (for orbit refinement),
extremely faint object detection (for orbit refinement),
phase function characterization,
compositional characterization,
rotational characterization,
activity detection,
activity characterization,
outburst and disruption detection,
system validation (for ensuring reliable activity detection),
alert brokering and follow-up observation management (for orbit refinement and additional physical characterization),
external data incorporation (for orbit refinement and additional physical characterization), and
data access and visualization.

\medskip\hangindent=.80cm\hangafter=1
\noindent 2. {\bf Estimation of the frequency of occurrences of active objects in various populations:} Will require tools for
orbital object and detection parameter computation (for activity detection),
orbital element and ephemeris uncertainty characterization (for orbit refinement and dynamical classification),
extended object astrometry (for orbit refinement and dynamical classification),
faint precovery and recovery identification (for orbit refinement and dynamical classification),
extremely faint object detection (for orbit refinement and dynamical classification),
phase function characterization (for activity detection),
rotational characterization (for activity detection),
activity detection,
activity characterization,
outburst and disruption detection,
system validation (for ensuring reliable activity detection),
survey efficiency characterization and debiasing,
alert brokering and follow-up observation management (for activity confirmation),
external data incorporation, and
data access and visualization.

\medskip\hangindent=.80cm\hangafter=1
\noindent  3. {\bf Estimation of the frequency of occurrences of outbursts and splitting events in various solar system populations:} Will require tools for
orbital object and detection parameter computation (for activity detection),
orbital element and ephemeris uncertainty characterization (for orbit refinement and dynamical classification),
extended object astrometry (for orbit refinement and dynamical classification),
faint precovery and recovery identification (for orbit refinement and dynamical classification),
extremely faint object detection (for orbit refinement and dynamical classification),
phase function characterization (for activity detection),
rotational characterization (for activity detection),
detection and characterization of multi-object systems,
activity detection,
activity characterization,
outburst and disruption detection,
advanced dynamical characterization,
dynamical clustering identification,
system validation (for ensuring reliable activity detection),
survey efficiency characterization and debiasing,
alert brokering and follow-up observation management (for activity confirmation and additional event characterization),
external data incorporation (for orbit refinement and additional event characterization), and
data access and visualization.

\medskip\hangindent=.80cm\hangafter=1
\noindent 4. {\bf Characterization of changes in colors, morphology, brightnesses, rotation periods, and other properties of active objects over time:} Will require tools for
orbital object and detection parameter computation (for activity detection and characterization),
faint precovery and recovery identification,
extremely faint object detection,
phase function characterization (for activity detection and characterization),
compositional characterization,
rotational characterization (for activity detection and characterization),
activity detection,
activity characterization,
outburst and disruption detection,
detection of changes in physical properties,
detection of changes in dynamical properties,
system validation (for ensuring reliable activity detection),
alert brokering and follow-up observation management (for activity characterization),
external data incorporation (for activity characterization), and
data access and visualization.
%orbital object and detection parameter computation (for providing context for changes in properties),
%compositional characterization (Appendix~\ref{subsection:appendix_compositional_characterization}),
%rotational property determination (Appendix~\ref{subsection:appendix_rotational_characterization}),
%faint object detection (Appendix~\ref{subsection:appendix_faint_object_detection}), and
%activity characterization (Appendix~\ref{subsection:appendix_activity_characterization}).

\medskip\hangindent=.80cm\hangafter=1
\noindent 5. {\bf Determination of rotational properties, spin angular momentum distribution, shape distribution, and binary frequency for active objects:} Will require tools for
orbital object and detection parameter computation (for activity detection),
phase function characterization (for activity detection),
rotational characterization,
detection and characterization of multi-object systems,
activity detection,
activity characterization (for potentially aiding in rotational characterization),
system validation (for ensuring reliable activity detection),
survey efficiency characterization and debiasing,
alert brokering and follow-up observation management (for additional physical characterization),
external data incorporation (for additional physical characterization), and
data access and visualization.

\medskip\hangindent=.80cm\hangafter=1
\noindent 6. {\bf Detection and characterization of non-gravitational forces acting on active bodies:} Will require tools for
orbital object and detection parameter computation (for activity characterization),
orbital element and ephemeris uncertainty characterization (for orbit refinement),
extended object astrometry (for orbit refinement),
faint precovery and recovery identification (for orbit refinement),
extremely faint object detection (for orbit refinement),
phase function characterization (for activity characterization),
rotational characterization (for activity characterization),
activity detection,
activity characterization,
outburst and disruption detection,
detection of changes in dynamical properties,
external data incorporation (for orbit refinement), and
data access and visualization.
%faint precovery/recovery identification (Appendix~\ref{subsection:appendix_lsst_precoveries_recoveries}; to help with orbit refinement), rotational property determination (Appendix~\ref{subsection:appendix_rotational_characterization}; for detection of changes in rotational properties), faint object detection (Appendix~\ref{subsection:appendix_faint_object_detection}; to help with orbit refinement), and detection and characterization of non-gravitational perturbations (Appendix~\ref{subsection:appendix_dynamical_evolution}).

%%%%%%%%%% SOFTWARE REQUIREMENTS FOR NEAR-EARTH OBJECTS %%%%%%%%%%

\bigskip
\subsection{Science Priorities and Software Requirements for Near-Earth Objects\label{subsection:appendix_sciencepriorities_neos}}

\medskip\hangindent=.80cm\hangafter=1
\noindent 1. {\bf Compilation of a NEO catalog with high completeness and orbit quality:} Will require tools for
orbital element and ephemeris uncertainty characterization (for orbit refinement),
extended object astrometry (for orbit refinement of active NEOs),
faint precovery and recovery identification (for orbit refinement),
extremely faint object detection (for orbit refinement),
advanced moving object detection,
system validation (for ensuring reliable NEO detection),
survey efficiency characterization and debiasing,
alert brokering and follow-up observation management (for orbit refinement),
external data incorporation (for orbit refinement), and
data access and visualization.

\medskip\hangindent=.80cm\hangafter=1
\noindent 2. {\bf Color and phase function measurements of NEOs to identify objects of probable cometary origin:} Will require tools for
orbital object and detection parameter computation (for activity detection to enable omission of active data),
phase function characterization,
compositional characterization,
activity detection (for enabling omission of active data),
outburst and disruption detection (for enabling omission of active data),
system validation (for ensuring reliable color and phase function measurements),
alert brokering and follow-up observation management (for additional physical characterization),
external data incorporation (for additional physical characterization), and
data access and visualization.

\medskip\hangindent=.80cm\hangafter=1
\noindent 3. {\bf Timely advance notice of close approaches or potential impacts to facilitate time-critical follow-up characterization efforts:} Will require tools for
orbital element and ephemeris uncertainty characterization (for orbit refinement),
extended object astrometry (for orbit refinement of active NEOs),
faint precovery and recovery identification (for orbit refinement),
extremely faint object detection (for orbit refinement),
advanced moving object detection,
advanced dynamical characterization,
system validation (for ensuring reliable NEO detection),
alert brokering and follow-up observation management,
external data incorporation (for orbit refinement), and
data access and visualization.

\medskip\hangindent=.80cm\hangafter=1
\noindent 4. {\bf Determination of orbital, absolute magnitude, and taxonomic distributions of NEOs to explore correlations between taxonomy and orbital properties:} Will require tools for
orbital object and detection parameter computation (for activity detection to enable omission of active data),
orbital element and ephemeris uncertainty characterization (for orbit refinement),
extended object astrometry (for orbit refinement of active NEOs),
faint precovery and recovery identification (for orbit refinement),
extremely faint object detection (for orbit refinement),
advanced moving object detection,
phase function characterization,
compositional characterization,
rotational characterization (for activity detection),
activity detection (for enabling omission of active data),
outburst and disruption detection (for enabling omission of active data),
advanced dynamical characterization,
dynamical clustering identification,
survey efficiency characterization and debiasing,
alert brokering and follow-up observation management (for additional physical characterization),
external data incorporation (for additional physical characterization), and
data access and visualization.

\medskip\hangindent=.80cm\hangafter=1
\noindent 5. {\bf Determination of the long-term impact flux of NEOs as a function of size:} Will require tools for
orbital object and detection parameter computation (for activity detection to enable omission of active data),
orbital element and ephemeris uncertainty characterization (for orbit refinement),
extended object astrometry (for orbit refinement of active NEOs),
faint precovery and recovery identification (for orbit refinement),
extremely faint object detection (for orbit refinement),
advanced moving object detection,
phase function characterization,
compositional characterization (for albedo estimation for size determination),
rotational characterization,
activity detection,
advanced dynamical characterization,
system validation (for ensuring reliable NEO detection),
survey efficiency characterization and debiasing,
alert brokering and follow-up observation management (for orbit refinement and additional physical characterization),
external data incorporation (for orbit refinement and additional physical characterization), and
data access and visualization.

\medskip\hangindent=.80cm\hangafter=1
\noindent 6. {\bf Discovery and estimation of the frequency of interstellar objects:} Will require tools for
orbital element and ephemeris uncertainty characterization (for orbit refinement),
extended object astrometry (for orbit refinement of active interstellar object candidates),
faint precovery and recovery identification (for orbit refinement),
extremely faint object detection (for orbit refinement),
advanced moving object detection,
advanced dynamical characterization,
system validation (for ensuring reliable interstellar object detection),
survey efficiency characterization and debiasing,
alert brokering and follow-up observation management (for orbit refinement),
external data incorporation (for orbit refinement), and
data access and visualization.

\medskip\hangindent=.80cm\hangafter=1
\noindent 7. {\bf Determination of rotational properties of NEOs to determine spin angular momentum and shape distributions, and binary frequency:} Will require tools for
orbital object and detection parameter computation (for activity detection to enable omission of active data),
phase function characterization,
rotational characterization,
detection and characterization of multi-object systems,
activity detection (for enabling omission of active data),
outburst and disruption detection (for enabling omission of active data),
system validation (for ensuring reliable rotational property measurement),
survey efficiency characterization and debiasing,
alert brokering and follow-up observation management (for additional rotational characterization),
external data incorporation (for additional rotational and physical characterization), and
data access and visualization.

\medskip\hangindent=.80cm\hangafter=1
\noindent 8. {\bf Detection and characterization of non-gravitational forces acting on NEOs:} Will require tools for
orbital element and ephemeris uncertainty characterization (for orbit refinement),
extended object astrometry (for orbit refinement of active NEOs),
faint precovery and recovery identification (for orbit refinement),
extremely faint object detection (for orbit refinement),
advanced moving object detection,
rotational characterization,
activity detection,
activity characterization,
outburst and disruption detection,
detection of changes in dynamical properties,
alert brokering and follow-up observation management (for orbit refinement),
external data incorporation (for orbit refinement), and
data access and visualization.

\medskip\hangindent=.80cm\hangafter=1
\noindent 9. {\bf Measurement of the absolute magnitude distribution of Temporarily Captured Orbiters:} Will require tools for
orbital object and detection parameter computation,
orbital element and ephemeris uncertainty characterization (for activity detection to enable omission of active data),
faint precovery and recovery identification (for orbit refinement and dynamical classification),
extremely faint object detection (for orbit refinement and dynamical classification),
advanced moving object detection,
phase function characterization,
rotational characterization (for activity detection to enable omission of active data),
activity detection (for enabling omission of active data),
outburst and disruption detection (for enabling omission of active data),
advanced dynamical characterization,
system validation (ensuring reliable detection of TCOs),
survey efficiency characterization and debiasing,
alert brokering and follow-up observation management (for orbit refinement and physical characterization),
external data incorporation (for orbit refinement and additional physical characterization), and
data access and visualization.

\medskip\hangindent=.80cm\hangafter=1
\noindent 10. {\bf Investigation of NEO disruption mechanisms at small perihelion distances:} Will require tools for
orbital object and detection parameter computation (for activity detection),
orbital element and ephemeris uncertainty characterization (for orbit refinement),
extended object astrometry (for orbit refinement of active NEOs),
faint precovery and recovery identification (for orbit refinement and physical characterization),
extremely faint object detection (for orbit refinement and physical characterization),
advanced moving object detection (for orbit refinement),
phase function characterization (for physical characterization),
compositional characterization (for physical characterization),
rotational characterization (for physical characterization),
activity detection (for direct detection of disruption events),
activity characterization (for constraining possible mechanisms for directly detected disruption events),
outburst and disruption detection,
survey efficiency characterization and debiasing (for enabling statistical studies of NEO disruption),
alert brokering and follow-up observation management (for orbit refinement and additional physical characterization),
external data incorporation (for orbit refinement and additional physical characterization), and
data access and visualization.

%%%%%%%%%% SOFTWARE REQUIREMENTS FOR INNER SOLAR SYSTEM OBJECTS %%%%%%%%%%

\bigskip
\subsection{Science Priorities and Software Requirements for Inner Solar System Objects\label{subsection:appendix_sciencepriorities_innerss}}

\medskip\hangindent=.80cm\hangafter=1
\noindent 1. {\bf Discovery and characterization of inner solar system objects to determine orbital distributions and size-frequency distributions of different taxonomic classes:} Will require tools for
orbital object and detection parameter computation (for activity detection to enable omission of active data),
orbital element and ephemeris uncertainty characterization (for orbit refinement and characterization),
faint precovery and recovery identification (for orbit refinement),
extremely faint object detection (for orbit refinement),
advanced moving object detection,
phase function characterization,
compositional characterization,
rotational characterization (for activity detection to enable omission of active data),
detection and characterization of multi-object systems,
activity detection (for enabling omission of active data),
outburst and disruption detection (for enabling omission of active data),
system validation (for ensuring reliable detection of inner solar system objects of different taxonomic types),
survey efficiency characterization and debiasing,
alert brokering and follow-up observation management (for additional physical characterization),
external data incorporation (for additional physical characterization), and
data access and visualization.

\medskip\hangindent=.80cm\hangafter=1
\noindent 2. {\bf Measurement of high quality astrometry for inner solar system objects to refine orbits and improve ephemerides for stellar occultation predictions:} Will require tools for
orbital element and ephemeris uncertainty characterization (for orbit refinement),
faint precovery and recovery identification (for orbit refinement),
extremely faint object detection (for orbit refinement),
occultation event prediction,
system validation (for ensuring reliable detection of inner solar system objects),
alert brokering and follow-up observation management (for orbit refinement),
external data incorporation (for orbit refinement), and
data access and visualization.

\medskip\hangindent=.80cm\hangafter=1
\noindent 3. {\bf Detection and characterization of impact events and exploration of space weathering processes:} Will require tools for
orbital object and detection parameter computation (for activity detection),
extended object astrometry (for orbit refinement of disrupted objects),
faint precovery and recovery identification (for orbit refinement and physical characterization),
extremely faint object detection (for orbit refinement and physical characterization),
phase function characterization,
compositional characterization,
rotational characterization,
detection and characterization of multi-object systems (for identification of impact-induced fission events),
activity detection,
activity characterization,
outburst and disruption detection,
detection of changes in physical properties,
system validation (for ensuring reliable detection of disruption events),
survey efficiency characterization and debiasing (for enabling statistical studies of space weathering processes),
alert brokering and follow-up observation management (for additional physical characterization),
external data incorporation (for additional physical characterization), and
data access and visualization.

\medskip\hangindent=.80cm\hangafter=1
\noindent 4. {\bf Determination of colors and compositions for inner solar system objects to identify correlations between taxonomy and dynamical properties:} Will require tools for
orbital object and detection parameter computation (for activity detection to enable omission of active data),
orbital element and ephemeris uncertainty characterization (for orbit refinement),
faint precovery and recovery identification (for orbit refinement),
extremely faint object detection (for orbit refinement),
phase function characterization,
compositional characterization,
rotational characterization (for activity detection to enable omission of active data),
activity detection (for enabling omission of active data),
outburst and disruption detection (for enabling omission of active data),
advanced dynamical characterization,
dynamical clustering identification,
system validation (for ensuring reliable color measurements of inner solar system objects),
survey efficiency characterization and debiasing (for ensuring unbiased searches for correlations between taxonomic and dynamical properties),
alert brokering and follow-up observation management (for additional physical characterization),
external data incorporation (for additional physical characterization), and
data access and visualization.

\medskip\hangindent=.80cm\hangafter=1
\noindent 5. {\bf Investigation of hydration states of C-complex objects and main-belt asteroids:} Will require tools for
orbital object and detection parameter computation (for activity detection to enable omission of active data),
phase function characterization (for compositional characterization),
compositional characterization,
rotational characterization (for activity detection to enable omission of active data),
activity detection (for enabling omission of active data),
outburst and disruption detection (for enabling omission of active data),
system validation (for ensuring reliable identification of hydration states),
alert brokering and follow-up observation management (for additional physical characterization),
external data incorporation (for additional physical characterization), and
data access and visualization.

\medskip\hangindent=.80cm\hangafter=1
\noindent 6. {\bf Determination of rotational properties for asteroids in different taxonomic classes to determine spin angular momentum and shape distributions, and binary frequency:} Will require tools for
orbital object and detection parameter computation (for activity detection to enable omission of active data),
phase function characterization,
compositional characterization,
rotational characterization (for activity detection to enable omission of active data),
detection and characterization of multi-object systems,
activity detection (for enabling omission of active data),
outburst and disruption detection (for enabling omission of active data),
system validation (for ensuring reliable measurement of rotational properties),
survey efficiency characterization and debiasing,
alert brokering and follow-up observation management (for additional physical characterization),
external data incorporation (for additional physical characterization), and
data access and visualization.

\medskip\hangindent=.80cm\hangafter=1
\noindent 7. {\bf Characterization of asteroid families, clusters, and pairs to study genetic relationships and homogeneity of families at small sizes:} Will require tools for
orbital object and detection parameter computation (for activity detection to enable omission of active data),
phase function characterization,
compositional characterization,
rotational characterization (for activity detection to enable omission of active data),
activity detection,
outburst and disruption detection,
advanced dynamical characterization,
dynamical clustering identification,
system validation (for ensuring reliable measurement of compositional properties),
survey efficiency characterization and debiasing (for ensuring unbiased measurement of compositional properties),
alert brokering and follow-up observation management (for additional physical characterization),
external data incorporation (for additional physical characterization), and
data access and visualization.

\medskip\hangindent=.80cm\hangafter=1
\noindent 8. {\bf Measurement of asteroid masses and bulk densities from mutual gravitational interactions:} Will require tools for
orbital object and detection parameter computation (for activity detection to enable omission of active data),
orbital element and ephemeris uncertainty characterization (for orbit refinement),
faint precovery and recovery identification (for orbit refinement),
extremely faint object detection (for orbit refinement),
phase function characterization,
compositional characterization (for albedo estimation for size and density determination),
rotational characterization (for activity detection to enable omission of active data),
detection and characterization of multi-object systems,
activity detection (for enabling omission of active data),
outburst and disruption detection (for enabling omission of active data),
detection of changes in dynamical properties,
alert brokering and follow-up observation management (for additional orbit refinement),
external data incorporation (for additional orbit refinement), and
data access and visualization.

\medskip\hangindent=.80cm\hangafter=1
\noindent 9. {\bf Detection and estimation of the frequency of rotational fission events among non-NEO asteroids:} Will require tools for
orbital object and detection parameter computation (for activity detection),
phase function characterization (for activity detection),
rotational characterization,
detection and characterization of multi-object systems,
activity detection,
activity characterization (for identification of fission events),
outburst and disruption detection,
dynamical clustering identification,
system validation (for ensuring reliable disruption event detection),
survey efficiency characterization and debiasing,
alert brokering and follow-up observation management (for additional physical characterization),
external data incorporation (for additional physical characterization), and
data access and visualization.

%%%%%%%%%% SOFTWARE REQUIREMENTS FOR OUTER SOLAR SYSTEM OBJECTS %%%%%%%%%%

\bigskip
\subsection{Science Priorities and Software Requirements for Outer Solar System Objects\label{subsection:appendix_sciencepriorities_outerss}}

\medskip\hangindent=.80cm\hangafter=1
\noindent 1. {\bf Discovery and orbital classification of outer solar system objects to characterize the size and orbital distributions of TNOs:} Will require tools for
orbital object and detection parameter computation (for dynamical classification and activity detection to enable omission of active data),
orbital element and ephemeris uncertainty characterization (for orbit refinement and dynamical classification),
faint precovery and recovery identification (for orbit refinement and dynamical classification),
extremely faint object detection (for orbit refinement and dynamical classification),
advanced moving object detection (for orbit refinement and dynamical classification),
phase function characterization,
compositional characterization (for albedo estimation for size determination),
rotational characterization (for activity detection to enable omission of active data),
activity detection (for enabling omission of active data),
outburst and disruption detection (for enabling omission of active data),
advanced dynamical characterization,
system validation (for ensuring reliable detection of different dynamical types of outer solar system objects),
survey efficiency characterization and debiasing,
alert brokering and follow-up observation management (for additional physical characterization),
external data incorporation (for additional physical characterization), and
data access and visualization.

\medskip\hangindent=.80cm\hangafter=1
\noindent 2. {\bf Discovery and orbital classification of objects on unusual or extreme orbits to place constraints on proposed origin scenarios:} Will require tools for
orbital object and detection parameter computation (for dynamical classification),
orbital element and ephemeris uncertainty characterization (for orbit refinement and dynamical classification),
faint precovery and recovery identification (for orbit refinement and dynamical classification),
extremely faint object detection (for orbit refinement and dynamical classification),
advanced moving object detection (for orbit refinement and dynamical classification),
advanced dynamical characterization,
system validation (for ensuring reliable detection of outer solar system objects on unusual or extreme orbits),
survey efficiency characterization and debiasing,
alert brokering and follow-up observation management (for additional orbit refinement),
external data incorporation (for additional orbit refinement), and
data access and visualization.

\medskip\hangindent=.80cm\hangafter=1
\noindent 3. {\bf Determination of colors for outer solar system objects to identify correlations with dynamical properties:} Will require tools for
orbital object and detection parameter computation (for dynamical classification and activity detection to enable omission of active data),
orbital element and ephemeris uncertainty characterization (for orbit refinement and dynamical classification),
faint precovery and recovery identification (for orbit refinement and dynamical classification),
extremely faint object detection (for orbit refinement and dynamical classification),
advanced moving object detection (for orbit refinement and dynamical classification),
phase function characterization,
compositional characterization,
rotational characterization (for activity detection to enable omission of active data),
activity detection (for enabling omission of active data),
outburst and disruption detection (for enabling omission of active data),
advanced dynamical characterization,
system validation (for ensuring reliable detection of different compositional types of outer solar system objects),
survey efficiency characterization and debiasing (for ensuring unbiased searches for correlations between colors and dynamical properties),
alert brokering and follow-up observation management (for additional physical characterization),
external data incorporation (for additional physical characterization), and
data access and visualization.

\medskip\hangindent=.80cm\hangafter=1
\noindent 4. {\bf Determination of rotational properties for outer solar system objects from different dynamical classes to determine spin angular momentum distributions and binary frequency:} Will require tools for
orbital object and detection parameter computation (for dynamical classification and activity detection to enable omission of active data),
orbital element and ephemeris uncertainty characterization (for orbit refinement and dynamical classification),
faint precovery and recovery identification (for orbit refinement and dynamical classification),
extremely faint object detection (for orbit refinement and dynamical classification),
advanced moving object detection (for orbit refinement and dynamical classification),
phase function characterization (for activity detection to enable omission of active data),
rotational characterization,
detection and characterization of multi-object systems,
activity detection (for enabling omission of active data),
outburst and disruption detection (for enabling omission of active data),
advanced dynamical characterization,
system validation (for ensuring reliable detection of different dynamical types of outer solar system objects),
survey efficiency characterization and debiasing,
alert brokering and follow-up observation management (for additional physical characterization),
external data incorporation (for additional physical characterization), and
data access and visualization.

\medskip\hangindent=.80cm\hangafter=1
\noindent 5. {\bf Discovery and orbital classification of outer solar system objects in resonance with the giant planets:} Will require tools for
orbital object and detection parameter computation (for dynamical classification),
orbital element and ephemeris uncertainty characterization (for orbit refinement and dynamical classification),
faint precovery and recovery identification (for orbit refinement and dynamical classification),
extremely faint object detection (for orbit refinement and dynamical classification),
advanced moving object detection (for orbit refinement and dynamical classification),
advanced dynamical characterization,
system validation (for ensuring reliable detection of resonant outer solar system objects),
survey efficiency characterization and debiasing,
alert brokering and follow-up observation management (for additional orbit refinement),
external data incorporation (for additional orbit refinement), and
data access and visualization.

\medskip\hangindent=.80cm\hangafter=1
\noindent 6. {\bf Discovery and characterization of resolved or partially resolved binary and other multi-object systems:} Will require tools for
orbital object and detection parameter computation (for activity detection to enable omission of active data),
phase function characterization,
compositional characterization,
rotational characterization,
detection and characterization of multi-object systems,
activity detection (for enabling omission of active data),
outburst and disruption detection (for enabling omission of active data),
dynamical clustering identification,
system validation (for ensuring reliable identification of resolved or partially resolved binary outer solar system objects),
alert brokering and follow-up observation management (for additional physical characterization),
external data incorporation (for additional physical characterization), and
data access and visualization.

\medskip\hangindent=.80cm\hangafter=1
\noindent 7. {\bf Measurement of precise astrometry for outer solar system objects to enable stellar occultation observations:} Will require tools for
orbital object and detection parameter computation (for enabling accurate ephemeris predictions),
orbital element and ephemeris uncertainty characterization,
faint precovery and recovery identification (for orbit refinement),
extremely faint object detection (for orbit refinement),
advanced moving object detection (for orbit refinement),
occultation event prediction,
system validation (for ensuring reliable measurement of precision astrometry),
alert brokering and follow-up observation management (for additional orbit refinement),
external data incorporation (for additional orbit refinement), and
data access and visualization.

\clearpage
\subsection{Summaries of Software Needs by Science Area\label{subsection:appendix_sciencepriorities_summarytables}}

\setlength{\tabcolsep}{4pt}
\renewcommand*{\arraystretch}{1.2}
\begin{table}[ht!]
\captionsetup{font=normalsize}
\centering
\caption{Summary of Software Needs for Active Object Science}
\label{table:swneeds_summary_active}
\begin{tabular}{p{3.35in}cccccc}
\hline\hline
 & \multicolumn{6}{c}{\vspace{-0.1cm}Active Object} \\
 & \multicolumn{6}{c}{Science Priorities$^a$} \\
Software Task & 1 & 2 & 3 & 4 & 5 & 6 \\
\hline\hline
\hangindent=.60cm\hangafter=1 1. Orbital parameter computation (\ref{subsection:appendix_orbital_parameters})
  & \checkmark & \checkmark & \checkmark & \checkmark & \checkmark & \checkmark \\
\hangindent=.60cm\hangafter=1 2. Orbital uncertainty characterization (\ref{subsection:appendix_uncertainty_characterization})
  & \checkmark & \checkmark & \checkmark & ---        & ---        & \checkmark \\
\hangindent=.60cm\hangafter=1 3. Extended object astrometry (\ref{subsection:appendix_extended_object_astrometry})
  & \checkmark & \checkmark & \checkmark & ---        & ---        & \checkmark \\
\hangindent=.60cm\hangafter=1 4. Faint precovery/recovery identification (\ref{subsection:appendix_lsst_precoveries_recoveries})
  & \checkmark & \checkmark & \checkmark & \checkmark & ---        & \checkmark \\
\hangindent=.60cm\hangafter=1 5. Extremely faint object detection (\ref{subsection:appendix_faint_object_detection})
  & \checkmark & \checkmark & \checkmark & \checkmark & ---        & \checkmark \\
\hangindent=.80cm\hangafter=1 6. Advanced moving object detection (\ref{subsection:appendix_advanced_moving_object_detection})
  & ---        & ---        & ---        & ---        & ---        & ---        \\
\hangindent=.60cm\hangafter=1 7. Phase function characterization (\ref{subsection:appendix_phase_function_characterization})
  & \checkmark & \checkmark & \checkmark & \checkmark & \checkmark & \checkmark \\
\hangindent=.60cm\hangafter=1 8. Compositional characterization (\ref{subsection:appendix_compositional_characterization})
  & \checkmark & ---        & ---        & \checkmark & ---        & ---        \\
\hangindent=.60cm\hangafter=1 9. Rotational characterization (\ref{subsection:appendix_rotational_characterization})
  & \checkmark & \checkmark & \checkmark & \checkmark & \checkmark & \checkmark \\
\hangindent=.60cm\hangafter=1 10. Multi-object system characterization (\ref{subsection:appendix_multi_object_systems})
  & ---        & ---        & \checkmark & ---        & \checkmark & ---        \\
\hangindent=.80cm\hangafter=1 11. Activity detection (\ref{subsection:appendix_activity_detection})
  & \checkmark & \checkmark & \checkmark & \checkmark & \checkmark & \checkmark \\
\hangindent=.80cm\hangafter=1 12. Activity characterization (\ref{subsection:appendix_activity_characterization})
  & \checkmark & \checkmark & \checkmark & \checkmark & \checkmark & \checkmark \\
\hangindent=.80cm\hangafter=1 13. Outburst and disruption detection (\ref{subsection:appendix_outburst_disruption_detection})
  & \checkmark & \checkmark & \checkmark & \checkmark & ---        & \checkmark \\
\hangindent=.80cm\hangafter=1 14. Advanced dynamical characterization (\ref{subsection:appendix_advanced_dynamical_characterization})
  & ---        & ---        & \checkmark & ---        & ---        & ---        \\
\hangindent=.80cm\hangafter=1 15. Dynamical clustering identification (\ref{subsection:appendix_dynamical_clustering_identification})
  & ---        & ---        & \checkmark & ---        & ---        & ---        \\
\hangindent=.80cm\hangafter=1 16. Occultation event prediction (\ref{subsection:appendix_occultation_prediction})
  & ---        & ---        & ---        & ---        & ---        & ---        \\
\hangindent=.80cm\hangafter=1 17. Physical property change detection (\ref{subsection:appendix_physical_evolution})
  & ---        & ---        & ---        & \checkmark & ---        & ---        \\
\hangindent=.80cm\hangafter=1 18. Dynamical property change detection (\ref{subsection:appendix_dynamical_evolution})
  & ---        & ---        & ---        & \checkmark & ---        & \checkmark \\
\hangindent=.80cm\hangafter=1 19. System validation (\ref{subsection:appendix_system_validation})
  & \checkmark & \checkmark & \checkmark & \checkmark & \checkmark & ---        \\
\hangindent=.80cm\hangafter=1 20. Survey debiasing (\ref{subsection:appendix_survey_debiasing})
  & ---        & \checkmark & \checkmark & ---        & \checkmark & ---        \\
\hangindent=.80cm\hangafter=1 21. Alert and follow-up management (\ref{subsection:appendix_alerts})
  & \checkmark & \checkmark & \checkmark & \checkmark & \checkmark & ---        \\
\hangindent=.80cm\hangafter=1 22. External data incorporation (\ref{subsection:appendix_externaldata})
  & \checkmark & \checkmark & \checkmark & \checkmark & \checkmark & \checkmark \\
\hangindent=.80cm\hangafter=1 23. Data access tools (\ref{subsection:appendix_dataaccess})
  & \checkmark & \checkmark & \checkmark & \checkmark & \checkmark & \checkmark \\
\hline\hline
\multicolumn{7}{l}{$^a$ Science roadmap priorities identified for active objects (Appendix~\ref{subsection:appendix_sciencepriorities_active})\vspace{-0.1cm}} \\
\end{tabular}
\end{table}

\setlength{\tabcolsep}{4pt}
\renewcommand*{\arraystretch}{1.2}
\begin{table}[ht!]
\captionsetup{font=normalsize}
\centering
\caption{Summary of Software Needs for Near-Earth Object Science}
\label{table:swneeds_summary_active}
\begin{tabular}{p{3.35in}cccccccccc}
\hline\hline
 & \multicolumn{10}{c}{NEO Science Priorities$^a$} \\
Software Task & 1 & 2 & 3 & 4 & 5 & 6 & 7 & 8 & 9 & 10 \\
\hline\hline
\hangindent=.60cm\hangafter=1 1. Orbital parameter computation (\ref{subsection:appendix_orbital_parameters})
  & ---        & \checkmark & ---        & \checkmark & \checkmark & ---        & \checkmark & ---        & \checkmark & \checkmark \\
\hangindent=.60cm\hangafter=1 2. Orbital uncertainty characterization (\ref{subsection:appendix_uncertainty_characterization})
  & \checkmark & ---        & \checkmark & \checkmark & \checkmark & \checkmark & ---        & \checkmark & \checkmark & \checkmark \\
\hangindent=.60cm\hangafter=1 3. Extended object astrometry (\ref{subsection:appendix_extended_object_astrometry})
  & \checkmark & ---        & \checkmark & \checkmark & \checkmark & \checkmark & ---        & \checkmark & ---        & \checkmark \\
\hangindent=.60cm\hangafter=1 4. Faint precovery/recovery identification (\ref{subsection:appendix_lsst_precoveries_recoveries})
  & \checkmark & ---        & \checkmark & \checkmark & \checkmark & \checkmark & ---        & \checkmark & \checkmark & \checkmark \\
\hangindent=.60cm\hangafter=1 5. Extremely faint object detection (\ref{subsection:appendix_faint_object_detection})
  & \checkmark & ---        & \checkmark & \checkmark & \checkmark & \checkmark & ---        & \checkmark & \checkmark & \checkmark \\
\hangindent=.80cm\hangafter=1 6. Advanced moving object detection (\ref{subsection:appendix_advanced_moving_object_detection})
  & \checkmark & ---        & \checkmark & \checkmark & \checkmark & \checkmark & ---        & \checkmark & \checkmark & \checkmark \\
\hangindent=.60cm\hangafter=1 7. Phase function characterization (\ref{subsection:appendix_phase_function_characterization})
  & ---        & \checkmark & ---        & \checkmark & \checkmark & ---        & \checkmark & ---        & \checkmark & \checkmark \\
\hangindent=.60cm\hangafter=1 8. Compositional characterization (\ref{subsection:appendix_compositional_characterization})
  & ---        & \checkmark & ---        & \checkmark & \checkmark & ---        & ---        & ---        & ---        & \checkmark \\
\hangindent=.60cm\hangafter=1 9. Rotational characterization (\ref{subsection:appendix_rotational_characterization})
  & ---        & ---        & ---        & \checkmark & \checkmark & ---        & \checkmark & \checkmark & \checkmark & \checkmark \\
\hangindent=.60cm\hangafter=1 10. Multi-object system characterization (\ref{subsection:appendix_multi_object_systems})
  & ---        & ---        & ---        & ---        & ---        & ---        & \checkmark & ---        & ---        & ---        \\
\hangindent=.80cm\hangafter=1 11. Activity detection (\ref{subsection:appendix_activity_detection})
  & ---        & \checkmark & ---        & \checkmark & \checkmark & ---        & \checkmark & \checkmark & \checkmark & \checkmark \\
\hangindent=.80cm\hangafter=1 12. Activity characterization (\ref{subsection:appendix_activity_characterization})
  & ---        & ---        & ---        & ---        & ---        & ---        & ---        & \checkmark & ---        & \checkmark \\
\hangindent=.80cm\hangafter=1 13. Outburst and disruption detection (\ref{subsection:appendix_outburst_disruption_detection})
  & ---        & ---        & ---        & \checkmark & \checkmark & ---        & \checkmark & \checkmark & \checkmark & \checkmark \\
\hangindent=.80cm\hangafter=1 14. Advanced dynamical characterization (\ref{subsection:appendix_advanced_dynamical_characterization})
  & ---        & ---        & \checkmark & \checkmark & \checkmark & ---        & ---        & ---        & \checkmark & ---        \\
\hangindent=.80cm\hangafter=1 15. Dynamical clustering identification (\ref{subsection:appendix_dynamical_clustering_identification})
  & ---        & ---        & ---        & \checkmark & ---        & \checkmark & ---        & ---        & ---        & ---        \\
\hangindent=.80cm\hangafter=1 16. Occultation event prediction (\ref{subsection:appendix_occultation_prediction})
  & ---        & ---        & ---        & ---        & ---        & ---        & ---        & ---        & ---        & ---        \\
\hangindent=.80cm\hangafter=1 17. Physical property change detection (\ref{subsection:appendix_physical_evolution})
  & ---        & ---        & ---        & ---        & ---        & ---        & ---        & ---        & ---        & ---        \\
\hangindent=.80cm\hangafter=1 18. Dynamical property change detection (\ref{subsection:appendix_dynamical_evolution})
  & ---        & ---        & ---        & ---        & ---        & ---        & ---        & \checkmark & ---        & ---        \\
\hangindent=.80cm\hangafter=1 19. System validation (\ref{subsection:appendix_system_validation})
  & \checkmark & \checkmark & \checkmark & ---        & \checkmark & \checkmark & \checkmark & ---        & \checkmark & ---        \\
\hangindent=.80cm\hangafter=1 20. Survey debiasing (\ref{subsection:appendix_survey_debiasing})
  & \checkmark & ---        & ---        & \checkmark & \checkmark & \checkmark & \checkmark & ---        & \checkmark & \checkmark \\
\hangindent=.80cm\hangafter=1 21. Alert and follow-up management (\ref{subsection:appendix_alerts})
  & \checkmark & \checkmark & \checkmark & \checkmark & \checkmark & \checkmark & \checkmark & \checkmark & \checkmark & \checkmark \\
\hangindent=.80cm\hangafter=1 22. External data incorporation (\ref{subsection:appendix_externaldata})
  & \checkmark & \checkmark & \checkmark & \checkmark & \checkmark & \checkmark & \checkmark & \checkmark & \checkmark & \checkmark \\
\hangindent=.80cm\hangafter=1 23. Data access tools (\ref{subsection:appendix_dataaccess})
  & \checkmark & \checkmark & \checkmark & \checkmark & \checkmark & \checkmark & \checkmark & \checkmark & \checkmark & \checkmark \\
\hline\hline
\multicolumn{11}{l}{$^a$ Science roadmap priorities identified for near-Earth objects (Appendix~\ref{subsection:appendix_sciencepriorities_neos})\vspace{-0.1cm}} \\
\end{tabular}
\end{table}

\setlength{\tabcolsep}{4pt}
\renewcommand*{\arraystretch}{1.2}
\begin{table}[ht!]
\captionsetup{font=normalsize}
\centering
\caption{Summary of Software Needs for Inner Solar System Science}
\label{table:swneeds_summary_active}
\begin{tabular}{p{3.35in}ccccccccc}
\hline\hline
 & \multicolumn{9}{c}{ISS Science Priorities$^a$} \\
Software Task & 1 & 2 & 3 & 4 & 5 & 6 & 7 & 8 & 9 \\
\hline\hline
\hangindent=.60cm\hangafter=1 1. Orbital parameter computation (\ref{subsection:appendix_orbital_parameters})
  & \checkmark & ---        & \checkmark & \checkmark & \checkmark & \checkmark & \checkmark & \checkmark & \checkmark \\
\hangindent=.60cm\hangafter=1 2. Orbital uncertainty characterization (\ref{subsection:appendix_uncertainty_characterization})
  & \checkmark & \checkmark & ---        & \checkmark & ---        & ---        & ---        & \checkmark & ---        \\
\hangindent=.60cm\hangafter=1 3. Extended object astrometry (\ref{subsection:appendix_extended_object_astrometry})
  & ---        & ---        & \checkmark & ---        & ---        & ---        & ---        & ---        & ---        \\
\hangindent=.60cm\hangafter=1 4. Faint precovery/recovery identification (\ref{subsection:appendix_lsst_precoveries_recoveries})
  & \checkmark & \checkmark & \checkmark & \checkmark & ---        & ---        & ---        & \checkmark & ---        \\
\hangindent=.60cm\hangafter=1 5. Extremely faint object detection (\ref{subsection:appendix_faint_object_detection})
  & \checkmark & \checkmark & \checkmark & \checkmark & ---        & ---        & ---        & \checkmark & ---        \\
\hangindent=.80cm\hangafter=1 6. Advanced moving object detection (\ref{subsection:appendix_advanced_moving_object_detection})
  & \checkmark & ---        & ---        & ---        & ---        & ---        & ---        & ---        & ---        \\
\hangindent=.60cm\hangafter=1 7. Phase function characterization (\ref{subsection:appendix_phase_function_characterization})
  & \checkmark & ---        & \checkmark & \checkmark & \checkmark & \checkmark & \checkmark & \checkmark & \checkmark \\
\hangindent=.60cm\hangafter=1 8. Compositional characterization (\ref{subsection:appendix_compositional_characterization})
  & \checkmark & ---        & \checkmark & \checkmark & \checkmark & \checkmark & \checkmark & \checkmark & ---        \\
\hangindent=.60cm\hangafter=1 9. Rotational characterization (\ref{subsection:appendix_rotational_characterization})
  & \checkmark & ---        & \checkmark & \checkmark & \checkmark & \checkmark & \checkmark & \checkmark & \checkmark \\
\hangindent=.60cm\hangafter=1 10. Multi-object system characterization (\ref{subsection:appendix_multi_object_systems})
  & \checkmark & ---        & \checkmark & ---        & ---        & \checkmark & ---        & \checkmark & \checkmark \\
\hangindent=.80cm\hangafter=1 11. Activity detection (\ref{subsection:appendix_activity_detection})
  & \checkmark & ---        & \checkmark & \checkmark & \checkmark & \checkmark & \checkmark & \checkmark & \checkmark \\
\hangindent=.80cm\hangafter=1 12. Activity characterization (\ref{subsection:appendix_activity_characterization})
  & ---        & ---        & \checkmark & ---        & ---        & ---        & ---        & ---        & \checkmark \\
\hangindent=.80cm\hangafter=1 13. Outburst and disruption detection (\ref{subsection:appendix_outburst_disruption_detection})
  & \checkmark & ---        & \checkmark & \checkmark & \checkmark & \checkmark & \checkmark & \checkmark & \checkmark \\
\hangindent=.80cm\hangafter=1 14. Advanced dynamical characterization (\ref{subsection:appendix_advanced_dynamical_characterization})
  & ---        & ---        & ---        & \checkmark & ---        & ---        & \checkmark & ---        & ---        \\
\hangindent=.80cm\hangafter=1 15. Dynamical clustering identification (\ref{subsection:appendix_dynamical_clustering_identification})
  & ---        & ---        & ---        & \checkmark & ---        & ---        & \checkmark & ---        & \checkmark \\
\hangindent=.80cm\hangafter=1 16. Occultation event prediction (\ref{subsection:appendix_occultation_prediction})
  & ---        & \checkmark & ---        & ---        & ---        & ---        & ---        & ---        & ---        \\
\hangindent=.80cm\hangafter=1 17. Physical property change detection (\ref{subsection:appendix_physical_evolution})
  & ---        & ---        & \checkmark & ---        & ---        & ---        & ---        & ---        & ---        \\
\hangindent=.80cm\hangafter=1 18. Dynamical property change detection (\ref{subsection:appendix_dynamical_evolution})
  & ---        & ---        & ---        & ---        & ---        & ---        & ---        & \checkmark & ---        \\
\hangindent=.80cm\hangafter=1 19. System validation (\ref{subsection:appendix_system_validation})
  & \checkmark & \checkmark & \checkmark & \checkmark & \checkmark & \checkmark & \checkmark & ---        & \checkmark \\
\hangindent=.80cm\hangafter=1 20. Survey debiasing (\ref{subsection:appendix_survey_debiasing})
  & \checkmark & ---        & \checkmark & \checkmark & ---        & \checkmark & \checkmark & ---        & \checkmark \\
\hangindent=.80cm\hangafter=1 21. Alert and follow-up management (\ref{subsection:appendix_alerts})
  & \checkmark & \checkmark & \checkmark & \checkmark & \checkmark & \checkmark & \checkmark & \checkmark & \checkmark \\
\hangindent=.80cm\hangafter=1 22. External data incorporation (\ref{subsection:appendix_externaldata})
  & \checkmark & \checkmark & \checkmark & \checkmark & \checkmark & \checkmark & \checkmark & \checkmark & \checkmark \\
\hangindent=.80cm\hangafter=1 23. Data access tools (\ref{subsection:appendix_dataaccess})
  & \checkmark & \checkmark & \checkmark & \checkmark & \checkmark & \checkmark & \checkmark & \checkmark & \checkmark \\
\hline\hline
\multicolumn{10}{l}{$^a$ Science roadmap priorities identified for inner solar system objects (Appendix~\ref{subsection:appendix_sciencepriorities_innerss})\vspace{-0.1cm}} \\
\end{tabular}
\end{table}

\setlength{\tabcolsep}{4pt}
\renewcommand*{\arraystretch}{1.2}
\begin{table}[ht!]
\captionsetup{font=normalsize}
\centering
\caption{Summary of Software Needs for Outer Solar System Science}
\label{table:swneeds_summary_active}
\begin{tabular}{p{3.35in}ccccccc}
\hline\hline
 & \multicolumn{7}{c}{OSS Science Priorities$^a$} \\
Software Task & 1 & 2 & 3 & 4 & 5 & 6 & 7 \\
\hline\hline
\hangindent=.60cm\hangafter=1 1. Orbital parameter computation (\ref{subsection:appendix_orbital_parameters})
  & \checkmark & \checkmark & \checkmark & \checkmark & \checkmark & \checkmark & \checkmark \\
\hangindent=.60cm\hangafter=1 2. Orbital uncertainty characterization (\ref{subsection:appendix_uncertainty_characterization})
  & \checkmark & \checkmark & \checkmark & \checkmark & \checkmark & ---        & \checkmark \\
\hangindent=.60cm\hangafter=1 3. Extended object astrometry (\ref{subsection:appendix_extended_object_astrometry})
  & ---        & ---        & ---        & ---        & ---        & ---        & ---        \\
\hangindent=.60cm\hangafter=1 4. Faint precovery/recovery identification (\ref{subsection:appendix_lsst_precoveries_recoveries})
  & \checkmark & \checkmark & \checkmark & \checkmark & \checkmark & ---        & \checkmark \\
\hangindent=.60cm\hangafter=1 5. Extremely faint object detection (\ref{subsection:appendix_faint_object_detection})
  & \checkmark & ---        & \checkmark & \checkmark & \checkmark & ---        & \checkmark \\
\hangindent=.80cm\hangafter=1 6. Advanced moving object detection (\ref{subsection:appendix_advanced_moving_object_detection})
  & \checkmark & \checkmark & \checkmark & \checkmark & \checkmark & ---        & \checkmark \\
\hangindent=.60cm\hangafter=1 7. Phase function characterization (\ref{subsection:appendix_phase_function_characterization})
  & \checkmark & ---        & \checkmark & \checkmark & ---        & \checkmark & ---        \\
\hangindent=.60cm\hangafter=1 8. Compositional characterization (\ref{subsection:appendix_compositional_characterization})
  & \checkmark & ---        & \checkmark & ---        & ---        & \checkmark & ---        \\
\hangindent=.60cm\hangafter=1 9. Rotational characterization (\ref{subsection:appendix_rotational_characterization})
  & \checkmark & ---        & \checkmark & \checkmark & ---        & \checkmark & ---        \\
\hangindent=.60cm\hangafter=1 10. Multi-object system characterization (\ref{subsection:appendix_multi_object_systems})
  & ---        & ---        & ---        & \checkmark & ---        & \checkmark & ---        \\
\hangindent=.80cm\hangafter=1 11. Activity detection (\ref{subsection:appendix_activity_detection})
  & \checkmark & ---        & \checkmark & \checkmark & ---        & \checkmark & ---        \\
\hangindent=.80cm\hangafter=1 12. Activity characterization (\ref{subsection:appendix_activity_characterization})
  & ---        & ---        & ---        & ---        & ---        & ---        & ---        \\
\hangindent=.80cm\hangafter=1 13. Outburst and disruption detection (\ref{subsection:appendix_outburst_disruption_detection})
  & \checkmark & ---        & \checkmark & \checkmark & ---        & \checkmark & ---        \\
\hangindent=.80cm\hangafter=1 14. Advanced dynamical characterization (\ref{subsection:appendix_advanced_dynamical_characterization})
  & \checkmark & \checkmark & \checkmark & \checkmark & \checkmark & ---        & ---        \\
\hangindent=.80cm\hangafter=1 15. Dynamical clustering identification (\ref{subsection:appendix_dynamical_clustering_identification})
  & ---        & ---        & ---        & ---        & ---        & \checkmark & ---        \\
\hangindent=.80cm\hangafter=1 16. Occultation event prediction (\ref{subsection:appendix_occultation_prediction})
  & ---        & ---        & ---        & ---        & ---        & ---        & \checkmark \\
\hangindent=.80cm\hangafter=1 17. Physical property change detection (\ref{subsection:appendix_physical_evolution})
  & ---        & ---        & ---        & ---        & ---        & ---        & ---        \\
\hangindent=.80cm\hangafter=1 18. Dynamical property change detection (\ref{subsection:appendix_dynamical_evolution})
  & ---        & ---        & ---        & ---        & ---        & ---        & ---        \\
\hangindent=.80cm\hangafter=1 19. System validation (\ref{subsection:appendix_system_validation})
  & \checkmark & \checkmark & \checkmark & \checkmark & \checkmark & \checkmark & \checkmark \\
\hangindent=.80cm\hangafter=1 20. Survey debiasing (\ref{subsection:appendix_survey_debiasing})
  & \checkmark & \checkmark & \checkmark & \checkmark & \checkmark & ---        & ---        \\
\hangindent=.80cm\hangafter=1 21. Alert and follow-up management (\ref{subsection:appendix_alerts})
  & \checkmark & \checkmark & \checkmark & \checkmark & \checkmark & \checkmark & \checkmark \\
\hangindent=.80cm\hangafter=1 22. External data incorporation (\ref{subsection:appendix_externaldata})
  & \checkmark & \checkmark & \checkmark & \checkmark & \checkmark & \checkmark & \checkmark \\
\hangindent=.80cm\hangafter=1 23. Data access tools (\ref{subsection:appendix_dataaccess})
  & \checkmark & \checkmark & \checkmark & \checkmark & \checkmark & \checkmark & \checkmark \\
\hline\hline
\multicolumn{8}{l}{$^a$ Science roadmap priorities identified for outer solar system objects\vspace{-0.1cm}} \\
\multicolumn{8}{l}{$~~~~$ (Appendix~\ref{subsection:appendix_sciencepriorities_outerss})\vspace{-0.1cm}} \\
\end{tabular}
\end{table}

\clearpage
\bibliographystyle{aasjournal}
\bibliography{lsst_sssb_swdev_refs}   % name your BibTeX data base

\begin{thebibliography}{}
\expandafter\ifx\csname natexlab\endcsname\relax\def\natexlab#1{#1}\fi
\providecommand{\url}[1]{\href{#1}{#1}}

\bibitem[{{Agarwal} {et~al.}(2017){Agarwal}, {Jewitt}, {Mutchler}, {Weaver}, \&
  {Larson}}]{agarwal2017_288p}
{Agarwal}, J., {Jewitt}, D., {Mutchler}, M., {Weaver}, H., \& {Larson}, S.
  2017, \nat, 549, 357

\bibitem[{{Agarwal} {et~al.}(2016){Agarwal}, {Jewitt}, {Weaver}, {Mutchler}, \&
  {Larson}}]{agarwal2016_288p}
{Agarwal}, J., {Jewitt}, D., {Weaver}, H., {Mutchler}, M., \& {Larson}, S.
  2016, \aj, 151, 12

\bibitem[{{A'Hearn} {et~al.}(1984){A'Hearn}, {Schleicher}, {Millis}, {Feldman},
  \& {Thompson}}]{ahearn1984_bowell}
{A'Hearn}, M.~F., {Schleicher}, D.~G., {Millis}, R.~L., {Feldman}, P.~D., \&
  {Thompson}, D.~T. 1984, \aj, 89, 579

\bibitem[{{A'Hearn} {et~al.}(2005){A'Hearn}, {Belton}, {Delamere}, {Kissel},
  {Klaasen}, {McFadden}, {Meech}, {Melosh}, {Schultz}, {Sunshine}, {Thomas},
  {Veverka}, {Yeomans}, {Baca}, {Busko}, {Crockett}, {Collins}, {Desnoyer},
  {Eberhardy}, {Ernst}, {Farnham}, {Feaga}, {Groussin}, {Hampton}, {Ipatov},
  {Li}, {Lindler}, {Lisse}, {Mastrodemos}, {Owen}, {Richardson}, {Wellnitz}, \&
  {White}}]{ahearn05}
{A'Hearn}, M.~F., {Belton}, M.~J.~S., {Delamere}, W.~A., {et~al.} 2005,
  Science, 310, 258

\bibitem[{{Akhlaghi} \& {Ichikawa}(2015)}]{akhlaghi2015_noisechisel}
{Akhlaghi}, M., \& {Ichikawa}, T. 2015, \apjs, 220, 1

\bibitem[{{Ashton}(2019)}]{ashton2019_ossos}
{Ashton}, E. e.~a. 2019, \icarus in review

\bibitem[{{Assafin} {et~al.}(2010){Assafin}, {Camargo}, {Vieira Martins},
  {Andrei}, {Sicardy}, {Young}, {da Silva Neto}, \&
  {Braga-Ribas}}]{assafin2010_occultations}
{Assafin}, M., {Camargo}, J.~I.~B., {Vieira Martins}, R., {et~al.} 2010, \aap,
  515, A32

\bibitem[{{Banda-Huarca} {et~al.}(2019){Banda-Huarca}, {Camargo}, {Desmars},
  {Ogand o}, {Vieira-Martins}, {Assafin}, {da Costa}, {Bernstein}, {Carrasco
  Kind}, {Drlica-Wagner}, {Gomes}, {Gysi}, {Braga-Ribas}, {Maia}, {Gerdes},
  {Hamilton}, {Wester}, {Abbott}, {Abdalla}, {Allam}, {Avila}, {Bertin},
  {Brooks}, {Buckley-Geer}, {Burke}, {Carnero Rosell}, {Carretero}, {Cunha},
  {Davis}, {De Vicente}, {Diehl}, {Doel}, {Fosalba}, {Frieman},
  {Garc{\'\i}a-Bellido}, {Gaztanaga}, {Gruen}, {Gruendl}, {Gschwend},
  {Gutierrez}, {Hartley}, {Hollowood}, {Honscheid}, {James}, {Kuehn},
  {Kuropatkin}, {Menanteau}, {Miller}, {Miquel}, {Plazas}, {Romer}, {Sanchez},
  {Scarpine}, {Schubnell}, {Serrano}, {Sevilla-Noarbe}, {Smith},
  {Soares-Santos}, {Sobreira}, {Suchyta}, {Swanson}, {Tarle}, \& {DES
  Collaboration}}]{bandahuarca2019_occultations}
{Banda-Huarca}, M.~V., {Camargo}, J.~I.~B., {Desmars}, J., {et~al.} 2019, \aj,
  157, 120

\bibitem[{Bannister(2014)}]{Bannister:2013}
Bannister, M.~T. 2014, PhD thesis, RSAA, the Australian National University.,
  Canberra

\bibitem[{{Bannister} {et~al.}(2016){Bannister}, {Kavelaars}, {Petit},
  {Gladman}, {Gwyn}, {Chen}, {Volk}, {Alexandersen}, {Benecchi}, \&
  {Delsanti}}]{Bannister:2016ossosdesign}
{Bannister}, M.~T., {Kavelaars}, J.~J., {Petit}, J.-M., {et~al.} 2016, \aj,
  152, 70

\bibitem[{{Bauer} {et~al.}(2003){Bauer}, {Fern{\'a}ndez}, \&
  {Meech}}]{bauer2003_c2001t4}
{Bauer}, J.~M., {Fern{\'a}ndez}, Y.~R., \& {Meech}, K.~J. 2003, \pasp, 115, 981

\bibitem[{{Belskaya} {et~al.}(2006){Belskaya}, {Ortiz}, {Rousselot}, {Ivanova},
  {Borisov}, {Shevchenko}, \& {Peixinho}}]{belskaya2006_tnooppositioneffect}
{Belskaya}, I.~N., {Ortiz}, J.~L., {Rousselot}, P., {et~al.} 2006, \icarus,
  184, 277

\bibitem[{{Belton} {et~al.}(2011){Belton}, {Meech}, {Chesley},
  {Pittichov{\'a}}, {Carcich}, {Drahus}, {Harris}, {Gillam}, {Veverka},
  {Mastrodemos}, {Owen}, {A'Hearn}, {Bagnulo}, {Bai}, {Barrera}, {Bastien},
  {Bauer}, {Bedient}, {Bhatt}, {Boehnhardt}, {Brosch}, {Buie}, {Candia},
  {Chen}, {Chiang}, {Choi}, {Cochran}, {Crockett}, {Duddy}, {Farnham},
  {Fern{\'a}ndez}, {Guti{\'e}rrez}, {Hainaut}, {Hampton}, {Herrmann}, {Hsieh},
  {Kadooka}, {Kaluna}, {Keane}, {Kim}, {Klaasen}, {Kleyna}, {Krisciunas},
  {Lara}, {Lauer}, {Li}, {Licandro}, {Lisse}, {Lowry}, {McFadden}, {Moskovitz},
  {Mueller}, {Polishook}, {Raja}, {Riesen}, {Sahu}, {Samarasinha}, {Sarid},
  {Sekiguchi}, {Sonnett}, {Suntzeff}, {Taylor}, {Thomas}, {Tozzi},
  {Vasundhara}, {Vincent}, {Wasserman}, {Webster-Schultz}, {Yang}, {Zenn}, \&
  {Zhao}}]{belton2011_tempel1}
{Belton}, M.~J.~S., {Meech}, K.~J., {Chesley}, S., {et~al.} 2011, \icarus, 213,
  345

\bibitem[{{B{\'e}rard} {et~al.}(2017){B{\'e}rard}, {Sicardy}, {Camargo},
  {Desmars}, {Braga-Ribas}, {Ortiz}, {Duffard}, {Morales}, {Meza}, \&
  {Leiva}}]{berard2017_chariklorings}
{B{\'e}rard}, D., {Sicardy}, B., {Camargo}, J.~I.~B., {et~al.} 2017, \aj, 154,
  144

\bibitem[{Bernstein \& Khushalani(2000)}]{bernstein2000_orbfit}
Bernstein, G., \& Khushalani, B. 2000, \aj, 120, 3323

\bibitem[{{Betzler} \& {Betzler}(2017)}]{betzler2017_haleboppoccultations}
{Betzler}, A.~S., \& {Betzler}, L.~B.~S. 2017, Earth Moon and Planets, 120, 1

\bibitem[{{Binzel} {et~al.}(2010){Binzel}, {Morbidelli}, {Merouane}, {DeMeo},
  {Birlan}, {Vernazza}, {Thomas}, {Rivkin}, {Bus}, \&
  {Tokunaga}}]{binzel2010_neoearthencounters}
{Binzel}, R.~P., {Morbidelli}, A., {Merouane}, S., {et~al.} 2010, \nat, 463,
  331

\bibitem[{{Bodewits} {et~al.}(2018){Bodewits}, {Farnham}, {Kelley}, \&
  {Knight}}]{bodewits2018_41Pspinslowdown}
{Bodewits}, D., {Farnham}, T.~L., {Kelley}, M.~S.~P., \& {Knight}, M.~M. 2018,
  \nat, 553, 186

\bibitem[{{Bodewits} {et~al.}(2011){Bodewits}, {Kelley}, {Li}, {Landsman},
  {Besse}, \& {A'Hearn}}]{bodewits2011_scheila}
{Bodewits}, D., {Kelley}, M.~S., {Li}, J.-Y., {et~al.} 2011, \apjl, 733, L3

\bibitem[{{Bodewits} {et~al.}(2014){Bodewits}, {Vincent}, \&
  {Kelley}}]{bodewits2014_scheila}
{Bodewits}, D., {Vincent}, J.-B., \& {Kelley}, M.~S.~P. 2014, \icarus, 229, 190

\bibitem[{{Boehnhardt} {et~al.}(2016){Boehnhardt}, {Riffeser}, {Kluge}, {Ries},
  {Schmidt}, \& {Hopp}}]{boehnhardt16}
{Boehnhardt}, H., {Riffeser}, A., {Kluge}, M., {et~al.} 2016, \mnras, 462, S376

\bibitem[{{Bolin} {et~al.}(2014){Bolin}, {Jedicke}, {Granvik}, {Brown},
  {Howell}, {Nolan}, {Jenniskens}, {Chyba}, {Patterson}, \&
  {Wainscoat}}]{Bolin2014}
{Bolin}, B., {Jedicke}, R., {Granvik}, M., {et~al.} 2014, \icarus, 241, 280

\bibitem[{{Bolin} {et~al.}(2017){Bolin}, {Delbo'}, {Morbidelli}, \&
  {Walsh}}]{Bolin2017}
{Bolin}, B.~T., {Delbo'}, M., {Morbidelli}, A., \& {Walsh}, K.~J. 2017,
  \icarus, 282, 290

\bibitem[{{Bowell} {et~al.}(1989){Bowell}, {Hapke}, {Domingue}, {Lumme},
  {Peltoniemi}, \& {Harris}}]{bowell1989_astphotmodels_ast2}
{Bowell}, E., {Hapke}, B., {Domingue}, D., {et~al.} 1989, in Asteroids II,
  524--556

\bibitem[{{Braga-Ribas} {et~al.}(2013){Braga-Ribas}, {Sicardy}, {Ortiz},
  {Lellouch}, {Tancredi}, {Lecacheux}, {Vieira-Martins}, {Camargo}, {Assafin},
  \& {Behrend}}]{bragaribas2013_quaoar}
{Braga-Ribas}, F., {Sicardy}, B., {Ortiz}, J.~L., {et~al.} 2013, \apj, 773, 26

\bibitem[{{Braga-Ribas} {et~al.}(2014){Braga-Ribas}, {Sicardy}, {Ortiz},
  {Snodgrass}, {Roques}, {Vieira-Martins}, {Camargo}, {Assafin}, {Duffard},
  {Jehin}, {Pollock}, {Leiva}, {Emilio}, {Machado}, {Colazo}, {Lellouch},
  {Skottfelt}, {Gillon}, {Ligier}, {Maquet}, {Benedetti-Rossi}, {Gomes},
  {Kervella}, {Monteiro}, {Sfair}, {El Moutamid}, {Tancredi}, {Spagnotto},
  {Maury}, {Morales}, {Gil-Hutton}, {Roland}, {Ceretta}, {Gu}, {Wang},
  {Harps{\o}e}, {Rabus}, {Manfroid}, {Opitom}, {Vanzi}, {Mehret}, {Lorenzini},
  {Schneiter}, {Melia}, {Lecacheux}, {Colas}, {Vachier}, {Widemann},
  {Almenares}, {Sandness}, {Char}, {Perez}, {Lemos}, {Martinez},
  {J{\o}rgensen}, {Dominik}, {Roig}, {Reichart}, {Lacluyze}, {Haislip},
  {Ivarsen}, {Moore}, {Frank}, \& {Lambas}}]{bragaribas2014_chariklo}
---. 2014, \nat, 508, 72

\bibitem[{{Brasser} \& {Lehto}(2002)}]{brasser2002_terrestrialplanettrojans}
{Brasser}, R., \& {Lehto}, H.~J. 2002, \mnras, 334, 241

\bibitem[{{Brown} {et~al.}(2007){Brown}, {Barkume}, {Ragozzine}, \&
  {Schaller}}]{brown2007_haumeafamily}
{Brown}, M.~E., {Barkume}, K.~M., {Ragozzine}, D., \& {Schaller}, E.~L. 2007,
  \nat, 446, 294

\bibitem[{{Brown} {et~al.}(2015){Brown}, {Bannister}, {Schmidt}, {Drake},
  {Djorgovski}, {Graham}, {Mahabal}, {Donalek}, {Larson}, \&
  {Christensen}}]{brown2015_catalina}
{Brown}, M.~E., {Bannister}, M.~T., {Schmidt}, B.~P., {et~al.} 2015, \aj, 149,
  69

\bibitem[{{Buie} \& {Keller}(2016)}]{buie2016_recon}
{Buie}, M.~W., \& {Keller}, J.~M. 2016, \aj, 151, 73

\bibitem[{{Bus} \& {Binzel}(2002)}]{bus2002_smasstaxonomy}
{Bus}, S.~J., \& {Binzel}, R.~P. 2002, \icarus, 158, 146

\bibitem[{{Bus} {et~al.}(1988){Bus}, {Bowell}, \& {French}}]{bus1988_chiron}
{Bus}, S.~J., {Bowell}, E., \& {French}, L.~M. 1988, \iaucirc, 4684

\bibitem[{{Carvano} \& {Davalos}(2015)}]{carvano2015_phaseangletaxonomy}
{Carvano}, J.~M., \& {Davalos}, J.~A.~G. 2015, \aap, 580, A98

\bibitem[{{Carvano} {et~al.}(2010){Carvano}, {Hasselmann}, {Lazzaro}, \&
  {Moth{\'e}-Diniz}}]{carvano2010_sdsstaxonomy}
{Carvano}, J.~M., {Hasselmann}, P.~H., {Lazzaro}, D., \& {Moth{\'e}-Diniz}, T.
  2010, \aap, 510, A43

\bibitem[{{Chandler} {et~al.}(2018){Chandler}, {Curtis}, {Mommert}, {Sheppard},
  \& {Trujillo}}]{chandler2018_safari}
{Chandler}, C.~O., {Curtis}, A.~M., {Mommert}, M., {Sheppard}, S.~S., \&
  {Trujillo}, C.~A. 2018, Publications of the Astronomical Society of the
  Pacific, 130, 114502

\bibitem[{{Chernitsov} {et~al.}(2017){Chernitsov}, {Tamarov}, \&
  {Barannikov}}]{chernitsov2017_orbitconfidenceregions}
{Chernitsov}, A.~M., {Tamarov}, V.~A., \& {Barannikov}, Y.~A. 2017, Solar
  System Research, 51, 400

\bibitem[{{Cibulkov{\'a}} {et~al.}(2018){Cibulkov{\'a}}, {Nortunen}, {{\v
  D}urech}, {Kaasalainen}, {Vere{\v s}}, {Jedicke}, {Wainscoat}, {Mommert},
  {Trilling}, {Schunov{\'a}-Lilly}, {Magnier}, {Waters}, \&
  {Flewelling}}]{cibulkova2018_mbashapes}
{Cibulkov{\'a}}, H., {Nortunen}, H., {{\v D}urech}, J., {et~al.} 2018, \aap,
  611, A86

\bibitem[{{Cikota} {et~al.}(2014){Cikota}, {Ortiz}, {Cikota}, {Morales}, \&
  {Tancredi}}]{cikota2014_activeasts}
{Cikota}, S., {Ortiz}, J.~L., {Cikota}, A., {Morales}, N., \& {Tancredi}, G.
  2014, \aap, 562, A94

\bibitem[{{DeMeo} {et~al.}(2014){DeMeo}, {Binzel}, \&
  {Lockhart}}]{demeo2014_marsresurfacing}
{DeMeo}, F.~E., {Binzel}, R.~P., \& {Lockhart}, M. 2014, \icarus, 227, 112

\bibitem[{{DeMeo} {et~al.}(2009){DeMeo}, {Binzel}, {Slivan}, \&
  {Bus}}]{demeo2009_taxonomy}
{DeMeo}, F.~E., {Binzel}, R.~P., {Slivan}, S.~M., \& {Bus}, S.~J. 2009,
  \icarus, 202, 160

\bibitem[{Denneau {et~al.}(2013)Denneau, Jedicke, Grav, Granvik, Kubica,
  Milani, Vere{\v{s}}, Wainscoat, Chang, Pierfederici, Kaiser, Chambers,
  Heasley, Magnier, Price, Myers, Kleyna, Hsieh, Farnocchia, Waters, Sweeney,
  Green, Bolin, Burgett, Morgan, Tonry, Hodapp, Chastel, Chesley, Fitzsimmons,
  Holman, Spahr, Tholen, Williams, Abe, Armstrong, Bressi, Holmes, Lister,
  McMillan, Micheli, Ryan, Ryan, \& Scotti}]{Denneau2013_panstarrsmops}
Denneau, L., Jedicke, R., Grav, T., {et~al.} 2013, Publications of the
  Astronomical Society of the Pacific, 125, 357

\bibitem[{{Denneau} {et~al.}(2015){Denneau}, {Jedicke}, {Fitzsimmons}, {Hsieh},
  {Kleyna}, {Granvik}, {Micheli}, {Spahr}, {Vere{\v s}}, {Wainscoat},
  {Burgett}, {Chambers}, {Draper}, {Flewelling}, {Huber}, {Kaiser}, {Morgan},
  \& {Tonry}}]{denneau2015_asteroiddisruptions}
{Denneau}, L., {Jedicke}, R., {Fitzsimmons}, A., {et~al.} 2015, \icarus, 245, 1

\bibitem[{{Desmars} {et~al.}(2015){Desmars}, {Camargo}, {Braga-Ribas},
  {Vieira-Martins}, {Assafin}, {Vachier}, {Colas}, {Ortiz}, {Duffard},
  {Morales}, {Sicardy}, {Gomes-J{\'u}nior}, \&
  {Benedetti-Rossi}}]{desmars2015_occultationprediction}
{Desmars}, J., {Camargo}, J.~I.~B., {Braga-Ribas}, F., {et~al.} 2015, \aap,
  584, A96

\bibitem[{{Desmars} {et~al.}(2019){Desmars}, {Meza}, {Sicardy}, {Assafin},
  {Camargo}, {Braga-Ribas}, {Benedetti-Rossi}, {Dias-Oliveira}, {Morgado}, \&
  {Gomes-J{\'u}nior}}]{desmars2019_plutooccultations}
{Desmars}, J., {Meza}, E., {Sicardy}, B., {et~al.} 2019, \aap, 625, A43

\bibitem[{{Durech} {et~al.}(2009){Durech}, {Kaasalainen}, {Warner},
  {Fauerbach}, {Marks}, {Fauvaud}, {Fauvaud}, {Vugnon}, {Pilcher},
  {Bernasconi}, \& {Behrend}}]{durech2009_sparsephotometry}
{Durech}, J., {Kaasalainen}, M., {Warner}, B.~D., {et~al.} 2009, \aap, 493, 291

\bibitem[{{Dybczy{\'n}ski} \&
  {Kr{\'o}likowska}(2016)}]{dybczynski2016_lpcorigin}
{Dybczy{\'n}ski}, P.~A., \& {Kr{\'o}likowska}, M. 2016, \planss, 123, 77

\bibitem[{{Eisner} {et~al.}(2017){Eisner}, {Knight}, \&
  {Schleicher}}]{eisner17}
{Eisner}, N., {Knight}, M.~M., \& {Schleicher}, D.~G. 2017, \aj, 154, 196

\bibitem[{{Farnocchia} {et~al.}(2015){Farnocchia}, {Chesley}, \&
  {Micheli}}]{farnocchia2015_ranging}
{Farnocchia}, D., {Chesley}, S.~R., \& {Micheli}, M. 2015, \icarus, 258, 18

\bibitem[{{Fedorets} {et~al.}(2019){Fedorets}, {Granvik}, {Jones}, {Juri\'{c}},
  \& {Jedicke}}]{fedorets2019}
{Fedorets}, G., {Granvik}, M., {Jones}, R.~L., {Juri\'{c}}, M., \& {Jedicke},
  R. 2019, \icarus, submitted.

\bibitem[{{Feierberg} {et~al.}(1985){Feierberg}, {Lebofsky}, \&
  {Tholen}}]{feierberg1985_3micronfeature}
{Feierberg}, M.~A., {Lebofsky}, L.~A., \& {Tholen}, D.~J. 1985, \icarus, 63,
  183

\bibitem[{{Finson} \& {Probstein}(1968)}]{finson1968_cometdustmodeling1}
{Finson}, M.~J., \& {Probstein}, R.~F. 1968, \apj, 154, 327

\bibitem[{{Fraser} {et~al.}(2013){Fraser}, {Gwyn}, {Trujillo}, {Stephens},
  {Kavelaars}, {Brown}, {Bianco}, {Boyle}, {Brucker}, {Hetherington}, {Joner},
  {Keel}, {Langill}, {Lister}, {McMillan}, \&
  {Young}}]{fraser2013_kboccultations}
{Fraser}, W.~C., {Gwyn}, S., {Trujillo}, C., {et~al.} 2013, \pasp, 125, 1000

\bibitem[{{Gil-Hutton}(2006)}]{gilhutton2006_highifamilies}
{Gil-Hutton}, R. 2006, \icarus, 183, 93

\bibitem[{{Gilbert} \& {Wiegert}(2009)}]{gilbert2009_cfhtmbcs}
{Gilbert}, A.~M., \& {Wiegert}, P.~A. 2009, \icarus, 201, 714

\bibitem[{{Gilbert} \& {Wiegert}(2010)}]{gilbert2010_cfhtmbcs}
---. 2010, \icarus, 210, 998

\bibitem[{{Gladman} \& {Kavelaars}(1997)}]{gladman1997_kbsearches}
{Gladman}, B., \& {Kavelaars}, J.~J. 1997, \aap, 317, L35

\bibitem[{{Gladman} {et~al.}(1998){Gladman}, {Kavelaars}, {Nicholson},
  {Loredo}, \& {Burns}}]{gladman1998_pencilbeamtnosurvey}
{Gladman}, B., {Kavelaars}, J.~J., {Nicholson}, P.~D., {Loredo}, T.~J., \&
  {Burns}, J.~A. 1998, \aj, 116, 2042

\bibitem[{{Gladman} {et~al.}(2008){Gladman}, {Marsden}, \&
  {Vanlaerhoven}}]{gladman2008_ossnomenclature}
{Gladman}, B., {Marsden}, B.~G., \& {Vanlaerhoven}, C. 2008, {Nomenclature in
  the Outer Solar System}, ed. M.~A. {Barucci}, H.~{Boehnhardt}, D.~P.
  {Cruikshank}, A.~{Morbidelli}, \& R.~{Dotson}, 43

\bibitem[{{Gladman} {et~al.}(2012){Gladman}, {Lawler}, {Petit}, {Kavelaars},
  {Jones}, {Parker}, {Van Laerhoven}, {Nicholson}, {Rousselot}, {Bieryla}, \&
  {Ashby}}]{gladman2012_resonanttnos}
{Gladman}, B., {Lawler}, S.~M., {Petit}, J.-M., {et~al.} 2012, \aj, 144, 23

\bibitem[{{Granvik} \& {Muinonen}(2005)}]{granvik2005}
{Granvik}, M., \& {Muinonen}, K. 2005, \icarus, 179, 109

\bibitem[{{Granvik} \& {Muinonen}(2008)}]{granvik2008}
---. 2008, \icarus, 198, 130

\bibitem[{{Granvik} {et~al.}(2009){Granvik}, {Virtanen}, {Oszkiewicz}, \&
  {Muinonen}}]{granvik2009}
{Granvik}, M., {Virtanen}, J., {Oszkiewicz}, D., \& {Muinonen}, K. 2009,
  Meteoritics and Planetary Science, 44, 1853

\bibitem[{{Grav} {et~al.}(2011){Grav}, {Jedicke}, {Denneau}, {Chesley},
  {Holman}, \& {Spahr}}]{grav2011_solarsystemmodel}
{Grav}, T., {Jedicke}, R., {Denneau}, L., {et~al.} 2011, \pasp, 123, 423

\bibitem[{{Hanu{\v{s}}} \& {{\v{D}}urech}(2012)}]{hanus2012_sparsephotometry}
{Hanu{\v{s}}}, J., \& {{\v{D}}urech}, J. 2012, Planetary and Space Science, 73,
  75

\bibitem[{{Hapke}(1981)}]{hapke1981_bidirectionalreflspec1_jgr}
{Hapke}, B. 1981, \jgr, 86, 3039

\bibitem[{{Hapke}(1984)}]{hapke1984_bidirectionalreflspec3}
---. 1984, \icarus, 59, 41

\bibitem[{{Hapke}(1986)}]{hapke1986_bidirectionalreflspec4}
---. 1986, \icarus, 67, 264

\bibitem[{{Hapke}(2002)}]{hapke2002_bidirectionalreflspec5}
---. 2002, \icarus, 157, 523

\bibitem[{{Hapke} \& {Wells}(1981)}]{hapke1981_bidirectionalreflspec2}
{Hapke}, B., \& {Wells}, E. 1981, \jgr, 86, 3055

\bibitem[{{Hartmann} {et~al.}(1990){Hartmann}, {Tholen}, {Meech}, \&
  {Cruikshank}}]{hartmann1990_chiron}
{Hartmann}, W.~K., {Tholen}, D.~J., {Meech}, K.~J., \& {Cruikshank}, D.~P.
  1990, \icarus, 83, 1

\bibitem[{{Heinze} {et~al.}(2015){Heinze}, {Metchev}, \&
  {Trollo}}]{heinze2015_digitaltracking}
{Heinze}, A.~N., {Metchev}, S., \& {Trollo}, J. 2015, \aj, 150, 125

\bibitem[{{Hirabayashi} {et~al.}(2015){Hirabayashi}, {S{\'a}nchez}, \&
  {Scheeres}}]{hirabayashi2015_rotationalshedding}
{Hirabayashi}, M., {S{\'a}nchez}, D.~P., \& {Scheeres}, D.~J. 2015, \apj, 808,
  63

\bibitem[{{Hirayama}(1918)}]{hirayama1918_astfam}
{Hirayama}, K. 1918, \aj, 31, 185

\bibitem[{{Hofmann} {et~al.}(2017){Hofmann}, {Sierks}, \&
  {Blum}}]{hofmann2017_impacttriggeredlandslides}
{Hofmann}, M., {Sierks}, H., \& {Blum}, J. 2017, \mnras, 469, S73

\bibitem[{{Holman} {et~al.}(2018){Holman}, {Payne}, {Blankley}, {Janssen}, \&
  {Kuindersma}}]{Holman2018_heliolinc}
{Holman}, M.~J., {Payne}, M.~J., {Blankley}, P., {Janssen}, R., \&
  {Kuindersma}, S. 2018, \aj, 156, 135

\bibitem[{Holman {et~al.}(2018)Holman, Payne, Fraser, Lacerda, Bannister,
  Lackner, Chen, Lin, Smith, Kokotanekova, Young, Chambers, Chastel, Denneau,
  Fitzsimmons, Flewelling, Grav, Huber, Induni, Kudritzki, Krolewski, Jedicke,
  Kaiser, Lilly, Magnier, Mark, Meech, Micheli, Murray, Parker, Protopapas,
  Ragozzine, Veres, Wainscoat, Waters, \& Weryk}]{Holman:2018_dwarf}
Holman, M.~J., Payne, M.~J., Fraser, W., {et~al.} 2018, The Astrophysical
  Journal Letters, 855, 0

\bibitem[{{Hsieh}(2009)}]{hsieh2009_htp}
{Hsieh}, H.~H. 2009, \aap, 505, 1297

\bibitem[{{Hsieh} \& {Jewitt}(2006)}]{hsieh2006_mbcs}
{Hsieh}, H.~H., \& {Jewitt}, D. 2006, Science, 312, 561

\bibitem[{{Hsieh} {et~al.}(2016){Hsieh}, {Schwamb}, {Zhang}, {Chen}, {Wang}, \&
  {Lintott}}]{hsieh2016_comethunters}
{Hsieh}, H.~H., {Schwamb}, M.~E., {Zhang}, Z.-W., {et~al.} 2016, in
  AAS/Division for Planetary Sciences Meeting Abstracts, Vol.~48, 406.03

\bibitem[{{Hsieh} \& {Sheppard}(2015)}]{hsieh2015_324p}
{Hsieh}, H.~H., \& {Sheppard}, S.~S. 2015, \mnras, 454, L81

\bibitem[{{Hsieh} {et~al.}(2012){Hsieh}, {Yang}, {Haghighipour}, {Kaluna},
  {Fitzsimmons}, {Denneau}, {Novakovi{\'c}}, {Jedicke}, {Wainscoat},
  {Armstrong}, {Duddy}, {Lowry}, {Trujillo}, {Micheli}, {Keane}, {Urban},
  {Riesen}, {Meech}, {Abe}, {Cheng}, {Chen}, {Granvik}, {Grav}, {Ip},
  {Kinoshita}, {Kleyna}, {Lacerda}, {Lister}, {Milani}, {Tholen}, {Vere{\v s}},
  {Lisse}, {Kelley}, {Fern{\'a}ndez}, {Bhatt}, {Sahu}, {Kaiser}, {Chambers},
  {Hodapp}, {Magnier}, {Price}, \& {Tonry}}]{hsieh2012_288p}
{Hsieh}, H.~H., {Yang}, B., {Haghighipour}, N., {et~al.} 2012, \apjl, 748, L15

\bibitem[{{Hsieh} {et~al.}(2015){Hsieh}, {Denneau}, {Wainscoat},
  {Sch{\"o}rghofer}, {Bolin}, {Fitzsimmons}, {Jedicke}, {Kleyna}, {Micheli},
  {Vere{\v s}}, {Kaiser}, {Chambers}, {Burgett}, {Flewelling}, {Hodapp},
  {Magnier}, {Morgan}, {Price}, {Tonry}, \& {Waters}}]{hsieh2015_ps1mbcs}
{Hsieh}, H.~H., {Denneau}, L., {Wainscoat}, R.~J., {et~al.} 2015, \icarus, 248,
  289

\bibitem[{{Hui} \& {Jewitt}(2017)}]{hui2017_activeastsnongrav}
{Hui}, M.-T., \& {Jewitt}, D. 2017, \aj, 153, 80

\bibitem[{{Ishiguro} {et~al.}(2011){Ishiguro}, {Hanayama}, {Hasegawa},
  {Sarugaku}, {Watanabe}, {Fujiwara}, {Terada}, {Hsieh}, {Vaubaillon}, {Kawai},
  {Yanagisawa}, {Kuroda}, {Miyaji}, {Fukushima}, {Ohta}, {Hamanowa}, {Kim},
  {Pyo}, \& {Nakamura}}]{ishiguro2011_scheila2}
{Ishiguro}, M., {Hanayama}, H., {Hasegawa}, S., {et~al.} 2011, \apjl, 741, L24

\bibitem[{{Ishiguro} {et~al.}(2016){Ishiguro}, {Kuroda}, {Hanayama}, {Kwon},
  {Kim}, {Lee}, {Watanabe}, {Akitaya}, {Kawabata}, {Itoh}, {Nakaoka},
  {Yoshida}, {Imai}, {Sarugaku}, {Yanagisawa}, {Ohta}, {Kawai}, {Miyaji},
  {Fukushima}, {Honda}, {Takahashi}, {Sato}, {Vaubaillon}, \&
  {Watanabe}}]{ishiguro16}
{Ishiguro}, M., {Kuroda}, D., {Hanayama}, H., {et~al.} 2016, \aj, 152, 169

\bibitem[{{Ivezi{\'c}} {et~al.}(2008){Ivezi{\'c}}, {Kahn}, {Tyson}, {Abel},
  {Acosta}, {Allsman}, {Alonso}, {AlSayyad}, {Anderson}, {Andrew}, \&
  et~al.}]{ivezic2008_lsst}
{Ivezi{\'c}}, {\v Z}., {Kahn}, S.~M., {Tyson}, J.~A., {et~al.} 2008, ArXiv
  e-prints, arXiv:0805.2366

\bibitem[{{Jedicke} {et~al.}(2016){Jedicke}, {Bolin}, {Granvik}, \&
  {Beshore}}]{Jedicke2016}
{Jedicke}, R., {Bolin}, B., {Granvik}, M., \& {Beshore}, E. 2016, \icarus, 266,
  173

\bibitem[{{Jedicke} {et~al.}(2018){Jedicke}, {Bolin}, {Bottke}, {Chyba},
  {Fedorets}, {Granvik}, {Jones}, \& {Urrutxua}}]{jedicke2018_minimoons}
{Jedicke}, R., {Bolin}, B.~T., {Bottke}, W.~F., {et~al.} 2018, Frontiers in
  Astronomy and Space Sciences, 5, 13

\bibitem[{{Jewitt}(2009)}]{jewitt2009_actvcentaurs}
{Jewitt}, D. 2009, \aj, 137, 4296

\bibitem[{{Jewitt} {et~al.}(2015){Jewitt}, {Hsieh}, \&
  {Agarwal}}]{jewitt2015_actvasts_ast4}
{Jewitt}, D., {Hsieh}, H., \& {Agarwal}, J. 2015, Asteroids IV (Tucson,
  University of Arizona Press), 221--241

\bibitem[{{Jewitt} {et~al.}(2010){Jewitt}, {Weaver}, {Agarwal}, {Mutchler}, \&
  {Drahus}}]{jewitt2010_p2010a2}
{Jewitt}, D., {Weaver}, H., {Agarwal}, J., {Mutchler}, M., \& {Drahus}, M.
  2010, \nat, 467, 817

\bibitem[{{Jewitt} {et~al.}(2011){Jewitt}, {Weaver}, {Mutchler}, {Larson}, \&
  {Agarwal}}]{jewitt2011_scheila}
{Jewitt}, D., {Weaver}, H., {Mutchler}, M., {Larson}, S., \& {Agarwal}, J.
  2011, \apjl, 733, L4

\bibitem[{Jones {et~al.}(2018)Jones, Slater, Moeyens, Allen, Axelrod, Cook,
  Ivezi{\'{c}}, Juri{\'{c}}, Myers, \& Petry}]{Jones2018_discoverymachine}
Jones, R.~L., Slater, C.~T., Moeyens, J., {et~al.} 2018, Icarus, 303, 181

\bibitem[{{Jones} {et~al.}(1990){Jones}, {Lebofsky}, {Lewis}, \&
  {Marley}}]{jones1990_cpdasteroids}
{Jones}, T.~D., {Lebofsky}, L.~A., {Lewis}, J.~S., \& {Marley}, M.~S. 1990,
  \icarus, 88, 172

\bibitem[{{Juri{\'c}} {et~al.}(2019){Juri{\'c}}, {Jones}, {Bryce Kalmbach},
  {Whidden}, {Bekte{\v{s}}evi{\'c}}, {Smotherman}, {Moeyens}, {Connolly},
  {Bannister}, {Fraser}, {Gerdes}, {Mommert}, {Ragozzine}, {Schwamb}, \&
  {Trilling}}]{juric2019_lsstdeepOSSOsearches}
{Juri{\'c}}, M., {Jones}, R.~L., {Bryce Kalmbach}, J., {et~al.} 2019, arXiv
  e-prints, arXiv:1901.08549

\bibitem[{{Kaasalainen}(2004)}]{kaasalainen2004_sparsephotometricmodels}
{Kaasalainen}, M. 2004, \aap, 422, L39

\bibitem[{{Kaasalainen} \&
  {Torppa}(2001)}]{kaasalainen2001_lightcurveinversion1}
{Kaasalainen}, M., \& {Torppa}, J. 2001, \icarus, 153, 24

\bibitem[{{Kaasalainen} {et~al.}(2001){Kaasalainen}, {Torppa}, \&
  {Muinonen}}]{kaasalainen2001_lightcurveinversion2}
{Kaasalainen}, M., {Torppa}, J., \& {Muinonen}, K. 2001, \icarus, 153, 37

\bibitem[{{Kammer} {et~al.}(2018){Kammer}, {Becker}, {Retherford}, {Stern},
  {Olkin}, {Buie}, {Spencer}, {Bosh}, \&
  {Wasserman}}]{kammer2018_MU69occultation}
{Kammer}, J.~A., {Becker}, T.~M., {Retherford}, K.~D., {et~al.} 2018, \aj, 156,
  72

\bibitem[{{Kne{\v z}evi{\'c}} \& {Milani}(2000)}]{knezevic2000_synthelements}
{Kne{\v z}evi{\'c}}, Z., \& {Milani}, A. 2000, Celestial Mechanics and
  Dynamical Astronomy, 78, 17

\bibitem[{{Kne{\v{z}}evi{\'c}} {et~al.}(2014){Kne{\v{z}}evi{\'c}}, {Milani},
  {Cellino}, {Novakovi{\'c}}, {Spoto}, \&
  {Paolicchi}}]{knezevic2014_automatedfamilyclassification}
{Kne{\v{z}}evi{\'c}}, Z., {Milani}, A., {Cellino}, A., {et~al.} 2014, in IAU
  Symposium, Vol. 310, Complex Planetary Systems, Proceedings of the
  International Astronomical Union, 130--133

\bibitem[{{Knight} {et~al.}(2011){Knight}, {Farnham}, {Schleicher}, \&
  {Schwieterman}}]{knight2011_tempel2}
{Knight}, M.~M., {Farnham}, T.~L., {Schleicher}, D.~G., \& {Schwieterman},
  E.~W. 2011, \aj, 141, 2

\bibitem[{{Knight} {et~al.}(2017){Knight}, {Snodgrass}, {Vincent}, {Conn},
  {Skiff}, {Schleicher}, \& {Lister}}]{knight17}
{Knight}, M.~M., {Snodgrass}, C., {Vincent}, J.-B., {et~al.} 2017, \mnras, 469,
  S661

\bibitem[{{Kramer} {et~al.}(2017){Kramer}, {Bauer}, {Fernandez}, {Stevenson},
  {Mainzer}, {Grav}, {Masiero}, {Nugent}, \& {Sonnett}}]{kramer2017_c2010l5}
{Kramer}, E.~A., {Bauer}, J.~M., {Fernandez}, Y.~R., {et~al.} 2017, \apj, 838,
  58

\bibitem[{{Kr{\'o}likowska}(2004)}]{krolikowska2004_lpcnongraveffects}
{Kr{\'o}likowska}, M. 2004, \aap, 427, 1117

\bibitem[{{Kr{\'o}likowska}(2006)}]{krolikowska2006_lpcnongraveffects}
---. 2006, \actaa, 56, 385

\bibitem[{{Kr{\'o}likowska}(2014)}]{krolikowska2014_warsawcatalogue}
---. 2014, \aap, 567, A126

\bibitem[{{Kr{\'o}likowska} {et~al.}(2014){Kr{\'o}likowska}, {Sitarski},
  {Pittich}, {Szutowicz}, {Zio{\l}kowski}, {Rickman}, {Gabryszewski}, \&
  {Rickman}}]{krolikowska2014_oortspikecomets}
{Kr{\'o}likowska}, M., {Sitarski}, G., {Pittich}, E.~M., {et~al.} 2014, \aap,
  571, A63

\bibitem[{{Kr{\'o}likowska} \& {Szutowicz}(2006)}]{krolikowska2006_81p}
{Kr{\'o}likowska}, M., \& {Szutowicz}, S. 2006, \aap, 448, 401

\bibitem[{Kubica {et~al.}(2007)Kubica, Denneau, Grav, Heasley, Jedicke,
  Masiero, Milani, Moore, Tholen, \& Wainscoat}]{Kubica2007_kdtreelinking}
Kubica, J., Denneau, L., Grav, T., {et~al.} 2007, Icarus, 189, 151

\bibitem[{{Lacerda}(2008)}]{lacerda2008_contactbinaries}
{Lacerda}, P. 2008, \apj, 672, L57

\bibitem[{Larsen {et~al.}(2007)Larsen, Roe, Albert, Descour, McMillan, Gleason,
  Jedicke, Block, Bressi, Cochran, Gehrels, Montani, Perry, Read, Scotti, \&
  Tubbiolo}]{larsen2007_spacewatch}
Larsen, J.~A., Roe, E.~S., Albert, C.~E., {et~al.} 2007, \aj, 133, 1247

\bibitem[{{Lawler} {et~al.}(2018){Lawler}, {Kavelaars}, {Alexandersen},
  {Bannister}, {Gladman}, {Petit}, \& {Shankman}}]{lawler2018_simulator}
{Lawler}, S.~M., {Kavelaars}, J.~J., {Alexandersen}, M., {et~al.} 2018,
  Frontiers in Astronomy and Space Sciences, 5, 14

\bibitem[{{Leiva} {et~al.}(2017){Leiva}, {Sicardy}, {Camargo}, {Ortiz},
  {Desmars}, {B{\'e}rard}, {Lellouch}, {Meza}, {Kervella}, {Snodgrass},
  {Duffard}, {Morales}, {Gomes-J{\'u}nior}, {Benedetti-Rossi},
  {Vieira-Martins}, {Braga-Ribas}, {Assafin}, {Morgado}, {Colas}, {De Witt},
  {Sickafoose}, {Breytenbach}, {Dauvergne}, {Schoenau}, {Maquet}, {Bath},
  {Bode}, {Cool}, {Lade}, {Kerr}, \& {Herald}}]{leiva2017_chariklo}
{Leiva}, R., {Sicardy}, B., {Camargo}, J.~I.~B., {et~al.} 2017, \aj, 154, 159

\bibitem[{{Levison} {et~al.}(2008){Levison}, {Morbidelli}, {Van Laerhoven},
  {Gomes}, \& {Tsiganis}}]{levison2008_kborigin}
{Levison}, H.~F., {Morbidelli}, A., {Van Laerhoven}, C., {Gomes}, R., \&
  {Tsiganis}, K. 2008, \icarus, 196, 258

\bibitem[{{Li} {et~al.}(2017){Li}, {Kelley}, {Samarasinha}, {Farnocchia},
  {Mutchler}, {Ren}, {Lu}, {Tholen}, {Lister}, \& {Micheli}}]{li2017_252p}
{Li}, J.-Y., {Kelley}, M.~S.~P., {Samarasinha}, N.~H., {et~al.} 2017, \aj, 154,
  136

\bibitem[{{Lintott} {et~al.}(2011){Lintott}, {Schawinski}, {Bamford}, {Slosar},
  {Land}, {Thomas}, {Edmondson}, {Masters}, {Nichol}, {Raddick}, {Szalay},
  {Andreescu}, {Murray}, \& {Vandenberg}}]{lintott2011_galaxyzoo}
{Lintott}, C., {Schawinski}, K., {Bamford}, S., {et~al.} 2011, \mnras, 410, 166

\bibitem[{{Malhotra}(1995)}]{malhotra1995_plutoorigin}
{Malhotra}, R. 1995, \aj, 110, 420

\bibitem[{{Malhotra}(2019)}]{malhotra2019_earthtrojans}
---. 2019, Nature Astronomy, 3, 193

\bibitem[{{Mann} {et~al.}(2007){Mann}, {Jewitt}, \&
  {Lacerda}}]{mann2007_contactbinarytrojans}
{Mann}, R.~K., {Jewitt}, D., \& {Lacerda}, P. 2007, \aj, 134, 1133

\bibitem[{{Marcus} {et~al.}(2011){Marcus}, {Ragozzine}, {Murray-Clay}, \&
  {Holman}}]{Marcus2011_kbofamilies}
{Marcus}, R.~A., {Ragozzine}, D., {Murray-Clay}, R.~A., \& {Holman}, M.~J.
  2011, \apj, 733, 40

\bibitem[{{Marzari} \& {Scholl}(2013)}]{marzari2013_earthtrojanstability}
{Marzari}, F., \& {Scholl}, H. 2013, Celestial Mechanics and Dynamical
  Astronomy, 117, 91

\bibitem[{{Masiero} {et~al.}(2009){Masiero}, {Jedicke}, {{\v D}urech}, {Gwyn},
  {Denneau}, \& {Larsen}}]{masiero2009_talcs}
{Masiero}, J., {Jedicke}, R., {{\v D}urech}, J., {et~al.} 2009, \icarus, 204,
  145

\bibitem[{{Masiero} {et~al.}(2013){Masiero}, {Mainzer}, {Bauer}, {Grav},
  {Nugent}, \& {Stevenson}}]{masiero2013_astfams_neowise}
{Masiero}, J.~R., {Mainzer}, A.~K., {Bauer}, J.~M., {et~al.} 2013, \apj, 770, 7

\bibitem[{{McLoughlin} {et~al.}(2015){McLoughlin}, {Fitzsimmons}, \&
  {McLoughlin}}]{mcloughlin2015_outburstphotometry}
{McLoughlin}, E., {Fitzsimmons}, A., \& {McLoughlin}, A. 2015, \icarus, 256, 37

\bibitem[{{McNeill} {et~al.}(2016){McNeill}, {Fitzsimmons}, {Jedicke},
  {Wainscoat}, {Denneau}, {Vere{\v s}}, {Magnier}, {Chambers}, {Kaiser}, \&
  {Waters}}]{mcneill2016_mbabrightnessvariations}
{McNeill}, A., {Fitzsimmons}, A., {Jedicke}, R., {et~al.} 2016, \mnras, 459,
  2964

\bibitem[{{Meech} \& {Belton}(1989)}]{meech1989_chiron}
{Meech}, K.~J., \& {Belton}, M.~J.~S. 1989, \iaucirc, 4770

\bibitem[{{Micheli} {et~al.}(2012){Micheli}, {Tholen}, \&
  {Elliott}}]{Micheli2012}
{Micheli}, M., {Tholen}, D.~J., \& {Elliott}, G.~T. 2012, New Astronomy, 17,
  446

\bibitem[{{Micheli} {et~al.}(2018){Micheli}, {Farnocchia}, {Meech}, {Buie},
  {Hainaut}, {Prialnik}, {Sch{\"o}rghofer}, {Weaver}, {Chodas}, \&
  {Kleyna}}]{micheli2018_oumuamua}
{Micheli}, M., {Farnocchia}, D., {Meech}, K.~J., {et~al.} 2018, \nat, 559, 223

\bibitem[{{Milani} {et~al.}(2014){Milani}, {Cellino}, {Kne{\v z}evi{\'c}},
  {Novakovi{\'c}}, {Spoto}, \& {Paolicchi}}]{milani2014_astfamilies}
{Milani}, A., {Cellino}, A., {Kne{\v z}evi{\'c}}, Z., {et~al.} 2014, \icarus,
  239, 46

\bibitem[{{Milani} \& {Knezevic}(1994)}]{milani1994_properelements}
{Milani}, A., \& {Knezevic}, Z. 1994, \icarus, 107, 219

\bibitem[{{Millis} \& {Schleicher}(1986)}]{millis1986_halleyrotation}
{Millis}, R.~L., \& {Schleicher}, D.~G. 1986, \nat, 324, 646

\bibitem[{{Mommert} {et~al.}(2018){Mommert}, {McNeill}, {Trilling},
  {Moskovitz}, \& {Delbo{\textquoteright}}}]{mommert2018_gaiaasteroidshapes}
{Mommert}, M., {McNeill}, A., {Trilling}, D.~E., {Moskovitz}, N., \&
  {Delbo{\textquoteright}}, M. 2018, \aj, 156, 139

\bibitem[{{Muinonen} {et~al.}(2010){Muinonen}, {Belskaya}, {Cellino},
  {Delb{\`o}}, {Levasseur-Regourd}, {Penttil{\"a}}, \&
  {Tedesco}}]{muinonen2010_threeparamphasefunction}
{Muinonen}, K., {Belskaya}, I.~N., {Cellino}, A., {et~al.} 2010, \icarus, 209,
  542

\bibitem[{{Murray-Clay} \& {Chiang}(2005)}]{murrayclay2005_neptuneresonance}
{Murray-Clay}, R.~A., \& {Chiang}, E.~I. 2005, \apj, 619, 623

\bibitem[{{Narayan} {et~al.}(2018){Narayan}, {Zaidi}, {Soraisam}, {Wang},
  {Lochner}, {Matheson}, {Saha}, {Yang}, {Zhao}, {Kececioglu}, {Scheidegger},
  {Snodgrass}, {Axelrod}, {Jenness}, {Maier}, {Ridgway}, {Seaman}, {Evans},
  {Singh}, {Taylor}, {Toeniskoetter}, {Welch}, {Zhu}, \& {The ANTARES
  Collaboration}}]{narayan2018_alertbrokers}
{Narayan}, G., {Zaidi}, T., {Soraisam}, M.~D., {et~al.} 2018, The Astrophysical
  Journal Supplement Series, 236, 9

\bibitem[{{Nesvorn{\'y}} {et~al.}(2008){Nesvorn{\'y}}, {Bottke},
  {Vokrouhlick{\'y}}, {Sykes}, {Lien}, \& {Stansberry}}]{nesvorny2008_beagle}
{Nesvorn{\'y}}, D., {Bottke}, W.~F., {Vokrouhlick{\'y}}, D., {et~al.} 2008,
  \apjl, 679, L143

\bibitem[{{Nesvorn{\'y}} {et~al.}(2002){Nesvorn{\'y}}, {Bottke}, {Dones}, \&
  {Levison}}]{nesvorny2002_karin}
{Nesvorn{\'y}}, D., {Bottke}, Jr., W.~F., {Dones}, L., \& {Levison}, H.~F.
  2002, \nat, 417, 720

\bibitem[{{Nesvorn{\'y}} {et~al.}(2015){Nesvorn{\'y}}, {Bro{\v z}}, \&
  {Carruba}}]{nesvorny2015_astfam_ast4}
{Nesvorn{\'y}}, D., {Bro{\v z}}, M., \& {Carruba}, V. 2015, Asteroids IV
  (Tucson, University of Arizona Press), 297--321

\bibitem[{{Nesvorn{\'y}} {et~al.}(2006){Nesvorn{\'y}}, {Enke}, {Bottke},
  {Durda}, {Asphaug}, \& {Richardson}}]{nesvorny2006_karin}
{Nesvorn{\'y}}, D., {Enke}, B.~L., {Bottke}, W.~F., {et~al.} 2006, \icarus,
  183, 296

\bibitem[{{Nesvorn{\'y}} \&
  {Vokrouhlick{\'y}}(2016)}]{nesvorny2016_neptunemigration}
{Nesvorn{\'y}}, D., \& {Vokrouhlick{\'y}}, D. 2016, \apj, 825, 94

\bibitem[{{Nortunen} \&
  {Kaasalainen}(2017)}]{nortunen2017_asteroidshapeelongation}
{Nortunen}, H., \& {Kaasalainen}, M. 2017, \aap, 608, A91

\bibitem[{{Nortunen} {et~al.}(2017){Nortunen}, {Kaasalainen}, {{\v D}urech},
  {Cibulkov{\'a}}, {Ali-Lagoa}, \& {Hanu{\v
  s}}}]{nortunen2017_asteroidshapespins}
{Nortunen}, H., {Kaasalainen}, M., {{\v D}urech}, J., {et~al.} 2017, \aap, 601,
  A139

\bibitem[{{Olsen} {et~al.}(2017){Olsen}, {Nidever}, {Fitzpatrick}, {Mighell},
  {SMASH Collaboration}, \& {NOAO Data Lab Team}}]{olsen2017_datalab}
{Olsen}, K.~A., {Nidever}, D.~L., {Fitzpatrick}, M.~J., {et~al.} 2017, in
  American Astronomical Society Meeting Abstracts, Vol. 229, American
  Astronomical Society Meeting Abstracts \#229, 154.25

\bibitem[{{Olsen} {et~al.}(2018){Olsen}, {Walker}, {Smith}, \& {NOAO Data Lab
  Team}}]{olsen2018_datalab_outreach}
{Olsen}, K.~A., {Walker}, C.~E., {Smith}, B., \& {NOAO Data Lab Team}. 2018, in
  American Astronomical Society Meeting Abstracts, Vol. 231, American
  Astronomical Society Meeting Abstracts \#231, 113.08

\bibitem[{{Ortiz} {et~al.}(2015){Ortiz}, {Duffard}, {Pinilla-Alonso},
  {Alvarez-Cand al}, {Santos-Sanz}, {Morales}, {Fern{\'a}ndez-Valenzuela},
  {Licandro}, {Campo Bagatin}, \& {Thirouin}}]{ortiz2015_chironrings}
{Ortiz}, J.~L., {Duffard}, R., {Pinilla-Alonso}, N., {et~al.} 2015, \aap, 576,
  A18

\bibitem[{{Ortiz} {et~al.}(2017){Ortiz}, {Santos-Sanz}, {Sicardy},
  {Benedetti-Rossi}, {B{\'e}rard}, {Morales}, {Duffard}, {Braga-Ribas}, {Hopp},
  \& {Ries}}]{ortiz2017_haumeaoccultation}
{Ortiz}, J.~L., {Santos-Sanz}, P., {Sicardy}, B., {et~al.} 2017, \nat, 550, 219

\bibitem[{{Orton} {et~al.}(1995){Orton}, {A'Hearn}, {Baines}, {Deming},
  {Dowling}, {Goguen}, {Griffith}, {Hammel}, {Hoffmann}, {Hunten}, {Jewitt},
  {Kostiuk}, {Miller}, {Noll}, {Zahnle}, {Achilleos}, {Dayal}, {Deutsch},
  {Espenak}, {Esterle}, {Friedson}, {Fast}, {Harrington}, {Hora}, {Joseph},
  {Kelly}, {Knacke}, {Lacy}, {Lisse}, {Rayner}, {Sprague}, {Shure}, {Wells},
  {Yanamandra-Fisher}, {Zipoy}, {Bjoraker}, {Buhl}, {Golisch}, {Griep},
  {Kaminski}, {Arden}, {Chaikin}, {Goldstein}, {Gilmore}, {Fazio}, {Kanamori},
  {Lam}, {Livengood}, {MacLow}, {Marley}, {Momary}, {Robertson}, {Romani},
  {Spitale}, {Sykes}, {Tennyson}, {Wellnitz}, \& {Ying}}]{orton1995_sl9irtf}
{Orton}, G., {A'Hearn}, M., {Baines}, K., {et~al.} 1995, Science, 267, 1277

\bibitem[{{Pravec} {et~al.}(2005){Pravec}, {Harris}, {Scheirich}, {Ku{\v
  s}nir{\'a}k}, {{\v S}arounov{\'a}}, {Hergenrother}, {Mottola}, {Hicks},
  {Masi}, {Krugly}, {Shevchenko}, {Nolan}, {Howell}, {Kaasalainen},
  {Gal{\'a}d}, {Brown}, {DeGraff}, {Lambert}, {Cooney}, \&
  {Foglia}}]{pravec2005_tumblingasteroids}
{Pravec}, P., {Harris}, A.~W., {Scheirich}, P., {et~al.} 2005, \icarus, 173,
  108

\bibitem[{{Pravec} {et~al.}(2006){Pravec}, {Scheirich}, {Ku{\v s}nir{\'a}k},
  {{\v S}arounov{\'a}}, {Mottola}, {Hahn}, {Brown}, {Esquerdo}, {Kaiser},
  {Krzeminski}, {Pray}, {Warner}, {Harris}, {Nolan}, {Howell}, {Benner},
  {Margot}, {Gal{\'a}d}, {Holliday}, {Hicks}, {Krugly}, {Tholen}, {Whiteley},
  {Marchis}, {DeGraff}, {Grauer}, {Larson}, {Velichko}, {Cooney}, {Stephens},
  {Zhu}, {Kirsch}, {Dyvig}, {Snyder}, {Reddy}, {Moore}, {Gajdo{\v s}},
  {Vil{\'a}gi}, {Masi}, {Higgins}, {Funkhouser}, {Knight}, {Slivan}, {Behrend},
  {Grenon}, {Burki}, {Roy}, {Demeautis}, {Matter}, {Waelchli}, {Revaz},
  {Klotz}, {Rieugn{\'e}}, {Thierry}, {Cotrez}, {Brunetto}, \&
  {Kober}}]{pravec2006_binarysurvey}
{Pravec}, P., {Scheirich}, P., {Ku{\v s}nir{\'a}k}, P., {et~al.} 2006, \icarus,
  181, 63

\bibitem[{Rabinowitz {et~al.}(2012)Rabinowitz, Schwamb, Hadjiyska, \&
  Tourtellotte}]{rabinowitz2012_lasilla}
Rabinowitz, D., Schwamb, M.~E., Hadjiyska, E., \& Tourtellotte, S. 2012, \aj,
  144, 140

\bibitem[{{Radovi{\'c}}(2017)}]{Radovic2017b}
{Radovi{\'c}}, V. 2017, \mnras, 471, 1321

\bibitem[{{Radovi{\'c}} {et~al.}(2017){Radovi{\'c}}, {Novakovi{\'c}},
  {Carruba}, \& {Mar{\v c}eta}}]{radovic2017_interlopers}
{Radovi{\'c}}, V., {Novakovi{\'c}}, B., {Carruba}, V., \& {Mar{\v c}eta}, D.
  2017, \mnras, 470, 576

\bibitem[{{Rivkin}(2012)}]{rivkin2012_hydratedasts}
{Rivkin}, A.~S. 2012, \icarus, 221, 744

\bibitem[{{Rivkin} {et~al.}(2003){Rivkin}, {Davies}, {Johnson}, {Ellison},
  {Trilling}, {Brown}, \& {Lebofsky}}]{rivkin2003_ctypeasteroids}
{Rivkin}, A.~S., {Davies}, J.~K., {Johnson}, J.~R., {et~al.} 2003, Meteoritics
  and Planetary Science, 38, 1383

\bibitem[{{Rosaev} \& {Pl{\'a}valov{\'a}}(2017)}]{rosaev2017_hobson}
{Rosaev}, A., \& {Pl{\'a}valov{\'a}}, E. 2017, \icarus, 282, 326

\bibitem[{{Rubincam}(2000)}]{rubincam2000_yorp}
{Rubincam}, D.~P. 2000, \icarus, 148, 2

\bibitem[{{Saha} {et~al.}(2014){Saha}, {Matheson}, {Snodgrass}, {Kececioglu},
  {Narayan}, {Seaman}, {Jenness}, \& {Axelrod}}]{saha2014_antares}
{Saha}, A., {Matheson}, T., {Snodgrass}, R., {et~al.} 2014, in Observatory
  Operations: Strategies, Processes, and Systems V, Vol. 9149, 914908

\bibitem[{{Saha} {et~al.}(2016){Saha}, {Wang}, {Matheson}, {Narayan},
  {Snodgrass}, {Kececioglu}, {Scheidegger}, {Axelrod}, {Jenness}, {Ridgway},
  {Seaman}, {Taylor}, {Toeniskoetter}, {Welch}, {Yang}, \&
  {Zaidi}}]{saha2016_antares}
{Saha}, A., {Wang}, Z., {Matheson}, T., {et~al.} 2016, in \procspie, Vol. 9910,
  Observatory Operations: Strategies, Processes, and Systems VI, 99100F

\bibitem[{{Samarasinha} \& {Mueller}(2015)}]{samarasinha2015_nparotation}
{Samarasinha}, N.~H., \& {Mueller}, B.~E.~A. 2015, \icarus, 248, 347

\bibitem[{{Samarasinha} {et~al.}(2011){Samarasinha}, {Mueller}, {A'Hearn},
  {Farnham}, \& {Gersch}}]{samarasinha2011_103p}
{Samarasinha}, N.~H., {Mueller}, B.~E.~A., {A'Hearn}, M.~F., {Farnham}, T.~L.,
  \& {Gersch}, A. 2011, \apjl, 734, L3

\bibitem[{{S{\'a}nchez-Lavega} {et~al.}(2010){S{\'a}nchez-Lavega}, {Wesley},
  {Orton}, {Hueso}, {Perez-Hoyos}, {Fletcher}, {Yanamand ra-Fisher},
  {Legarreta}, {de Pater}, {Hammel}, {Simon-Miller}, {Gomez-Forrellad},
  {Ortiz}, {Garc{\'\i}a-Melendo}, {Puetter}, \&
  {Chodas}}]{sanchezlavega2010_jupiterimpact}
{S{\'a}nchez-Lavega}, A., {Wesley}, A., {Orton}, G., {et~al.} 2010, \apj, 715,
  L155

\bibitem[{{Scheirich} \& {Pravec}(2009)}]{scheirich2009_binarylightcures}
{Scheirich}, P., \& {Pravec}, P. 2009, \icarus, 200, 531

\bibitem[{{Scheirich} {et~al.}(2015){Scheirich}, {Pravec}, {Jacobson},
  {{\v{D}}urech}, {Ku{\v{s}}nir{\'a}k}, {Hornoch}, {Mottola}, {Mommert},
  {Hellmich}, {Pray}, {Polishook}, {Krugly}, {Inasaridze}, {Kvaratskhelia},
  {Ayvazian}, {Slyusarev}, {Pittichov{\'a}}, {Jehin}, {Manfroid}, {Gillon},
  {Gal{\'a}d}, {Pollock}, {Licandro}, {Al{\'\i}-Lagoa}, {Brinsfield}, \&
  {Molotov}}]{scheirich2015_binaryNEO}
{Scheirich}, P., {Pravec}, P., {Jacobson}, S.~A., {et~al.} 2015, \icarus, 245,
  56

\bibitem[{{Schleicher} {et~al.}(2019){Schleicher}, {Knight}, {Eisner}, \&
  {Thirouin}}]{schleicher2019_41protation}
{Schleicher}, D.~G., {Knight}, M.~M., {Eisner}, N.~L., \& {Thirouin}, A. 2019,
  \aj, 157, 108

\bibitem[{{Scholl} {et~al.}(2005){Scholl}, {Marzari}, \&
  {Tricarico}}]{scholl2005_marstrojans}
{Scholl}, H., {Marzari}, F., \& {Tricarico}, P. 2005, \icarus, 175, 397

\bibitem[{{Schwamb} {et~al.}(2010){Schwamb}, {Brown}, {Rabinowitz}, \&
  {Ragozzine}}]{schwamb2010_palomar}
{Schwamb}, M.~E., {Brown}, M.~E., {Rabinowitz}, D.~L., \& {Ragozzine}, D. 2010,
  \apj, 720, 1691

\bibitem[{{Schwamb} {et~al.}(2017){Schwamb}, {Hsieh}, {Zhang}, {Chen},
  {Lintott}, {Wang}, \& {Mishra}}]{schwamb2017_comethunters}
{Schwamb}, M.~E., {Hsieh}, H.~H., {Zhang}, Z.-W., {et~al.} 2017, in American
  Astronomical Society Meeting Abstracts, Vol. 229, American Astronomical
  Society Meeting Abstracts \#229, 112.04

\bibitem[{{Schwamb} {et~al.}(2018){Schwamb}, {Jones}, {Chesley}, {Fitzsimmons},
  {Fraser}, {Holman}, {Hsieh}, {Ragozzine}, {Thomas}, {Trilling}, {Brown},
  {Bannister}, {Bodewits}, {de Val-Borro}, {Gerdes}, {Granvik}, {Kelley},
  {Knight}, {Seaman}, {Ye}, \& {Young}}]{schwamb2018_ssroadmap}
{Schwamb}, M.~E., {Jones}, R.~L., {Chesley}, S.~R., {et~al.} 2018, ArXiv
  e-prints, arXiv:1802.01783

\bibitem[{{Sheppard} \& {Jewitt}(2002)}]{sheppard2002_kbophotometry}
{Sheppard}, S.~S., \& {Jewitt}, D.~C. 2002, \aj, 124, 1757

\bibitem[{{Siltala} \& {Granvik}(2017)}]{siltala2017_asteroidmassestimation}
{Siltala}, L., \& {Granvik}, M. 2017, \icarus, 297, 149

\bibitem[{{Smith} {et~al.}(2019){Smith}, {Williams}, {Young}, {Ibsen},
  {Smartt}, {Lawrence}, {Morris}, {Voutsinas}, \& {Nicholl}}]{smith2019_lasair}
{Smith}, K.~W., {Williams}, R.~D., {Young}, D.~R., {et~al.} 2019, Research
  Notes of the American Astronomical Society, 3, 26

\bibitem[{{Solontoi} {et~al.}(2010){Solontoi}, {Ivezi{\'c}}, {West}, {Claire},
  {Juri{\'c}}, {Becker}, {Jones}, {Hall}, {Kent}, {Lupton}, {Knapp}, {Quinn},
  {Gunn}, {Schneider}, \& {Loomis}}]{solontoi2010_sdsscomets}
{Solontoi}, M., {Ivezi{\'c}}, {\v Z}., {West}, A.~A., {et~al.} 2010, \icarus,
  205, 605

\bibitem[{{Sonnett} {et~al.}(2011){Sonnett}, {Kleyna}, {Jedicke}, \&
  {Masiero}}]{sonnett2011_cfhtmbcs}
{Sonnett}, S., {Kleyna}, J., {Jedicke}, R., \& {Masiero}, J. 2011, \icarus,
  215, 534

\bibitem[{{Sonnett} {et~al.}(2015){Sonnett}, {Mainzer}, {Grav}, {Masiero}, \&
  {Bauer}}]{sonnett2015_binaries}
{Sonnett}, S., {Mainzer}, A., {Grav}, T., {Masiero}, J., \& {Bauer}, J. 2015,
  \apj, 799, 191

\bibitem[{{Spoto} {et~al.}(2015){Spoto}, {Milani}, \& {Kne{\v
  z}evi{\'c}}}]{Spoto2015}
{Spoto}, F., {Milani}, A., \& {Kne{\v z}evi{\'c}}, Z. 2015, \icarus, 257, 275

\bibitem[{{Steckloff} {et~al.}(2016){Steckloff}, {Graves}, {Hirabayashi},
  {Melosh}, \& {Richardson}}]{steckloff2016_hartley2avalanches}
{Steckloff}, J.~K., {Graves}, K., {Hirabayashi}, M., {Melosh}, H.~J., \&
  {Richardson}, J.~E. 2016, \icarus, 272, 60

\bibitem[{{Stellingwerf}(1978)}]{stellingwerf1978_pdm}
{Stellingwerf}, R.~F. 1978, \apj, 224, 953

\bibitem[{{Szab{\'o}} {et~al.}(2016){Szab{\'o}}, {P{\'a}l}, {S{\'a}rneczky},
  {Szab{\'o}}, {Moln{\'a}r}, {Kiss}, {Hanyecz}, {Plachy}, \&
  {Kiss}}]{szabo2016_keplerasteroidlightcurves}
{Szab{\'o}}, R., {P{\'a}l}, A., {S{\'a}rneczky}, K., {et~al.} 2016, \aap, 596,
  A40

\bibitem[{{The Gaia Collaboration} {et~al.}(2016){The Gaia Collaboration},
  {Prusti}, {de Bruijne}, {Brown}, {Vallenari}, {Babusiaux}, {Bailer-Jones},
  {Bastian}, {Biermann}, {Evans}, {Eyer}, {Jansen}, {Jordi}, {Klioner},
  {Lammers}, {Lindegren}, {Luri}, {Mignard}, {Milligan}, {Panem}, {Poinsignon},
  {Pourbaix}, {Randich}, {Sarri}, {Sartoretti}, {Siddiqui}, {Soubiran},
  {Valette}, {van Leeuwen}, {Walton}, {Aerts}, {Arenou}, {Cropper}, {Drimmel},
  {H{\o}g}, {Katz}, {Lattanzi}, {O'Mullane}, {Grebel}, {Holland}, {Huc},
  {Passot}, {Bramante}, {Cacciari}, {Casta{\~n}eda}, {Chaoul}, {Cheek}, {De
  Angeli}, {Fabricius}, {Guerra}, {Hern{\'a}ndez}, {Jean-Antoine-Piccolo},
  {Masana}, {Messineo}, {Mowlavi}, {Nienartowicz}, {Ord{\'o}{\~n}ez-Blanco},
  {Panuzzo}, {Portell}, {Richards}, {Riello}, {Seabroke}, {Tanga},
  {Th{\'e}venin}, {Torra}, {Els}, {Gracia-Abril}, {Comoretto},
  {Garcia-Reinaldos}, {Lock}, {Mercier}, {Altmann}, {Andrae}, {Astraatmadja},
  {Bellas-Velidis}, {Benson}, {Berthier}, {Blomme}, {Busso}, {Carry},
  {Cellino}, {Clementini}, {Cowell}, {Creevey}, {Cuypers}, {Davidson}, {De
  Ridder}, {de Torres}, {Delchambre}, {Dell'Oro}, {Ducourant}, {Fr{\'e}mat},
  {Garc{\'\i}a-Torres}, {Gosset}, {Halbwachs}, {Hambly}, {Harrison}, {Hauser},
  {Hestroffer}, {Hodgkin}, {Huckle}, {Hutton}, {Jasniewicz}, {Jordan},
  {Kontizas}, {Korn}, {Lanzafame}, {Manteiga}, {Moitinho}, {Muinonen},
  {Osinde}, {Pancino}, {Pauwels}, {Petit}, {Recio-Blanco}, {Robin}, {Sarro},
  {Siopis}, {Smith}, {Smith}, {Sozzetti}, {Thuillot}, {van Reeven}, {Viala},
  {Abbas}, {Abreu Aramburu}, {Accart}, {Aguado}, {Allan}, {Allasia},
  {Altavilla}, {{\'A}lvarez}, {Alves}, {Anderson}, {Andrei}, {Anglada Varela},
  {Antiche}, {Antoja}, {Ant{\'o}n}, {Arcay}, {Atzei}, {Ayache}, {Bach},
  {Baker}, {Balaguer-N{\'u}{\~n}ez}, {Barache}, {Barata}, {Barbier}, {Barblan},
  {Baroni}, {Barrado y Navascu{\'e}s}, {Barros}, {Barstow}, {Becciani},
  {Bellazzini}, {Bellei}, {Bello Garc{\'\i}a}, {Belokurov}, {Bendjoya},
  {Berihuete}, {Bianchi}, {Bienaym{\'e}}, {Billebaud}, {Blagorodnova},
  {Blanco-Cuaresma}, {Boch}, {Bombrun}, {Borrachero}, {Bouquillon}, {Bourda},
  {Bouy}, {Bragaglia}, {Breddels}, {Brouillet}, {Br{\"u}semeister},
  {Bucciarelli}, {Budnik}, {Burgess}, {Burgon}, {Burlacu}, {Busonero}, {Buzzi},
  {Caffau}, {Cambras}, {Campbell}, {Cancelliere}, {Cantat-Gaudin}, {Carlucci},
  {Carrasco}, {Castellani}, {Charlot}, {Charnas}, {Charvet}, {Chassat},
  {Chiavassa}, {Clotet}, {Cocozza}, {Collins}, {Collins}, {Costigan}, {Crifo},
  {Cross}, {Crosta}, {Crowley}, {Dafonte}, {Damerdji}, {Dapergolas}, {David},
  {David}, {De Cat}, {de Felice}, {de Laverny}, {De Luise}, {De March}, {de
  Martino}, {de Souza}, {Debosscher}, {del Pozo}, {Delbo}, {Delgado},
  {Delgado}, {di Marco}, {Di Matteo}, {Diakite}, {Distefano}, {Dolding}, {Dos
  Anjos}, {Drazinos}, {Dur{\'a}n}, {Dzigan}, {Ecale}, {Edvardsson}, {Enke},
  {Erdmann}, {Escolar}, {Espina}, {Evans}, {Eynard Bontemps}, {Fabre},
  {Fabrizio}, {Faigler}, {Falc{\~a}o}, {Farr{\`a}s Casas}, {Faye}, {Federici},
  {Fedorets}, {Fern{\'a}ndez-Hern{\'a}ndez}, {Fernique}, {Fienga}, {Figueras},
  {Filippi}, {Findeisen}, {Fonti}, {Fouesneau}, {Fraile}, {Fraser}, {Fuchs},
  {Furnell}, {Gai}, {Galleti}, {Galluccio}, {Garabato}, {Garc{\'\i}a-Sedano},
  {Gar{\'e}}, {Garofalo}, {Garralda}, {Gavras}, {Gerssen}, {Geyer}, {Gilmore},
  {Girona}, {Giuffrida}, {Gomes}, {Gonz{\'a}lez-Marcos},
  {Gonz{\'a}lez-N{\'u}{\~n}ez}, {Gonz{\'a}lez-Vidal}, {Granvik}, {Guerrier},
  {Guillout}, {Guiraud}, {G{\'u}rpide}, {Guti{\'e}rrez-S{\'a}nchez}, {Guy},
  {Haigron}, {Hatzidimitriou}, {Haywood}, {Heiter}, {Helmi}, {Hobbs},
  {Hofmann}, {Holl}, {Holland}, {Hunt}, {Hypki}, {Icardi}, {Irwin}, {Jevardat
  de Fombelle}, {Jofr{\'e}}, {Jonker}, {Jorissen}, {Julbe}, {Karampelas},
  {Kochoska}, {Kohley}, {Kolenberg}, {Kontizas}, {Koposov}, {Kordopatis},
  {Koubsky}, {Kowalczyk}, {Krone-Martins}, {Kudryashova}, {Kull}, {Bachchan},
  {Lacoste-Seris}, {Lanza}, {Lavigne}, {Le Poncin-Lafitte}, {Lebreton},
  {Lebzelter}, {Leccia}, {Leclerc}, {Lecoeur-Taibi}, {Lemaitre}, {Lenhardt},
  {Leroux}, {Liao}, {Licata}, {Lindstr{\o}m}, {Lister}, {Livanou}, {Lobel},
  {L{\"o}ffler}, {L{\'o}pez}, {Lopez-Lozano}, {Lorenz}, {Loureiro},
  {MacDonald}, {Magalh{\~a}es Fernandes}, {Managau}, {Mann}, {Mantelet},
  {Marchal}, {Marchant}, {Marconi}, {Marie}, {Marinoni}, {Marrese},
  {Marschalk{\'o}}, {Marshall}, {Mart{\'\i}n-Fleitas}, {Martino}, {Mary},
  {Matijevi{\v{c}}}, {Mazeh}, {McMillan}, {Messina}, {Mestre}, {Michalik},
  {Millar}, {Miranda}, {Molina}, {Molinaro}, {Molinaro}, {Moln{\'a}r},
  {Moniez}, {Montegriffo}, {Monteiro}, {Mor}, {Mora}, {Morbidelli}, {Morel},
  {Morgenthaler}, {Morley}, {Morris}, {Mulone}, {Muraveva}, {Musella},
  {Narbonne}, {Nelemans}, {Nicastro}, {Noval}, {Ord{\'e}novic},
  {Ordieres-Mer{\'e}}, {Osborne}, {Pagani}, {Pagano}, {Pailler}, {Palacin},
  {Palaversa}, {Parsons}, {Paulsen}, {Pecoraro}, {Pedrosa}, {Pentik{\"a}inen},
  {Pereira}, {Pichon}, {Piersimoni}, {Pineau}, {Plachy}, {Plum}, {Poujoulet},
  {Pr{\v{s}}a}, {Pulone}, {Ragaini}, {Rago}, {Rambaux}, {Ramos-Lerate},
  {Ranalli}, {Rauw}, {Read}, {Regibo}, {Renk}, {Reyl{\'e}}, {Ribeiro},
  {Rimoldini}, {Ripepi}, {Riva}, {Rixon}, {Roelens}, {Romero-G{\'o}mez},
  {Rowell}, {Royer}, {Rudolph}, {Ruiz-Dern}, {Sadowski}, {Sagrist{\`a}
  Sell{\'e}s}, {Sahlmann}, {Salgado}, {Salguero}, {Sarasso}, {Savietto},
  {Schnorhk}, {Schultheis}, {Sciacca}, {Segol}, {Segovia}, {Segransan},
  {Serpell}, {Shih}, {Smareglia}, {Smart}, {Smith}, {Solano}, {Solitro},
  {Sordo}, {Soria Nieto}, {Souchay}, {Spagna}, {Spoto}, {Stampa}, {Steele},
  {Steidelm{\"u}ller}, {Stephenson}, {Stoev}, {Suess}, {S{\"u}veges}, {Surdej},
  {Szabados}, {Szegedi-Elek}, {Tapiador}, {Taris}, {Tauran}, {Taylor},
  {Teixeira}, {Terrett}, {Tingley}, {Trager}, {Turon}, {Ulla}, {Utrilla},
  {Valentini}, {van Elteren}, {Van Hemelryck}, {van Leeuwen}, {Varadi},
  {Vecchiato}, {Veljanoski}, {Via}, {Vicente}, {Vogt}, {Voss}, {Votruba},
  {Voutsinas}, {Walmsley}, {Weiler}, {Weingrill}, {Werner}, {Wevers},
  {Whitehead}, {Wyrzykowski}, {Yoldas}, {{\v{Z}}erjal}, {Zucker}, {Zurbach},
  {Zwitter}, {Alecu}, {Allen}, {Allende Prieto}, {Amorim},
  {Anglada-Escud{\'e}}, {Arsenijevic}, {Azaz}, {Balm}, {Beck}, {Bernstein},
  {Bigot}, {Bijaoui}, {Blasco}, {Bonfigli}, {Bono}, {Boudreault}, {Bressan},
  {Brown}, {Brunet}, {Bunclark}, {Buonanno}, {Butkevich}, {Carret}, {Carrion},
  {Chemin}, {Ch{\'e}reau}, {Corcione}, {Darmigny}, {de Boer}, {de Teodoro}, {de
  Zeeuw}, {Delle Luche}, {Domingues}, {Dubath}, {Fodor}, {Fr{\'e}zouls},
  {Fries}, {Fustes}, {Fyfe}, {Gallardo}, {Gallegos}, {Gardiol}, {Gebran},
  {Gomboc}, {G{\'o}mez}, {Grux}, {Gueguen}, {Heyrovsky}, {Hoar}, {Iannicola},
  {Isasi Parache}, {Janotto}, {Joliet}, {Jonckheere}, {Keil}, {Kim},
  {Klagyivik}, {Klar}, {Knude}, {Kochukhov}, {Kolka}, {Kos}, {Kutka}, {Lainey},
  {LeBouquin}, {Liu}, {Loreggia}, {Makarov}, {Marseille}, {Martayan},
  {Martinez-Rubi}, {Massart}, {Meynadier}, {Mignot}, {Munari}, {Nguyen},
  {Nordlander}, {Ocvirk}, {O'Flaherty}, {Olias Sanz}, {Ortiz}, {Osorio},
  {Oszkiewicz}, {Ouzounis}, {Palmer}, {Park}, {Pasquato}, {Peltzer}, {Peralta},
  {P{\'e}turaud}, {Pieniluoma}, {Pigozzi}, {Poels}, {Prat}, {Prod'homme},
  {Raison}, {Rebordao}, {Risquez}, {Rocca-Volmerange}, {Rosen}, {Ruiz-Fuertes},
  {Russo}, {Sembay}, {Serraller Vizcaino}, {Short}, {Siebert}, {Silva},
  {Sinachopoulos}, {Slezak}, {Soffel}, {Sosnowska}, {Strai{\v{z}}ys}, {ter
  Linden}, {Terrell}, {Theil}, {Tiede}, {Troisi}, {Tsalmantza}, {Tur},
  {Vaccari}, {Vachier}, {Valles}, {Van Hamme}, {Veltz}, {Virtanen}, {Wallut},
  {Wichmann}, {Wilkinson}, {Ziaeepour}, \&
  {Zschocke}}]{gaiacollaboration2016_gaia}
{The Gaia Collaboration}, {Prusti}, T., {de Bruijne}, J.~H.~J., {et~al.} 2016,
  \aap, 595, A1

\bibitem[{{The Gaia Collaboration} {et~al.}(2018){The Gaia Collaboration},
  {Brown}, {Vallenari}, {Prusti}, {de Bruijne}, {Babusiaux}, {Bailer-Jones},
  {Biermann}, {Evans}, {Eyer}, {Jansen}, {Jordi}, {Klioner}, {Lammers},
  {Lindegren}, {Luri}, {Mignard}, {Panem}, {Pourbaix}, {Randich}, {Sartoretti},
  {Siddiqui}, {Soubiran}, {van Leeuwen}, {Walton}, {Arenou}, {Bastian},
  {Cropper}, {Drimmel}, {Katz}, {Lattanzi}, {Bakker}, {Cacciari},
  {Casta{\~n}eda}, {Chaoul}, {Cheek}, {De Angeli}, {Fabricius}, {Guerra},
  {Holl}, {Masana}, {Messineo}, {Mowlavi}, {Nienartowicz}, {Panuzzo},
  {Portell}, {Riello}, {Seabroke}, {Tanga}, {Th{\'e}venin}, {Gracia-Abril},
  {Comoretto}, {Garcia-Reinaldos}, {Teyssier}, {Altmann}, {Andrae}, {Audard},
  {Bellas-Velidis}, {Benson}, {Berthier}, {Blomme}, {Burgess}, {Busso},
  {Carry}, {Cellino}, {Clementini}, {Clotet}, {Creevey}, {Davidson}, {De
  Ridder}, {Delchambre}, {Dell'Oro}, {Ducourant},
  {Fern{\'a}ndez-Hern{\'a}ndez}, {Fouesneau}, {Fr{\'e}mat}, {Galluccio},
  {Garc{\'\i}a-Torres}, {Gonz{\'a}lez-N{\'u}{\~n}ez}, {Gonz{\'a}lez-Vidal},
  {Gosset}, {Guy}, {Halbwachs}, {Hambly}, {Harrison}, {Hern{\'a}ndez},
  {Hestroffer}, {Hodgkin}, {Hutton}, {Jasniewicz}, {Jean-Antoine-Piccolo},
  {Jordan}, {Korn}, {Krone-Martins}, {Lanzafame}, {Lebzelter}, {L{\"o}ffler},
  {Manteiga}, {Marrese}, {Mart{\'\i}n-Fleitas}, {Moitinho}, {Mora}, {Muinonen},
  {Osinde}, {Pancino}, {Pauwels}, {Petit}, {Recio-Blanco}, {Richards},
  {Rimoldini}, {Robin}, {Sarro}, {Siopis}, {Smith}, {Sozzetti}, {S{\"u}veges},
  {Torra}, {van Reeven}, {Abbas}, {Abreu Aramburu}, {Accart}, {Aerts},
  {Altavilla}, {{\'A}lvarez}, {Alvarez}, {Alves}, {Anderson}, {Andrei},
  {Anglada Varela}, {Antiche}, {Antoja}, {Arcay}, {Astraatmadja}, {Bach},
  {Baker}, {Balaguer-N{\'u}{\~n}ez}, {Balm}, {Barache}, {Barata}, {Barbato},
  {Barblan}, {Barklem}, {Barrado}, {Barros}, {Barstow}, {Bartholom{\'e}
  Mu{\~n}oz}, {Bassilana}, {Becciani}, {Bellazzini}, {Berihuete}, {Bertone},
  {Bianchi}, {Bienaym{\'e}}, {Blanco-Cuaresma}, {Boch}, {Boeche}, {Bombrun},
  {Borrachero}, {Bossini}, {Bouquillon}, {Bourda}, {Bragaglia}, {Bramante},
  {Breddels}, {Bressan}, {Brouillet}, {Br{\"u}semeister}, {Brugaletta},
  {Bucciarelli}, {Burlacu}, {Busonero}, {Butkevich}, {Buzzi}, {Caffau},
  {Cancelliere}, {Cannizzaro}, {Cantat-Gaudin}, {Carballo}, {Carlucci},
  {Carrasco}, {Casamiquela}, {Castellani}, {Castro-Ginard}, {Charlot},
  {Chemin}, {Chiavassa}, {Cocozza}, {Costigan}, {Cowell}, {Crifo}, {Crosta},
  {Crowley}, {Cuypers}, {Dafonte}, {Damerdji}, {Dapergolas}, {David}, {David},
  {de Laverny}, {De Luise}, {De March}, {de Martino}, {de Souza}, {de Torres},
  {Debosscher}, {del Pozo}, {Delbo}, {Delgado}, {Delgado}, {Di Matteo},
  {Diakite}, {Diener}, {Distefano}, {Dolding}, {Drazinos}, {Dur{\'a}n},
  {Edvardsson}, {Enke}, {Eriksson}, {Esquej}, {Eynard Bontemps}, {Fabre},
  {Fabrizio}, {Faigler}, {Falc{\~a}o}, {Farr{\`a}s Casas}, {Federici},
  {Fedorets}, {Fernique}, {Figueras}, {Filippi}, {Findeisen}, {Fonti},
  {Fraile}, {Fraser}, {Fr{\'e}zouls}, {Gai}, {Galleti}, {Garabato},
  {Garc{\'\i}a-Sedano}, {Garofalo}, {Garralda}, {Gavel}, {Gavras}, {Gerssen},
  {Geyer}, {Giacobbe}, {Gilmore}, {Girona}, {Giuffrida}, {Glass}, {Gomes},
  {Granvik}, {Gueguen}, {Guerrier}, {Guiraud}, {Guti{\'e}rrez-S{\'a}nchez},
  {Haigron}, {Hatzidimitriou}, {Hauser}, {Haywood}, {Heiter}, {Helmi}, {Heu},
  {Hilger}, {Hobbs}, {Hofmann}, {Holland}, {Huckle}, {Hypki}, {Icardi},
  {Jan{\ss}en}, {Jevardat de Fombelle}, {Jonker}, {Juh{\'a}sz}, {Julbe},
  {Karampelas}, {Kewley}, {Klar}, {Kochoska}, {Kohley}, {Kolenberg},
  {Kontizas}, {Kontizas}, {Koposov}, {Kordopatis}, {Kostrzewa-Rutkowska},
  {Koubsky}, {Lambert}, {Lanza}, {Lasne}, {Lavigne}, {Le Fustec}, {Le
  Poncin-Lafitte}, {Lebreton}, {Leccia}, {Leclerc}, {Lecoeur-Taibi},
  {Lenhardt}, {Leroux}, {Liao}, {Licata}, {Lindstr{\o}m}, {Lister}, {Livanou},
  {Lobel}, {L{\'o}pez}, {Managau}, {Mann}, {Mantelet}, {Marchal}, {Marchant},
  {Marconi}, {Marinoni}, {Marschalk{\'o}}, {Marshall}, {Martino}, {Marton},
  {Mary}, {Massari}, {Matijevi{\v{c}}}, {Mazeh}, {McMillan}, {Messina},
  {Michalik}, {Millar}, {Molina}, {Molinaro}, {Moln{\'a}r}, {Montegriffo},
  {Mor}, {Morbidelli}, {Morel}, {Morris}, {Mulone}, {Muraveva}, {Musella},
  {Nelemans}, {Nicastro}, {Noval}, {O'Mullane}, {Ord{\'e}novic},
  {Ord{\'o}{\~n}ez-Blanco}, {Osborne}, {Pagani}, {Pagano}, {Pailler},
  {Palacin}, {Palaversa}, {Panahi}, {Pawlak}, {Piersimoni}, {Pineau}, {Plachy},
  {Plum}, {Poggio}, {Poujoulet}, {Pr{\v{s}}a}, {Pulone}, {Racero}, {Ragaini},
  {Rambaux}, {Ramos-Lerate}, {Regibo}, {Reyl{\'e}}, {Riclet}, {Ripepi}, {Riva},
  {Rivard}, {Rixon}, {Roegiers}, {Roelens}, {Romero-G{\'o}mez}, {Rowell},
  {Royer}, {Ruiz-Dern}, {Sadowski}, {Sagrist{\`a} Sell{\'e}s}, {Sahlmann},
  {Salgado}, {Salguero}, {Sanna}, {Santana-Ros}, {Sarasso}, {Savietto},
  {Schultheis}, {Sciacca}, {Segol}, {Segovia}, {S{\'e}gransan}, {Shih},
  {Siltala}, {Silva}, {Smart}, {Smith}, {Solano}, {Solitro}, {Sordo}, {Soria
  Nieto}, {Souchay}, {Spagna}, {Spoto}, {Stampa}, {Steele},
  {Steidelm{\"u}ller}, {Stephenson}, {Stoev}, {Suess}, {Surdej}, {Szabados},
  {Szegedi-Elek}, {Tapiador}, {Taris}, {Tauran}, {Taylor}, {Teixeira},
  {Terrett}, {Teyssand ier}, {Thuillot}, {Titarenko}, {Torra Clotet}, {Turon},
  {Ulla}, {Utrilla}, {Uzzi}, {Vaillant}, {Valentini}, {Valette}, {van Elteren},
  {Van Hemelryck}, {van Leeuwen}, {Vaschetto}, {Vecchiato}, {Veljanoski},
  {Viala}, {Vicente}, {Vogt}, {von Essen}, {Voss}, {Votruba}, {Voutsinas},
  {Walmsley}, {Weiler}, {Wertz}, {Wevers}, {Wyrzykowski}, {Yoldas},
  {{\v{Z}}erjal}, {Ziaeepour}, {Zorec}, {Zschocke}, {Zucker}, {Zurbach}, \&
  {Zwitter}}]{gaiacollaboration2018_gaiadr2}
{The Gaia Collaboration}, {Brown}, A.~G.~A., {Vallenari}, A., {et~al.} 2018,
  \aap, 616, A1

\bibitem[{{Thirouin} {et~al.}(2014){Thirouin}, {Noll}, {Ortiz}, \&
  {Morales}}]{thirouin2014_tnobinaries}
{Thirouin}, A., {Noll}, K.~S., {Ortiz}, J.~L., \& {Morales}, N. 2014, \aap,
  569, A3

\bibitem[{{Tholen} {et~al.}(1988){Tholen}, {Hartmann}, {Cruikshank}, {Lilly},
  {Bowell}, \& {Hewitt}}]{tholen1988_chiron}
{Tholen}, D.~J., {Hartmann}, W.~K., {Cruikshank}, D.~P., {et~al.} 1988,
  \iaucirc, 4554, 2

\bibitem[{{{\v D}urech} {et~al.}(2018){{\v D}urech}, {Hanu{\v s}}, \&
  {Al{\'{\i}}-Lagoa}}]{durech2018_asteroidmodels}
{{\v D}urech}, J., {Hanu{\v s}}, J., \& {Al{\'{\i}}-Lagoa}, V. 2018, \aap, 617,
  A57

\bibitem[{{{\v{D}}urech} {et~al.}(2015){{\v{D}}urech}, {Hanu{\v{s}}}, \&
  {Van{\v{c}}o}}]{durech2015_asteroidsathome}
{{\v{D}}urech}, J., {Hanu{\v{s}}}, J., \& {Van{\v{c}}o}, R. 2015, Astronomy and
  Computing, 13, 80

\bibitem[{{Vere{\v{s}}} {et~al.}(2017){Vere{\v{s}}}, {Farnocchia}, {Chesley},
  \& {Chamberlin}}]{veres2017_astrometricerrors}
{Vere{\v{s}}}, P., {Farnocchia}, D., {Chesley}, S.~R., \& {Chamberlin}, A.~B.
  2017, \icarus, 296, 139

\bibitem[{{Vilas}(1994)}]{vilas1994_hydrationfeatures}
{Vilas}, F. 1994, \icarus, 111, 456

\bibitem[{{Vilas} \& {Gaffey}(1989)}]{vilas1989_phyllosilicatefeatures}
{Vilas}, F., \& {Gaffey}, M.~J. 1989, Science, 246, 790

\bibitem[{{Vincent} {et~al.}(2016){Vincent}, {A'Hearn}, {Lin}, {El-Maarry},
  {Pajola}, {Sierks}, {Barbieri}, {Lamy}, {Rodrigo}, {Koschny}, {Rickman},
  {Keller}, {Agarwal}, {Barucci}, {Bertaux}, {Bertini}, {Besse}, {Bodewits},
  {Cremonese}, {Da Deppo}, {Davidsson}, {Debei}, {De Cecco}, {Deller},
  {Fornasier}, {Fulle}, {Gicquel}, {Groussin}, {Guti{\'e}rrez},
  {Guti{\'e}rrez-Marquez}, {G{\"u}ttler}, {H{\"o}fner}, {Hofmann}, {Hviid},
  {Ip}, {Jorda}, {Knollenberg}, {Kovacs}, {Kramm}, {K{\"u}hrt}, {K{\"u}ppers},
  {Lara}, {Lazzarin}, {Lopez Moreno}, {Marzari}, {Massironi}, {Mottola},
  {Naletto}, {Oklay}, {Preusker}, {Scholten}, {Shi}, {Thomas}, {Toth}, \&
  {Tubiana}}]{vincent16}
{Vincent}, J.-B., {A'Hearn}, M.~F., {Lin}, Z.-Y., {et~al.} 2016, \mnras, 462,
  S184

\bibitem[{{Vokrouhlick{\'y}} \&
  {Nesvorn{\'y}}(2008)}]{vokrouhlicky2008_asteroidpairs}
{Vokrouhlick{\'y}}, D., \& {Nesvorn{\'y}}, D. 2008, \aj, 136, 280

\bibitem[{{Volk} {et~al.}(2018){Volk}, {Murray-Clay}, {Gladman}, {Lawler},
  {Yu}, {Alexandersen}, {Bannister}, {Chen}, {Dawson}, {Greenstreet}, {Gwyn},
  {Kavelaars}, {Lin}, {Lykawka}, \& {Petit}}]{volk2018_neptuneresonantobjects}
{Volk}, K., {Murray-Clay}, R.~A., {Gladman}, B.~J., {et~al.} 2018, \aj, 155,
  260

\bibitem[{{Waszczak} {et~al.}(2013){Waszczak}, {Ofek}, {Aharonson}, {Kulkarni},
  {Polishook}, {Bauer}, {Levitan}, {Sesar}, {Laher}, {Surace}, \& {PTF
  Team}}]{waszczak2013_ptfmbcs}
{Waszczak}, A., {Ofek}, E.~O., {Aharonson}, O., {et~al.} 2013, \mnras, 433,
  3115

\bibitem[{{Waszczak} {et~al.}(2015){Waszczak}, {Chang}, {Ofek}, {Laher},
  {Masci}, {Levitan}, {Surace}, {Cheng}, {Ip}, {Kinoshita}, {Helou}, {Prince},
  \& {Kulkarni}}]{waszczak2015_ptflcs}
{Waszczak}, A., {Chang}, C.-K., {Ofek}, E.~O., {et~al.} 2015, \aj, 150, 75

\bibitem[{{Weaver} {et~al.}(1995){Weaver}, {A'Hearn}, {Arpigny}, {Boice},
  {Feldman}, {Larson}, {Lamy}, {Levy}, {Marsden}, {Meech}, {Noll}, {Scotti},
  {Sekanina}, {Shoemaker}, {Shoemaker}, {Smith}, {Stern}, {Storrs}, {Trauger},
  {Yeomans}, \& {Zellner}}]{weaver1995_sl9hubble}
{Weaver}, H.~A., {A'Hearn}, M.~F., {Arpigny}, C., {et~al.} 1995, Science, 267,
  1282

\bibitem[{{Whidden} {et~al.}(2019){Whidden}, {Bryce Kalmbach}, {Connolly},
  {Jones}, {Smotherman}, {Bektesevic}, {Slater}, {Becker}, {Ivezi{\'c}},
  {Juri{\'c}}, {Bolin}, {Moeyens}, {F{\"o}rster}, \&
  {Golkhou}}]{whidden2019_tnosearchalgorithm}
{Whidden}, P.~J., {Bryce Kalmbach}, J., {Connolly}, A.~J., {et~al.} 2019, \aj,
  157, 119

\bibitem[{{Wood} {et~al.}(2018){Wood}, {Horner}, {Hinse}, \&
  {Marsden}}]{wood2018_chironrings}
{Wood}, J., {Horner}, J., {Hinse}, T.~C., \& {Marsden}, S.~C. 2018, \aj, 155, 2

\bibitem[{{Zappala} {et~al.}(1990){Zappala}, {Cellino}, {Farinella}, \&
  {Knezevic}}]{zappala1990_hcm}
{Zappala}, V., {Cellino}, A., {Farinella}, P., \& {Knezevic}, Z. 1990, \aj,
  100, 2030

\bibitem[{{Zappala} {et~al.}(1994){Zappala}, {Cellino}, {Farinella}, \&
  {Milani}}]{zappala1994_hcm}
{Zappala}, V., {Cellino}, A., {Farinella}, P., \& {Milani}, A. 1994, \aj, 107,
  772

\bibitem[{{Zhai} {et~al.}(2014){Zhai}, {Shao}, {Nemati}, {Werne}, {Zhou},
  {Turyshev}, {Sandhu}, {Hallinan}, \& {Harding}}]{zhai2014_synthetictracking}
{Zhai}, C., {Shao}, M., {Nemati}, B., {et~al.} 2014, \apj, 792, 60

\end{thebibliography}


\end{document}